\documentclass[prd, twocolumn,showpacs,floats,floatfix,nofootinbib]{revtex4}

\usepackage{lscape}
\usepackage{lipsum}
\usepackage{calc}
\usepackage{longtable,array}

\usepackage{bm}
\usepackage{latexsym}
\usepackage{amsmath}
\usepackage{amsfonts}
\usepackage{amssymb}
\usepackage{graphicx,epsfig}
\usepackage{psfrag}
\usepackage{subfigure}
\usepackage{ulem}
\usepackage{color}
\usepackage{appendix} 
\usepackage[utf8]{inputenc} 
\usepackage{enumitem}
\usepackage{cancel}
\usepackage{pifont}
\usepackage{color}
\usepackage{xcolor}

\usepackage{hyperref}
\hypersetup{
    unicode=false,          
    pdftoolbar=true,        
    pdfmenubar=true,        
    pdffitwindow=false,     
    pdfstartview={FitH},    
    pdftitle={My title},    
    pdfauthor={Author},     
    pdfsubject={Subject},   
    pdfcreator={Creator},   
    pdfproducer={Producer}, 
    pdfkeywords={keyword1} {key2} {key3}, 
    pdfnewwindow=true,      
    colorlinks=true,       
    linkcolor=red,          
    citecolor=darkblue,        
    filecolor=magenta,      
    urlcolor=darkblue,           
    linktocpage=true
}

\newcommand{\cmark}{\ding{51}}%
\newcommand{\xmark}{\ding{55}}%
\definecolor{blue}{rgb}{0.19,0.64,0.54}
\definecolor{red}{rgb}{0.7,0.3,0.3}
\definecolor{darkblue}{rgb}{0.3,0.40,0.48}
\definecolor{gray}{rgb}{.8,.8,.8}

\newcommand{\ex}{\mathrm{e}}
\newcommand{\dd}{\mathrm{d}}

\newcommand{\lsim}{\lesssim}
\def\spose#1{\hbox to 0pt{#1\hss}}

\def\lta{\mathrel{\spose{\lower 3pt\hbox{$\mathchar"218$}}
     \raise 2.0pt\hbox{$\mathchar"13C$}}}
\def\gta{\mathrel{\spose{\lower 3pt\hbox{$\mathchar"218$}}
     \raise 2.0pt\hbox{$\mathchar"13E$}}}

\def\setN{\mathbb{N}}

\newcommand{\ie}{\textsl{i.e.~}}

\newcommand{\eg}{\textsl{e.g.~}}
\newcommand{\etal}{\textsl{et al.~}}

\newcommand{\etc}{\textsl{etc.~}}

\newcommand{\Hu}{\mathcal{H}}
\newcommand{\Ka}{\mathcal{K}}

\newcommand{\GN}{G_{_\mathrm{N}}}

\newcommand{\nT}{n_{_\mathrm{T}}}

\newcommand{\Mp}{M_{_\mathrm{Pl}}}

\begin{document}

\date{\today}
\title{A Critical Review of Classical Bouncing Cosmologies}

\author{Diana Battefeld} \email[ ]{dbattefeld08@gmail.com}
\affiliation{Institut for Astrophysics, University of Goettingen,\\
 Friedrich-Hund Platz 1, D-37077, Germany}

\author{Patrick Peter} \email[ ]{peter@iap.fr}
\affiliation{Institut d'Astrophysique de Paris, UMR 7095-CNRS,
Universit\'e Pierre et Marie Curie, 98 bis boulevard Arago, 75014
Paris, France}


\begin{abstract}
  Given the proliferation of bouncing models in recent years, we
  gather and critically assess these proposals in a comprehensive
  review.  The P{\footnotesize LANCK} data shows an unmistakably red,
  quasi scale-invariant, purely adiabatic primordial power spectrum
  and no primary non-Gaussianities.  While these observations are
  consistent with inflationary predictions, bouncing cosmologies
  aspire to provide an alternative framework to explain them.  Such
  models face many problems, both of the purely theoretical kind, such
  as the necessity of violating the NEC and instabilities, and at the
  cosmological application level, as exemplified by the possible
  presence of shear. We provide a pedagogical introduction to these
  problems and also assess the fitness of different proposals with
  respect to the data. For example, many models predict a slightly
  blue spectrum and must be fine-tuned to generate a red spectral
  index; as a side effect, large non-Gaussianities often result.

  We highlight several promising attempts to violate the NEC without
  introducing dangerous instabilities at the classical and/or quantum
  level.   If primordial gravitational waves are observed, certain
  bouncing cosmologies, such as the cyclic scenario, are in trouble,
  while others remain valid. We conclude that, while most bouncing
  cosmologies are far from providing an alternative to the
  inflationary paradigm, a handful of interesting proposals have
  surfaced, which warrant further research. The constraints and
  lessons learned as laid out in this review might guide future
  research.
\end{abstract}


\maketitle

\tableofcontents 
\makeatletter



\section{Introduction}

It is often stated nowadays that cosmology has entered a regime of
precision somewhat comparable to particle physics
\cite{Mukhanov:2005sc,PeterUzan2009}. Although probably an
exaggeration, there is a grain of truth in such a statement: based on
the P{\footnotesize LANCK} data\footnote{We also take into
  consideration the B{\footnotesize ICEP2} data
  \protect\cite{Ade:2014xna}, pending independent confirmation.},
researchers have begun not only to discriminate between frameworks,
but also to argue in favor of specific mechanisms. Indeed, with a
purely Gaussian signal and a scalar spectral index strictly less than
unity (at the $5\sigma$ level), but close to scale invariance, and no
isocurvature contribution at any detectable level, it is hard to
imagine any mechanism not based on quantum vacuum fluctuations of a
single effective scalar field. Inflation \cite{Guth:1980zm} thus
appears, contrary to what has been stated
\cite{Ijjas:2013vea,Ijjas:2014nta}, the most fashionable
\cite{Linde:2007fr,Lemoine:2008zz,Baumann:2009ds,Martin:2014vha,Martin:2013nzq,Baumann:2014nda}
and some might say natural \cite{Guth:2013sya,Linde:2014nna} candidate
to explain the data (it has been argued that inflation with its wide
range of possible predictions is unverifiable and thus untenable and
not in the realm of science; if such an argument were valid, the same
could be said about the quantum field theory paradigm, which also
needs a specific implementation to be put to the test experimentally
-- see also \cite{Barrow:1997fp}).

Being the most fashionable candidate, however, does not make inflation
true, and before we can confidently say that a phase of inflation took
place, assuming there will ever be such a time, we need to make sure
that all other possibilities are ruled out. To our knowledge, this is
the case for most of the other proposals to generate large scale
structures, such as seeding fluctuations by a cosmic string network
\cite{Ringeval:2005kr,Peter:2013jj,Ade:2013xla}.  Apart from models of
an emerging universe in string gas cosmology
\cite{Battefeld:2005av,Brandenberger:2014faa}, all viable nonsingular
alternatives to date replace the primordial singularity by a bounce
connecting a contracting phase to the currently expanding one.

We concentrate on pure alternatives to inflation, \ie we do not
consider the otherwise well-justified models in which a bounce is
followed by a phase of inflation
\cite{Piao:2003zm,Falciano:2008gt,Lilley:2011ag,Liu:2013kea,Biswas:2013dry}.
Such models naturally get the best of both ideas and should be
considered in view of addressing the question of the primordial
singularity in the inflationary paradigm.

The purpose of this work is to review models that aim to explain
observations by mechanisms in a bouncing universe and to provide a
critical assessment. We believe it is useful to know not only the
strength but also the weaknesses of a given approach to find possible
cures and to yield a better understanding of these models. If all
possible alternatives turn out to be irreconcilable with the given
data, the inflationary paradigm would not be proven, but our
confidence in it would increase considerably; if the primordial
gravitational wave background level is sufficiently high, as would be
the case if \cite{Ade:2014xna} and its interpretation is confirmed,
there is hope to verify the inflationary consistency relation between
the tensor spectral index $\nT$ and the tensor-to-scalar ratio $r$;
this would put all models featured in this review in difficulty.

\subsection{Why is there a necessity for alternative models to inflation?}

How can we accurately describe the  13.8  billion year
evolution of our Universe?  The standard model of the early Universe
can be traced back to several seminal observations: galaxies are
receding faster the further away they are, indicating an expanding
universe \cite{Hubble:1929ig}; the cosmic microwave background
radiation (CMBR) is highly isotropic and the expansion is accelerating
\cite{Perlmutter:1998np,Riess:1998cb}; this acceleration is attributed
to an unknown component, dark energy.  Big bang cosmology accounts for
the Hubble expansion and predicts the existence of the CMBR.  The
abundance of light elements can be computed, and their values agree
with observations, with the possible exception of $^7$Li (see
\cite{Coc:2014oia} for a recent review). Moreover, numerical
simulations \cite{Teyssier:2008zd}
of large scale structure formation based on what we
believe to be the relevant initial conditions, as deduced from the
properties of the CMBR, reproduce well the observed features of the
actual distribution: at first sight, the standard hot big bang model
successfully provides a description of the Universe back to a fraction
of a second after its birth until today with amazing precision; it is
hard to overestimate the success that such a model represents in a
science that a century ago did not exist.

However successful from a fraction of a second onward, the simple hot
big bang model is plagued by several problems when extrapolated
backwards in time: it begins with an initial singularity leading to a
tiny horizon, without an explanation for the vanishingly small spatial
curvature, it does not explain why baryons should have been formed in
an asymmetric way (with respect to antibaryons), why exotic relics are
absent or how the density fluctuations, from which large-scale
structures developed, are seeded. Most of these problems can be
addressed by postulating an inflationary phase \cite{Guth:1980zm}, \ie
a period of accelerated expansion taking place during the early stages
of our Universe; however, the existence of a primeval singularity is
not modified in the inflationary framework, which remains geodesically
incomplete \cite{Borde:1996pt}.

Originally conceived in order to rid the hot big bang model from the
Grand Unified Theory (GUT) monopole problem, inflation has rapidly
been developed to become a
paradigm of modern theoretical cosmology. The simplest models of
inflation not only solve the horizon and flatness problems, but they
also predict, as an initially unexpected bonus, the statistical
properties of temperature fluctuations in the CMBR, in full agreement
with the most recent observations. However, from a theoretical point
of view, inflation is not free of problems. First, in large field
models of inflation, the inflaton has to traverse a distance in field
space larger than the Planck mass $\Mp$ in natural units. This has
been argued to be problematic, since non-renormalizable quantum
corrections to the field's action arise.  In the absence of functional
fine-tuning or additional symmetries, inflation would be spoiled; this
is known as the $\eta$ problem of inflation. Small field models are
more appealing, but also fine-tuned, for instance to account for the
proper amplitude of the power spectrum. An exhaustive review and
comparison of single field models with the P{\footnotesize LANCK} data
is given in \cite{Martin:2014vha,Martin:2013nzq}.  However, if the
B{\footnotesize ICEP2} detection of gravitational waves were
confirmed, all of these small field models would be in trouble
\cite{Martin:2014lra}. Foreground emission studies using the
B{\footnotesize ICEP1} and B{\footnotesize ICEP2} data suggest that
the background and a gravitational wave signal are indistinguishable
in this region \cite{Flauger:2014qra,Mortonson:2014bja}. See
Sec.\ref{bicep2} for more details and  \cite{Baumann:2014nda} for a
recent review of inflation in string theory post B{\footnotesize
  ICEP2}. Second, the presence of eternal inflation in almost all
proposals has been argued to lead to a possible loss of predictability
due to our inability to prescribe a unique measure
\cite{Ijjas:2013vea,Ijjas:2014nta}: this is the so-called measure
problem (see however \cite{Guth:2013sya,Linde:2014nna}). Third,
inflation does not provide a theory of initial conditions that would
explain why the inflaton field starts out high in its potential. A
related issue is the low initial entropy of the initial state that has
to be assumed, just as in big bang cosmology; this is known as the
entropy problem. Fourth, the initial singularity, as visible in
curvature invariants, does not disappear, but is merely pushed into
the past; this may stem from the strict use of General Relativity
(GR). Some of these problems could have an environmental solution in
terms of the anthropic principle in a wide landscape of otherwise
uninhabitable solutions. See \cite{Linde:2014nna} for a recent
review on these topics; it should be noted that the measure problem
may render a quantification of anthropic arguments challenging.

These problems of inflation have fueled the search for alternatives,
most of which have not passed the CMBR constraints. A seemingly viable
alternative, which also provides a GR-compatible solution to the
singularity problem, relies on a nonsingular bouncing cosmology
\cite{Brandenberger:1993ef,Novello:2008ra}, whereby an initially
contracting phase connects with the currently expanding one through
some minimal scale factor (and hence a vanishing Hubble rate). These
models have a history that predates inflationary solutions by many
decades, as they were proposed shortly after the first observations of
the expansion \cite{Lemaitre:1927zz,Hubble:1929ig} by Tolman
\cite{Tolman31} and Lema\^{\i}tre \cite{Lemaitre33,Lemaitre:1933gd}
(see also \cite{Barrow:1995cfa} for a more modern viewpoint concerning
Tolman's cyclic approach): at this time, the expansion appeared to
imply that Einstein's theory of gravity was doomed to fail, as the
scale factor reaches infinitesimally small values, such that the
Universe emerges from a primordial singularity. This singularity
problem was ignored for many years as interest in cosmology faded
among physicists, until it reemerged in the early 1980s
\cite{Starobinsky78,Starobinsky:1980te}, when GR was again perceived
as not only a mathematically entertaining theory, but also as a
physically relevant description on large scales. Shortly thereafter,
cosmological inflation was proposed \cite{Guth:1980zm}, see
\cite{Bassett:2005xm,Lemoine:2008zz,Baumann:2014nda} for reviews, and
bouncing cosmologies faded again into oblivion as researchers focused
on developing the inflationary framework.

 In parallel, string theorists investigated cosmological solutions
in dilaton gravity, leading to the pre-big bang (PBB) scenario, which was
the first attempt to implement a non singular bounce within this framework;
Ref.~\cite{Gasperini:2002bn} presents a comprehensive review of this
model. The universe
starts out empty and flat, with the dilaton in the weak coupling
regime. As the dilaton evolves towards strong coupling, a transition
from pre- to post big bang was though to occur in the strong coupling
regime, which appears as a bounce in the Jordan frame (but not in the
Einstein frame). While ultimately not a successful model of the early
universe, as detailed in Ref. \cite{Gasperini:2002bn}, the pre-big bang
scenario paved the way for bouncing scenarios, which employ many of
the ideas and techniques of the PBB. Thus, bouncing cosmologies
resurfaced to provide a challenge or merely a working alternative to
the inflationary paradigm, see \eg
\cite{Durrer:1995mz,Peter:2001fy,Khoury:2001wf} among many other
proposals. Over the last years, considerable effort has been made in
developing well-behaved, nonsingular and singular bouncing models.
Our goal is to critically review these new developments. A prior
review \cite{Novello:2008ra} concentrated on quite different
categories of models, while this work aims at discussing more widely
held views.

\subsection{What is used to get a bounce?}


 To achieve a bounce, the Hubble rate $H$, which emerges from the
contracting phase with a negative value, must increase, since it is
positive during the subsequent expanding phase. There are two options
to increase the Hubble rate from negative to positive: the first one
operates within General Relativity and hence usually requires the
violation of the null energy condition, NEC, $\rho+P\geq 0$
\cite{Peter:2001fy}: Einstein equations (\ref{FriedCosm}), as provided
in the next section, indeed imply that the time derivative of the
Hubble rate reads
\begin{equation}
\dot{H} = \frac{\Ka}{a^2} - \frac12 \left( \rho + P \right),
\end{equation}
so that when the spatial sections are flat ($\Ka\to 0$), $\dot{H}>0$
definitely demands $\rho+P<0$.  A generic consequence of violating the
null energy condition is the appearance of fields with negative
kinetic energy: ghosts; a crucial point in bouncing models is actually
to construct a regular model in which such ghosts are absent while
still having a bouncing phase. It is possible to generate a bounce in
the presence of curvature $\mathcal K=1$ without violating the NEC,
but only the strong energy condition, SEC, which demands $\rho+P\geq
0$ and $\rho+3P\geq 0$, see \cite{Martin:2003sf,Falciano:2008gt} for
concrete models. Such a bounce could leave some amount of
spatial curvature in the expanding phase, whose amplitude may require
a subsequent inflationary phase to dilute it, hence possibly ruining
the alternative-to-inflation program (as emphasized above,
we shall not be concerned here
with the mixed models in which a bounce permits to avoid a primordial
singularity while a subsequent inflation phase solves the other
puzzles of the standard hot big-bang model).  

The second option is to allow for a classically singular bounce. Here
the scale factor actually vanishes and as such, four-dimensional
General Relativity ceases to be valid close to the bounce.
Pragmatically, the contracting phase is often matched to the expanding
one within GR under the assumption that the actual bounce leaves
observables unaffected. In the words of \cite{Xue:2010ux}: ``{\it
  [...] the Universe contracts towards a ‘‘big crunch’’ until the
  scale factor $a(t)$ is so small that quantum gravity effects become
  important. The presumption is that these quantum gravity effects
  introduce deviations from conventional general relativity and
  produce a bounce that preserves the smooth, flat conditions achieved
  during the ultra-slow contraction phase}''. One thus assumes all
goes roughly unchanged on the cosmologically interesting scales
through the otherwise quantum gravity dominated phase.

This matching procedure is not as easy as it appears at first sight,
because ambiguities arise when trying to impose the Deruelle-Mukanov
matching conditions to cosmological perturbations
\cite{Deruelle:1995kd}; see Sec.~\ref{modemixing}. Attempts have been
made to employ methods akin to the AdS/CFT correspondence to a
singular bounce \cite{Craps:2007ch,Craps:2009qc,Smolkin:2012er}, see
Sec.~\ref{sbst}, with limited success. An intriguing proposal by Bars
\etal in
\cite{Bars:2011mh,Bars:2011th,Bars:2011aa,Bars:2013yba,Bars:2013vba,Bars:2012fq,Bars:2013qna}
allows to trace the evolution of the universe unambiguously through a
singular bounce via a brief antigravity phase, see
Sec.~\ref{Antigravity}; however, a computation of observables in this
framework has not been performed yet. Thus, a non-perturbative
treatment of singular bounces within string theory is desirable to
assess not only the viability of the bounce itself, but also to
unambiguously compute observables in the subsequent expanding phase.

To obtain a nonsingular bounce without introducing ghosts is
challenging, but phenomenologically, it appears possible to produce an
instability-free bounce by introducing new matter fields, such as
ghost condensates \cite{Peter:2002cn,Lin:2010pf}, galileons
\cite{Qiu:2011cy}, quintom fields \cite{Cai:2007qw}, S-branes
\cite{Kounnas:2011fk}, a gravitational action that allows higher
derivative terms \cite{Biswas:2005qr,Biswas:2006bs} or change the way
gravity couples to matter \cite{Langlois:2013cya}, among other
proposals. An implementation of these proposals within string theory
is desirable, but still missing. For example, trying to implement
ghost condensates into a supersymmetric setting appears to generically
re-introduce ghosts via the superpartners
\cite{Koehn:2013hk}. However, a nonsingular cosmic {\it{super-bounce}}
in $\mathcal N=1$ supergravity, based on a ghost condensate and
galileon scalar field theories, was found in \cite{Koehn:2013upa},
where it was shown that perturbative ghost instabilities can be
avoided; further, perturbations are well-behaved and nonsingular so
that the pre-bounce spectrum is unaffected on large scales by the
bounce \cite{Battarra:2014tga}. Such models appear promising.

A final word of caution: all bouncing cosmological models, as most
inflationary ones, come from theories whose motivation is usually
unrelated with its capability to produce a bounce. An example is
provided by the Ho\v{r}ava-Lifshitz theory whose bounce implementation
is described in Sec.~\ref{Horava}: the goal of this proposal was to
provide a renormalizable version of quantum gravity. We shall not
expand on those external motivations, but concentrate on the relevant
bouncing models they induce; nevertheless, we provide the relevant
references so that the reader may critically assess the viability of
the respective framework.

\subsection{Notation and conventions}

Unless explicitly stated otherwise, we use the
Friedmann-Lema\^{\i}tre-Robertson-Walker (FLRW) metric, given by the
line element
\begin{equation}
\dd s^2 = -\dd t^2 + a^2(t) \gamma_{ij} \dd x^i \dd x^j,
\label{FLRW}
\end{equation}
where the spatial part takes the form
\begin{equation}
\gamma_{ij} \equiv\frac{\delta_{ij}}{\left(1
+\displaystyle\frac{\Ka}{4}\delta_{mn} x^m x^n\right)^2},
\label{gij}
\end{equation}
depending on the constant $\Ka$ (the spatial curvature). This constant
can be rescaled to $\Ka=-1,0,1$ for an open, flat or closed universe
respectively.

We work in natural units where
\begin{equation}
\hbar=c=8\pi \GN\equiv 1,
\end{equation}
so that the Planck mass $\Mp \equiv \GN^{-1/2}$ is dimensionless;
occasionally, we shall write it explicitly to emphasize quantum
gravity points.
 
In the presence of a fluid with energy density $\rho$, pressure $P$,
and stress-energy tensor
\begin{equation}
T_{\mu\nu} = \left(\rho + P\right) u_\mu u_\nu + P g_{\mu\nu},
\label{Tmn}
\end{equation}
with $u_\mu$ a timelike vector, the Einstein equations read
\begin{equation}
H^2+\frac{\Ka}{a^2} = \frac13\rho \ \ \ \hbox{and} \ \ \ \ \dot H+H^2
= \frac{\ddot a}{a} =-\frac16\left(\rho + 3 P\right),
\label{FriedCosm}
\end{equation}
where the Hubble rate $H$ is defined by $H\equiv \dot a/a$, and an
overdot denotes a derivative w.r.t.~cosmic time
$t$. Eqs.~(\ref{FriedCosm}) can also be written in the equivalent form
\begin{equation}
\Hu^2+\Ka = \frac13\rho a^2 \ \ \ \hbox{and} \ \ \ \ \Hu'
=-\frac16a^2 \left(\rho + 3 P\right),
\label{FriedConf}
\end{equation}
obtained from the transformation to conformal time $\eta$, defined
through $\dd t = a \dd \eta$; derivatives w.r.t.~$\eta$ are denoted by
a prime and the conformal Hubble rate is $\Hu\equiv
a'/a$. Conservation of (\ref{Tmn}), \ie $\nabla_\mu T^{\mu\nu} = 0$,
entails
\begin{equation}
\dot \rho + 3 H\left(\rho + P\right) = 0 \ \ \ 
\Leftrightarrow \ \ \ \rho'+3\Hu\left(\rho + P\right) = 0.
\label{DT0}
\end{equation}
The usual Lagrangian for a scalar field with canonical kinetic term
and potential reads
\begin{equation}
\mathcal{L}_\mathrm{can}\left[ \phi\left( x\right) \right]  = -\frac12 \left(
\partial\phi\right)^2 - V(\phi),
\label{canLagphi}
\end{equation}
leading to 
\begin{eqnarray}
\rho &=& \frac12 \dot \phi^2 + V(\phi) = \frac{\phi'^2}{2a^2}+V(\phi),
\nonumber\\
P &=& \frac12 \dot \phi^2 - V(\phi) =
\frac{\phi'^2}{2a^2} - V(\phi)
\label{rhoPscalar}
\end{eqnarray}
for a homogeneous and isotropic field.  These relations are used
extensively for describing inflationary phases as well as bouncing
epochs.

\section{Overview of bouncing models}\label{overviewofbounces}

In the literature, one can find many models that are based on
well-tested physics (semi-classical scalar fields in the framework of
4D General Relativity) and string theory (the only known
self-consistent theory of all interactions including quantum gravity);
these are the models we shall restrict attention to in this review, so
let us mention briefly in Sec.~\ref{QGM} the other direction in which
quantization of GR is used explicitly as an important ingredient to
implement the bouncing phase, namely Quantum Cosmology, be it through
Loop Quantum Gravity (LQG), a supposedly background-independent
attempt at quantizing General Relativity, or by using well-controlled
matter fields (fluids or scalars) in conjunction with the Wheeler-De
Witt equation (canonical quantum gravity). Because the former, Loop
Quantum Cosmology (LQC), can be argued to be in demand of technical
improvements, the latter appears more conservative.

 After this brief excursion, we follow with the above mentioned
scenarios. All models are introduced briefly with references to the
original literature to provide an encyclopedic overview; we follow
with a more cohesive in depth discussion of the requirements for a
successful bounce, the computation of cosmological perturbations and
potential fatal effects undermining nonsingular models in subsequent
sections. It should be noted that most bouncing models are modular:
the process whereby the bounce is achieved is a priori independent of
the process whereby scale invariant cosmological perturbations are
generated. For this reason, we clearly separate these two key
ingredients in Sec.~\ref{requirements} and
Sec.~\ref{Perturbations}. Nevertheless, in this section, and in
particular in Table \ref{checklist} and \ref{table:}, we combine
particular bounce models with the generation mechanism for
fluctuations that has been associate with it in the literature. For
example, the new ekpyrotic scenario entails a ghost condensate bounce
and an entropic two-field mechanism to produce a scale-invariant
spectrum. Our reasoning for this approach is two-fold: firstly, we
would like to highlight which combinations have been already
considered to serve as a guide for future research to go beyond the
status quo, particularly in those models that are in tension with
observations. Secondly, not every bounce mechanism may be combined
with every pre-bounce phase in a consistent manner. For instance, in
the cyclic scenario, which is based on string theory, multi-field
models as well as an entropic mechanism appear well-motivated. Yet,
introducing a galileon into the scenario would go against the string
theoretical underpinnings, since it has not been shown that galileons
can arise in string theory. For this reason, we decided deliberately
not to speculate on possible combinations one might want to
investigate in the future.

\subsection{Quantum gravity based models}\label{QGM}

Quantum gravity based models sometimes appear to be not as developed
as GR-based ones, because a bouncing phase is induced in a regime that
is less well understood. They are however natural in the following sense:
the very existence of a primordial singularity stems from the use of a
classical theory of gravitation, GR, extrapolated to its very limit, precisely
where it is expected not to be valid anymore. Taking this fact into account,
LQC relies on LQG to avoid the singularity, in much the same way that
quantum mechanics avoids the ultraviolet catastrophe\footnote{
This originally motivated the argument invoked for the PBB scenario,
which predates most of the models discussed below.}:
the Universe naturally goes through a maximum of the
curvature, after which the latter can only decrease; this is achieved
with the scale factor passing through a minimal value, and hence a
bounce. Similarly, canonical QG provides a wave function which
vanishes for vanishing values of the scale factor, thereby again
spontaneously avoiding the singularity and in most instance
yielding a bounce.  Here we briefly review both mechanisms.

\subsubsection{Loop quantum cosmology}
\label{LQG}

Loop quantum gravity is a non-perturbative attempt at a background
independent quantization of General Relativity, reviewed in
\cite{Smolin:2004sx,Rovelli:2011eq}.  This proposal has been argued
\cite{Nicolai:2005mc} to have internal inconsistencies (see however
Refs. \cite{Henderson:2012ie,Henderson:2012cu,Tomlin:2012qz} for
recent attempts of addressing anomalies in $2+1$ dimensions), and to
be in violation with current observations such as tests of Lorentz
Invariance (LI).  Stringent constraints on LI violation have been
placed via observations of Gamma-Ray Bursts (GRB) by the
F{\footnotesize ERMI} Large Area Telescope, LAT \cite{Atwood:2009ez},
which is sensitive to MeV-to-GeV GRBs, and the Gamma-Ray Burst Monitor
collaborations that use GRB $080916$C \cite{Shao:2009bv} and GRB
$090510$ \cite{Ackermann:2009aa}. In addition, competitive results can
be achieved by observations of flares of Active Galactive Nuclei by
M{\footnotesize AGIC}, or the H.E.S.S. analysis of the exceptional
flare PKS $2155-304$ \cite{Aharonian:2008kz,HESS:2011aa}.  In essence,
the attempt to combine quantum mechanics and gravity in LQG entails
the presence of a {\it natural length scale}, implying a ``quantum
gravity energy scale" $E_{_\mathrm{QG}}$; this scale is expected to be
of order of the Planck scale, $E_{_\mathrm{Pl}}\equiv \sqrt{(\hbar
  c^5)/\GN}\simeq1.22\times 10^{19}$ GeV ($\equiv\sqrt{8\pi}$ in the
natural units used here), and it is actually lower in the case of LQG,
$E_{_\mathrm{LQG}} \lesssim E_{_\mathrm{Pl}}$
\cite{Ackermann:2009aa}. At this scale, the physics of space-time
predicted by General Relativity breaks down.  Introducing such a scale
violates LI since relativity prohibits an invariant length.
 
The high photon energies and large distances of GRBs can test a
prediction of LQG that, since energy dispersion in the speed of the
photons exists, high energy photons should arrive later than low
energy photons. In the linear approximation, this arrival-time
difference $\Delta t$ is proportional to the ratio of the photon
energy difference to the quantum gravity mass $\Delta
E/E_{_\mathrm{QG}}$ and depends on the photons' traveled distance
\cite{AmelinoCamelia:1997gz}.  Going beyond the linear order, one
finds possible Lorentz violation energies
at linear and quadratic energy dependence are $E_{_{\mathrm{QG},1}}$
and $E_{_{\mathrm{QG},2}}$ respectively, \ie $\Delta t_\mathrm{lin}
\propto E/E_{_{\mathrm{QG},1}}$ and $\Delta t_\mathrm{quad} \propto
\left( E/E_{_{\mathrm{QG},2}}\right)^2$.
The constrains placed by the
F{\footnotesize ERMI} collaboration read $E_{_{\mathrm{QG},1}}>3.5$
$E_{_\mathrm{Pl}}$ at $95\%$CL and $E_{_{\mathrm{QG},2}}>6.4\times
10^{10}$GeV by the H.E.S.S. collaboration.  A recent, independent
combined analysis in \cite{vasileiou:2013vra} confirms and improves
these bounds by a factor of $\sim 2$, namely,
$E_{_{\mathrm{QG},1}}>7.6$ $E_{_\mathrm{Pl}}$ and
$E_{_{\mathrm{QG},2}}>1.3\times 10^{11}$GeV; thus, any theory that
requires $E_{_{\mathrm{QG},1}}\lsim E_{_\mathrm{Pl}}$ is strongly
disfavored. It has however been claimed that a linear dispersion
relation may not be generic, in a sense to be further elaborated.

Loop quantum cosmology \cite{Bojowald:2012xy} is an attempt to
use the same quantization techniques employed in LQG in a homogeneous
and isotropic universe.  If one takes this framework as a working
hypothesis, ignoring possible observational and theoretical
shortcomings, it was shown that the initial singularity is resolved
\cite{Ashtekar:2011ni} and inflationary as well as bouncing
cosmologies may be achieved
\cite{Vereshchagin:2004uc,Singh:2006im,Cailleteau:2009fv,Linsefors:2013cd,Amoros:2013nxa,Mielczarek:2012qs,Gazeau:2013iya,Barrau:2013ula,Wilson-Ewing:2013bla,Gupt:2013swa,Cai:2014zga}
(see also \cite{Corichi:2012bg} for a related approach involving a
minimal length).  A consistent treatment of perturbations in LQC has
been proposed in
Refs. \cite{Bojowald:2008jv,Bojowald:2008gz,Cailleteau:2013kqa}.  The
most common approach consists in taking a modified Friedmann equation
containing a $-\rho^2$ contribution to the right hand side
\cite{Singh:2006im,Ashtekar:2006es,Ashtekar:2006wn,Ashtekar:2006rx,WilsonEwing:2012pu}
(see also, \cite{Vereshchagin:2004uc}). Such modifications have been
known in the literature for a long time
\cite{Shtanov:2002mb,Brown:2004cs,Battefeld:2005cj} and were
originally motivated by brane world set-ups in string theory
\cite{Randall:1999ee}.  However, the negative sign in front of
$\rho^2$ would correspond to an extra timelike dimension, which has
never been considered in string theory, although there is, as far as
we know, no fully established no-go theorem that would prevent it (see
Sec.~\ref{brane_world}).

Since there is no, as of now, accepted particle physics approach to
LQC (in \cite{Bojowald:2012xy} the current status of this point is
explained), it is overall unknown whether or not ghosts and/or fatal
instabilities are present (see \cite{Bojowald:2011aa}, which indicates
that fatal instabilities are indeed present; however, in
\cite{Singh:2011gp} it was shown at the homogeneous level that shear
and curvature invariants are usually bounded).  Several attempts have
been made to incorporate fluctuations into the framework of a bouncing
LQC setup (\cite{Mielczarek:2012qs,Gazeau:2013iya} and references
therein). It is possible to accommodate a scale-invariant spectrum if
at least one scalar field and either a matter phase
\cite{WilsonEwing:2012pu}, or a second scalar field combined with an
entropic mechanism, is introduced.  While phenomenologically
acceptable, if ghosts were absent, the introduction of space-time
dependent fluctuations into the mini-superspace approach used in LQC
appears questionable: if one is interested in deviations of
homogeneity and isotropy, one should use the full framework of LQG to
perform the quantization at the background and perturbed level.  It
has been argued in the LQG literature that LQC is not the homogeneous
and isotropic limit of LQG \cite{Thiemann:2006cf}, and thus, the
operation of quantization and taking the mini-superspace approximation
might not commute. For recent works on perturbations, which aim to go
beyond the mini-superspace approximation, see
\cite{Bojowald:2008jv,Bojowald:2008gz,Cailleteau:2013kqa}.

Given these theoretical uncertainties, combined with yet-unanswered
questions regarding ghosts and instabilities, comparisons of these
models' predictions with observation may be too early, improvements on
the foundations of this framework being called for first.

\subsubsection{Canonical quantum cosmology}
  
The cosmological singularity in a Universe dominated by a perfect
fluid with positive-definite energy and pressure is a consequence of
Einstein's field equations. In order to avoid it, one can modify these
classical field equations, either by modifying gravity itself, or by
including a material content with unusual properties. One could also
try to quantize gravity directly.  Indeed, the typical maximal energy
at which one expects a bounce to take place is of the order of
$\sim 10^{-3}\Mp$, so that using the ADM formalism
(canonical quantum gravity), the Wheeler-De Witt equation in
mini-superspace is expected to yield a good approximation of the
quantum effects taking place during these early stages. To complete
the model, one then needs to add a universe-filling matter component,
which can be taken in the form of a perfect fluid, a choice that also
naturally provides a preferred time variable.

Solving for the wave function of the universe is not the whole story
as it can at most provide an average value for the scale factor as a
function of time, the scale factor being an operator in this
formulation.  A proposal to circumvent this problem consists of
assuming a trajectory formulation of quantum mechanics
\cite{Holland:1993ee,Sanz:2012bct} in which the scale factor follows
specific trajectory values \cite{Pinto-Neto:2013toa}. Applying this
formalism and assuming regular boundary conditions, one finds that all
possible trajectories are nonsingular and include a bounce
\cite{AcaciodeBarros:1997gy} (see also \cite{Casadio:1998yr} for a
different but related approach).   Of course, all known
formulations of quantum mechanics being strictly equivalent, the fact
that the universe underwent a regular bouncing phase or not should not
depend on which formulation one picks, so it is reasonable to expect
that the results obtained in
Refs. \cite{Pinto-Neto:2013toa,Casadio:1998yr} generically indicate
that it is canonical quantum gravity itself which allows for a bounce
to take place.  

On top of these trajectories, a perturbative expansion can be done
consistently, with the meaning that both the background and the
perturbations are quantised
\cite{Pinto-Neto:2013toa,Peter:2006id,Pinho:2006ym}.  However, more
work is needed to assess the compatibility of such models with
available data \cite{Peter:2001fy}.

\hskip6mm

Evidently, both models discussed above need more work to be compared
with currently available and forthcoming data, because both require
quantum gravity as a central ingredient. On the other hand, models
based on GR often make use of much more speculative ingredients, such
as ghost-condensates, galileons or massive gravity. The legacy of past
bouncing models has fueled the use of such unconventional
ingredients. It is interesting though that, would the universe have
chosen to use such ingredients as to permit a classical theory of
gravity (GR or otherwise) to implement a bounce, the question of
quantum gravity would remain forever bound to the interior of black
holes, and hence possibly merely philosophical until one finds a way
to accelerate particles to reach Planck energy collisions.

In the remainder of this section we provide a brief
overview of those proposals that are quoted as reasonably fashionable.

\subsection{Ekpyrotic and cyclic scenarios}
\label{ekcyclic}

The ekpyrotic scenario
\cite{Khoury:2001wf,Donagi:2001fs,Khoury:2001iy} is based on
five-dimensional heterotic string theory, where the fifth dimension
ends at two boundary branes, one of which is identified with our
Universe. The branes, on which matter and forces other than gravity
are localized, can only interact with one another via gravity as long
as they are widely separated.  During the ekpyrotic phase the branes
are attracted to each other and eventually collide, producing matter
and radiation on the branes. This collision does not occur everywhere
at the same time on the brane: quantum fluctuations produce ripples on
the brane so that the collision occurs earlier in some places than in
others; regions that collide earlier provide the universe with
additional time to cool and expand, while regions where the collision
occurs later, stay relatively hotter; such a collision represents the
big bang \cite{Khoury:2001wf}. Thus, fluctuations in the CMBR can be
traced back to these geometric fluctuations, which can also be
interpreted in terms of an effective scalar field in a 4d theory.
This is the picture of
the {\it old ekpyrotic scenario} \cite{Khoury:2001wf}; it purportedly
solves the isotropy problem of the big bang by having the universe
undergo a period of slow contraction, the ekpyrotic phase, superseded
by a bounce to the standard expanding phase. This proposal was
criticized in \cite{Kallosh:2001ai} primarily for fine-tuning. These
points were addressed in \cite{Donagi:2001fs,Khoury:2001iy}.  In
\cite{Kallosh:2001du}, following earlier work in \cite{Lukas:1998tt}
and follow-up papers, it is argued that the predicted big bang is
instead a big crunch and that computations in the ekpyrotic scenario
need to be performed in the full $5$d setup; more importantly, setting
aside such potential theoretical concerns, the scenario was shown to
be observationally problematic \cite{Martin:2001ue,Martin:2002ar},
since density fluctuations do not inherit a scale invariant spectrum,
see below. 

The {\it cyclic}\footnote{For a historical account of cyclic
  oscillating models dating back to the 1920's see
  \cite{Kragh:2013dva}.}  {\it scenario} is an extension of the old
ekpyrotic scenario.  It can, as the previous scenario, be described
by means of an effective 4d scalar field whose potential is represented
schematically on Fig.~\ref{CyclicP}.  It was introduced in
\cite{Steinhardt:2001st,Steinhardt:2001vw,Khoury:2001bz} and critized
in \cite{Felder:2002jk,Linde:2007fr}. This cyclic extension with a
singular bounce continues to be investigated.  The idea is that after
the brane collision, the inter-brane distance grows again, but since
the branes continue to attract each other, the distance between them
reaches an apex, before turning around. This quasi-static phase of the
internal space is associated with the late time FLRW Universe of dark
energy domination and flattens out the branes.  Ultimately, the
branes' attraction wins and a new ekpyrotic phase takes place.

\begin{figure*}[tb]
\includegraphics[scale=0.435]{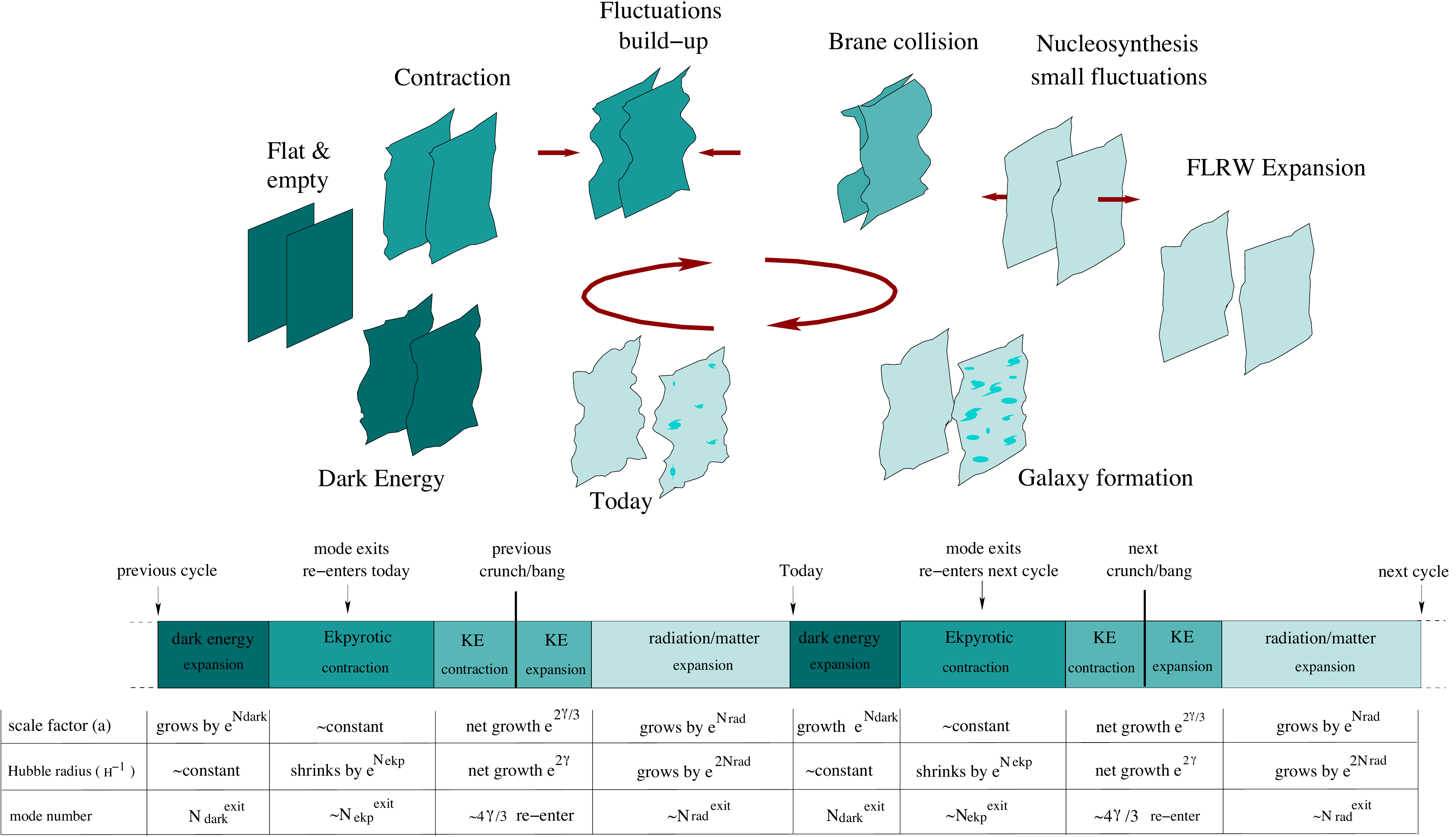} 
 \caption{ \label{branes} Schematic of the cyclic universe as
   initially envisioned in \cite{Steinhardt:2001vw,Steinhardt:2001st}:
   expansion and contraction correspond to the growing and shrinking
   of the orbifold in M-theory. The collision of the boundary branes
   is identified with the big bang, a singular bounce since the scale
   factor of the orbifold vanishes. Fluctuations in the distance
   between branes can be identified with density fluctuations, which
   are imprinted onto density fluctuations during the collision, which
   also reheats the matter content on our brane. During each cycle the
   Universe is rendered flat and empty via a phase of dark energy
   domination. Whilst this model was not practically working, it has
   provided a strong motivation for subsequent developments of the
   cyclic universe.  In the table, the parameter $\gamma$ is given by
   $\gamma=\ln(-V_\mathrm{end})^{1/4}T_\mathrm{rh}$, where
   $T_\mathrm{rh}$ is the temperature of radiation when it
   dominates. To be compatible with observations, cyclic models
   require $\gamma\approx 10-20$ \cite{Khoury:2003rt}. }
  \end{figure*}

\begin{figure}[tb]
\begin{center}
 \includegraphics[scale=0.42]{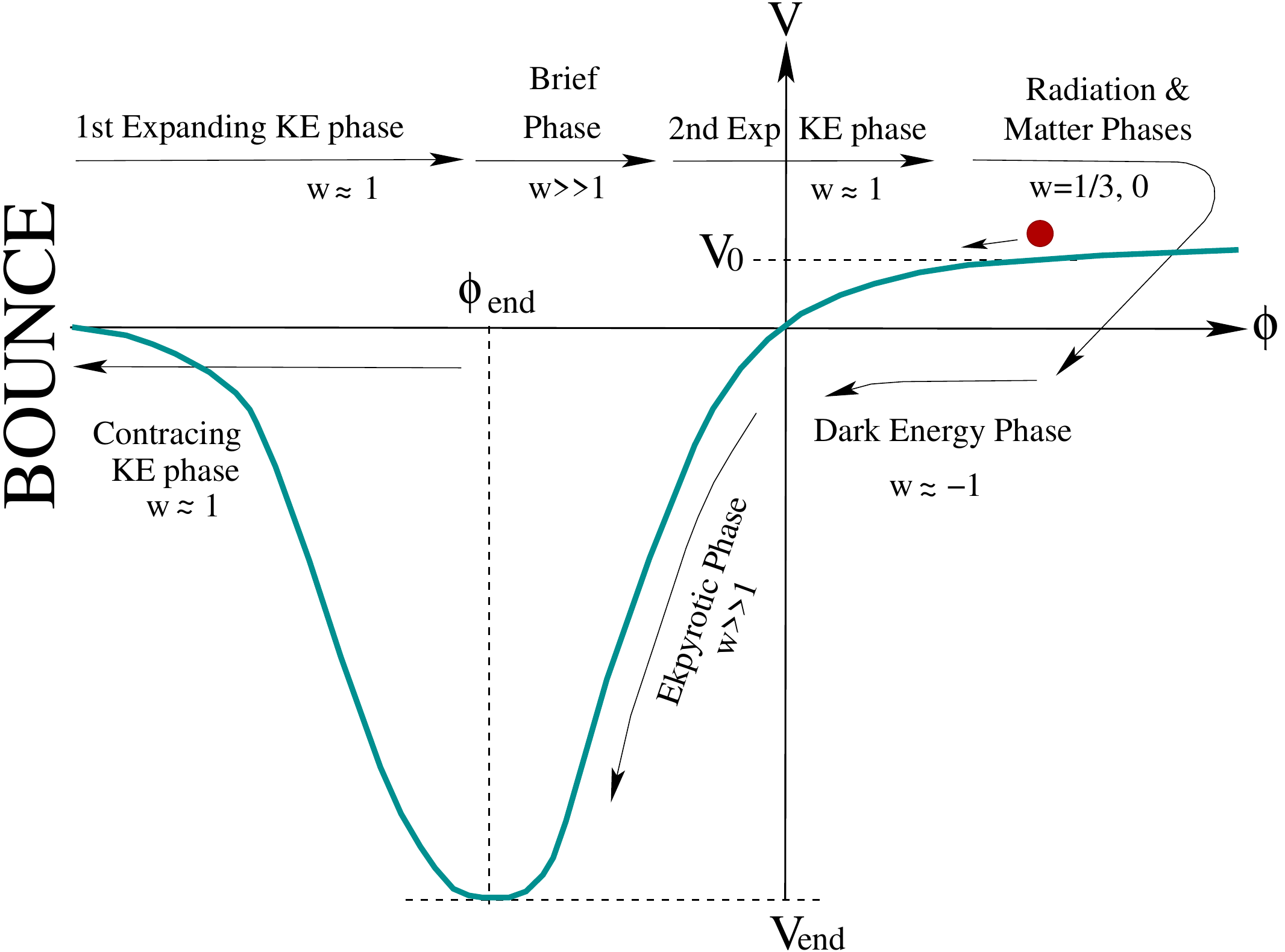}
   \caption{Schematic of the potential in the ekpyrotic/cyclic
     scenario \cite{Lehners:2008vx}.
   \label{CyclicP}}
   \end{center}
\end{figure}

In this model, the current dark energy dominated Universe will be
superseded by a contracting ekpyrotic phase, followed by a bounce, an
expansion phase, and a subsequent phase of radiation and matter
domination, succeeded by another dark energy dominated phase and
continuing so in a cyclic manner. Fig.~\ref{branes} shows a schematic
representation of the cyclic model based on colliding branes in
M-theory. A conceptual advantage of this model is the apparent lack of
need for a specified microphysical origin of time, making the problem
of initial conditions inconsequential, see Sec.~\ref{cyclic_ic} for
details and Table~\ref{checklist} for general properties of singular
bouncing models.

However, it should be noted that the cyclic universe is not past
eternal, similar in that regard to eternal inflation
\cite{Mithani:2012ii}.  Unfortunately, each singular bounce requires
the use of non-perturbative techniques in string theory and is
therefore ill-understood, if at all.
 
The spectrum of curvature fluctuations in the old ekpyrotic scenario
was found to be deeply blue
\cite{Brandenberger:2001bs,Lyth:2001pf,Durrer:2002jn,Creminelli:2004jg}
(an additional problem is that these modes do not become classical
\cite{Tseng:2012qd} as opposed to the ones resulting from the entropic
mechanism \cite{Battarra:2013cha} described below). As a result,
two-field models \cite{Notari:2002yc} were introduced to overcome this
problem
\cite{Lehners:2007ac,Buchbinder:2007ad,Creminelli:2007aq,Buchbinder:2007tw}.
One realization is the {\it new ekpyrotic scenario}, a nonsingular
setup, which makes use of the entropic mechanism to generate a nearly
scale-invariant spectrum of primordial density fluctuations in an
isocurvature field.  If seen as a fundamental theory, the ghost
condensate employed in the new Ekpyrotic scenario contains ghosts due
to the higher derivative equations of motion, as shown in
\cite{KKLM08}. However, from an effective field theory (EFT) point of
view, ghosts are absent below the energy scale demarcating the
validity range of the EFT \cite{Woodard:2006nt}. Thus, the description
in the new ekpyrotic scenario is self-consistent, as long as the
energy scale during the bounce remains below that cut-off, such that
the degrees of freedom associated with the higher derivatives do not
get excited.

\label{sec:QC}

\begin{table*}[hbt]
\setlength{\tabcolsep}{0.5 pc}
\newlength{\digitwidth} \settowidth{\digitwidth}{\rm 0}
\begin{tabular*}{17.9cm}{@{}l@{\extracolsep{\fill}}cccccccccccc}
\hline\hline
&&&&&\multicolumn{4}{c}{Instabilities}&&             
\\ [0.09ex] 
Model                                        & Bounce           & no tuned
i.c.         &no ghosts                 & A                            & B      
                          & C &   D&   $n_\mathrm{s}$  &
$f_{_\mathrm{NL}}^\mathrm{local}$&\\\hline
Ekpyrotic \cite{Khoury:2001zk} &singular brane 
&{\color{red}\xmark}&{\color{green}\cmark}   &{\color{green}\cmark}&
{\color{green}\cmark}                                 
&{\color{green}\cmark}&{\color{green}\cmark}&  blue \cite{Khoury:2001zk}  
&{\bf{\color{darkblue} ?}}&\\
Cyclic \cite{Steinhardt:2001vw,Steinhardt:2001st}           &quant.grav.eff.
&{\color{red}\xmark}&{\color{green}\cmark}
&{\color{green}\cmark}&{\color{green}\cmark}    & {\color{green}\cmark}     &
{\color{green}\cmark}&HZ            &{\bf{\color{darkblue} ?}}&\\
Ph\oe{}nix \cite{Lehners:2008qe}  &brane collision
&{\color{green}\cmark}&{\color{green}\cmark} 
&{\color{green}\cmark}&{\color{green}\cmark}&{\color{green}\cmark}&
{\color{green}\cmark} &HZ          &$\mathcal O(\pm10)$ \cite{Lehners:2011ig}&\\
Bars \etal
\cite{Bars:2011mh,Bars:2011th,Bars:2011aa,Bars:2013yba,Bars:2013vba,Bars:2012fq,Bars:2013qna}
&antigravity &{\bf{\color{darkblue} ?}}&{\color{green}\cmark} 
&{\color{green}\cmark}&{\color{green}\cmark}&{\color{green}\cmark}&
{\color{green}\cmark} &{\bf{\color{darkblue} ?}} &{\bf{\color{darkblue} ?}}&\\
\hline\hline
\end{tabular*}
\caption{Singular bouncing models.  Instabilities: A -- Curvature
  pertubation; B -- Quantum induced anisotropy; C -- Gravitational
  instability; D -- Initial anisotropy, see
  Sec.~{\ref{Fatal_effects}}. Fine-tuned initial conditions, i.c.,
  entail: a) how to get the brane flat, and b) how to get both fields
  near the top of the ridge as in Fig.~\ref{traj2}. The notation HZ
  indicates a power spectrum close to the Harrison-Zeldovich one with
  $n_\mathrm{s}=1$; in the cyclic/Ph\oe{}nix universe, the index can
  be made red by changing the potential slightly from the exponential
  one used in \eg (\ref{Pekpyrotic}).  The first three models lack an
  analytic understanding of the singular bounce and rely on matching
  conditions; see section \ref{sbst} for a brief review of
  non-perturbative attempts based on the AdS/CFT correspondence and
  Sec.~\ref{Antigravity} for the singular antigravity bounce.
  Gravitational waves on CMBR scales are generically not generated,
  see Sec.~\ref{rcyclic}. }
\label{checklist}
\end{table*}

A similar extension of the singular cyclic model to a two-field setup
is given in \cite{Lehners:2007ac}, which subsequently led to the
proposal of the {\it Ph\oe{}nix universe}
\cite{Lehners:2013cka,Lehners:2009eg,Lehners:2011kr,Johnson:2011aa}.
The conversion from isocurvature to adiabatic modes, first proposed in
\cite{Notari:2002yc} to counter the problems encountered in the old
ekpyrotic scenario\footnote{An idea taken from the curvaton mechanism
  \cite{Enqvist:2001zp,Lyth:2001nq}, which utilizes isocurvature
  perturbations to that effect.}, can occur before the bounce via the
movement of fields away from the scaling solution towards an ekpyrotic
attractor \cite{Koyama:2007ag,Koyama:2007mg,Koyama:2007if} see also
\cite{Buchbinder:2007tw} and Sec.~\ref{bending} for
details. Alternatively, a reflection of fields from a sharp boundary
of field space can result in a different conversion
\cite{Lehners:2006pu,Lehners:2006ir}, see Sec.~\ref{reflection}; one
may also use the curvaton mechanism or modulated (p)reheating
\cite{Enqvist:2001zp,Lyth:2001nq,Battefeld:2007st} after the bounce,
Sec.~\ref{modulatedp}.  These entropic mechanisms are constrained by
P{\footnotesize LANCK} \cite{Ade:2013uln,Ade:2013nlj} due to their
generic prediction of large non-Gaussianities. In that regard, it
should be noted that different aspects are highlighted in the
literature: before the improved constraints by P{\footnotesize LANCK},
Lehners \etal \cite{Lehners:2007wc,Lehners:2008my,Lehners:2010fy}
highlighted the generic prediction of observably large
non-Gaussianities of $f_{_\mathrm{NL}}^\mathrm{local}$ of order $10$
or bigger for the conversion mechanism in
\cite{Lehners:2006pu,Lehners:2006ir}. However, after the publication
of P{\footnotesize LANCK}, the emphasis was put onto the possibility
to counterbalance different contributions to non-Gaussianities to
enable $f_{_\mathrm{NL}}^\mathrm{local}$ of order $1$
\cite{Lehners:2013cka}. To this end, the focus shifted to potentials
approximately symmetric transverse to the adiabatic direction, as well
as non-minimal entropic models
\cite{Qiu:2013eoa,Li:2013hga,Fertig:2013kwa,Ijjas:2014fja}. All these
models entail an unobservable primordial gravitational wave spectrum;
they are therefore ruled out if the B{\footnotesize ICEP2} detection
of $r\sim 0.2$ is confirmed to be a signal of primordial
origin by the future Keck Array observations at $100$ GHZ and
P{\footnotesize LANCK} observations at higher frequency
\cite{Flauger:2014qra,Mortonson:2014bja}, see Sec.~\ref{rcyclic}.

In a recent publication applicable to singular models
\cite{Xue:2014oea}, Xue \etal studied the classical dynamics of the
universe experiencing a transition from a contracting phase, which is
dominated by a scalar field with a time-varying equation of state
parameter, to an expanding one through a big bang singularity. It was
found that the evolution of a bouncing universe through such a
singularity lacks a continuous classical limit except when the
equation of state is highly fine-tuned; this result implies that a
transition from contraction to expansion is contingent on quantum
processes and not on a simple classical limit.
 
Other studies pertaining to the cylic universe, which are not reviewed
here, include: 5D dynamics of general braneworld models
\cite{Saridakis:2007cf}, past-shrinking cycles that spend more time in
an entropy conserving Hagedorn phase \cite{Biswas:2008ti}, a cyclic
magnetic universe \cite{Novello:2008xp,Medeiros:2012hh}, phantom
accretion onto black holes \cite{Sun:2008si}, deformed
Ho\v{r}ava-Lifshitz gravity \cite{Son:2010ci}, a string-inspired model
via a scalar-tachyon coupling and a contribution from curvature in a
closed universe \cite{Li:2011nj}, cosmological hysteresis
\cite{Sahni:2012er}, Finsler-like gravity theories constructed on
tangent bundles to Lorentz manifolds \cite{Stavrinos:2012ty}, a
combination of cyclic and inflationary phases with quintessence
\cite{Ivanov:2012hq} and a cyclic model with a chameleon field
\cite{Gao:2014rsa}, among others
\cite{Barrow:2004ad,Clifton:2007tn,Barrow:2013qfa}.

\subsection{String gas cosmology}

The presence of a maximal temperature in string theory, the Hagedorn
temperature, as well as T-duality led to the hope of constructing a
nonsingular cosmological setup by invoking these intrinsically stringy
phenomena
\cite{Brandenberger:1988aj,Tseytlin:1991xk,Alexander:2000xv,Kripfganz:1987rh,Brandenberger:2005nz}.
Early work on string thermodynamics can be found in
\cite{Hagedorn:1965st,Frautschi:1971ij,Bowick:1985az,Sundborg:1984uk,Alvarez:1985fw,Polchinski:1985zf,Salomonson:1985eq,Mitchell:1987th,O'Brien:1987pn}. As
the Hagedorn temperature is approached, new massless degrees of
freedom arise, indicating a phase transition. This thermal component
can be modeled by a string gas \cite{Brandenberger:2013zea}. String
gas cosmology is an attempt to incorporate strings and branes into a
cosmological setting by means of a gas approximation, see
\cite{Battefeld:2005av,Brandenberger:2008nx} for reviews.  While
attempts to construct alternative proposals to inflation in string gas
cosmology, such as in \cite{Brandenberger:2006vv}, are still subject
to unsolved problems\footnote{Although it is possible to generate a
  nearly scale-invariant spectrum and gravitational waves, this
  proposals is still hampered by the flatness and relic problems; this
  is discussed in Sec.~\ref{relics}.}
\cite{Battefeld:2004xw,Battefeld:2005av,Kaloper:2007pw}, it was shown
in a series of recent articles
\cite{Florakis:2010is,Kounnas:2011fk,Kounnas:2011gz,Kounnas:2013yda}
that a string gas can be used successfully to describe the matter
content in a Hagedorn phase, while providing the possibility of
violating the NEC in a controlled manner in string theory. Based on
this idea, a cosmological model has been constructed and cosmological
perturbations were computed in \cite{Brandenberger:2013zea}. There it
was shown that a nearly scale-invariant spectrum can be transferred
through this nonsingular bounce, if it has been previously
generated. The violation of the NEC mediated by an S-brane is under
computational control and most instabilities can be avoided, with the
notable exception of the Belinsky, Khalatnikov and Lifshitz (BKL)
instability, see Sec.~\ref{BKL}; the model, dubbed {\it {S-brane
    bounce}} in this review, provides a promising avenue for future
research. We discuss its ingredients in more detail in
Sec.~\ref{stringy}.

\subsection{A nonsingular bounce in string theory}
\label{sbst}

An attempt to circumvent the initial singularity using methods akin to
the AdS/CFT correspondence \cite{Maldacena:1997re} was proposed in
\cite{Turok:2007ry} following prior work in \cite{Hertog:2004rz}. The
AdS/CFT correspondence provides a non-perturbative definition of
string theory in anti de Sitter (AdS) spacetimes in terms of conformal
field theories (CFT) \cite{Maldacena:2003nj}. In \cite{Turok:2007ry}
Turok \etal suggest the possibility of not only attaining a healthy
nonsingular bounce, but also propose a new mechanism for generating
nearly scale-invariant cosmological perturbations
\cite{Craps:2007ch}. The cosmological set-up considered in
\cite{Craps:2007ch} is a toy model and not compatible with the
necessary ingredients for the ekpyrotic scenario to take place.  This
line of research was subsequently followed in
\cite{Craps:2009qc,Smolkin:2012er}. At the time of writing, a
cosmological model ready to be compared with observations has not been
constructed.
  
\subsection{Antigravity}
\label{Antigravity}  

In a series of papers, Bars \etal
\cite{Bars:2011mh,Bars:2011th,Bars:2011aa,Bars:2013yba,Bars:2013vba,Bars:2012fq,Bars:2013qna}
showed that theories motivated by the minimal conformal extension of
the standard model with scalar fields coupled to gravity can be lifted
to a Weyl-invariant theory that allows the cosmological evolution to
be unambiguously traced through a big-crunch/big-bang (singular)
transition.  Here the classical evolution can be followed in a
homogeneous, but potentially anisotropic (\eg Bianchi IX), universe
through a brief antigravity phase.  Early work on antigravity can be
found in \cite{Linde:1979kf}. As pointed out in
\cite{Carrasco:2013hua,Linde:2014nna} and acknowledged in
\cite{Bars:2013qna}, this Weyl-invariant extension does not resolve
the singularity: for example, the Weyl-invariant curvature squared
diverges \cite{Carrasco:2013hua,Kallosh:2013oma}. Because of the
presence of a curvature singularity, the use of classical General
Relativity methods throughout \cite{Bars:2013qna} is therefore questionable.

Bars \etal argue that a geodesically complete, unambiguous solution
arises, because the cosmic evolution becomes smoothly ultra-local so
that density perturbations and spatial gradients become negligible
\cite{Linde:2014nna}.  The presence of an unambiguous classical
evolution through said singularity is intriguing and warrants further
study\footnote{A resolution of cosmological singularities has been
  attempted in string theory repeatedly
  \cite{Cornalba:2002fi,Liu:2002ft,Horowitz:2002mw,Cornalba:2003kd,Berkooz:2007nm}
  and is currently an active field of research.}, since it is unknown,
at the time of writing, whether or not quantum gravity corrections
leave the smooth transition found in
\cite{Bars:2011th,Bars:2011aa,Bars:2013yba,Bars:2013vba} unaffected.
A debate on this topic can be found in
\cite{Ijjas:2014nta,Linde:2014nna}, and in particular it was found in
\cite{Carrasco:2013hua} that the curvature invariants diverge.  If it
can be shown that the considered Weyl-invariant quantities remain
unscathed, one can employ this type of bounce in the cyclic scenario
in lieu of the complicated ghost condensate/galileon models that we
focus on subsequently in this review.   Most recently,
\cite{Oltean:2014bua} considered two scalar fields, the dilaton and
the Higgs, coupled to Einstein gravity and showed that the isotropic
cosmological solutions deep in the antigravity regime are stable at
the level of scalar perturbations~\cite{Oltean:2014bua}. A full
analysis is still an open research topic.

\begin{table*}[hbt]
\setlength{\tabcolsep}{0.6pc}
\begin{tabular*}{17.9cm}{@{}l@{\extracolsep{\fill}}cccccccccccc}
\hline\hline
&&P.~inv.~vac.&&\multicolumn{4}{c}{Instabilities}\\
Model                                                        & Bounce   
&$\&$ sublum.       & BKL            
& A & B & C & D &   $n_\mathrm{s}$  & $f_{_\mathrm{NL}}^\mathrm{local}$
\\[0.5ex]\hline 
New ekpyrotic \cite{Buchbinder:2007ad,Khoury:2001zk}       &ghost cond.  
           &{\color{red}\xmark} &{\color{green}\cmark}     
&{\color{red}\xmark}&{\color{red}\xmark}&{\color{red}\xmark}&{\color{red}\xmark}&
HZ       &$\mathcal O(50)$ \cite{Koyama:2007if,Buchbinder:2007at}\\
Matter bounce
\cite{Brandenberger:2012zb,Cai:2012va,Cai:2013vm,Cai:2013kja}&ghost.cond/gal.
&{\bf
{\color{darkblue} ?}} &{\color{green}\cmark}  
&{\color{green}\cmark}&{\color{green}\cmark}&{\color{green}\cmark}&{\color{green}\cmark}&
HZ &$-35/8$ \cite{Cai:2009fn}\\
G-bounce \cite{Easson:2011zy}                                 &KGB/galileons
&{\bf
{\color{darkblue} ?}} &{\color{green}\cmark} 
&{\color{green}\cmark}&{\color{green}\cmark}&{\color{green}\cmark}&{\color{green}\cmark}&
blue &{\bf {\color{darkblue} ?}}\\ 
Non-min entr. \cite{Qiu:2013eoa,Li:2013hga}            &galileon/other   
   &{\bf
{\color{darkblue} ?}}&{\color{green}\cmark}
&{\color{green}\cmark}&{\color{green}\cmark}&{\color{green}\cmark}&{\color{green}\cmark}&red
 &$\mathcal O(1)$ \cite{Fertig:2013kwa,Ijjas:2014fja}\\  
Cosm.~super-bounce \cite{Koehn:2013upa}                   
&ghost.cond/gal.&{\color{green}\cmark}
&{\color{green}\cmark}&{\color{green}\cmark}&{\color{green}\cmark}&{\color{green}\cmark}&{\color{green}\cmark}&{\bf
{\color{darkblue} ?}}&{\bf {\color{darkblue} ?}}\\                              
S-brane bounce \cite{Brandenberger:2013zea}&S-brane        &{\color{green}\cmark}
&{\color{red}\xmark}
&{\color{green}\cmark}&{\color{green}\cmark}&{\color{green}\cmark}&{\color{green}\cmark}&
 HZ   &{\bf {\color{darkblue} ?}}\\ 
\hline\hline
\end{tabular*}
\caption{Comparison of several promising nonsingular bouncing models.
  Instabilities: A -- Curvature pertubation; B -- Quantum induced
  anisotropy; C -- Gravitational Instability; D -- Initial anisotropy,
  see Sec.~\ref{Fatal_effects}. The notation HZ indicates a power
  spectrum close to the Harrison-Zeldovich one with $n_\mathrm{s}=1$;
  a slightly red spectrum can be achieved by a slight change of the
  potential used in the new-ekpyrotic scenario; for models employing a
  matter phase, such as the matter bounce or the S-brane bounce, a red
  spectrum can be attained by a small deviation of $w=0$, which is
  easily achieved. KGB stands for kinetic gravity braiding. Galileon
  models often suffer from superluminality for the Poincar\'e
  invariant vacuum (abbreviated P.~inv.~vac., even though the NEC
  violating solution maybe subluminal), but a detailed analysis for
  bouncing cosmologies is missing; see Sec.~\ref{galileon_genesis} for
  discussion in inflationary cosmology. If a nearly scale-invariant
  spectrum is achieved via the entropic mechanism, observable
  non-Gaussianities commonly result from the conversion mechanism. If
  the spectrum is generated in a matter phase, non-Gaussianities
  result due to the growth of fluctuations after Hubble radius
  crossing in the contracting phase, see
  Sec.~\ref{absenceNG}. Observables in the {\it{super bounce}} model
  are in line with other ekpyrotic models according to
  \cite{Battarra:2014tga}.}
\label{table:}
\end{table*}
 
\subsection{Nonsingular bounces via  a galileon}
\label{galileon} 
Nonsingular scenarios often include a combination of a
contracting matter dominated phase (ordinary dust or mimicked by a
scalar field) to yield a nearly scale-invariant power spectrum, an
ekpyrotic phase to dilute the curvature and shear contributions,
followed by a bounce phase\footnote{  An exception is the new
  ekpyrotic scenario, which is nonsingular and does not contain a
  contracting, matter dominated phase. }. Almost all of the
hitherto mentioned bounce mechanisms have problems, such as the growth
of instabilities and the presence of ghosts.  In this section, we
provide an overview of mechanisms based on galileon fields: these
non-canonical scalar fields can induce a bounce while preserving the
smooth, flat conditions achieved during the contracting phase and
avoiding instabilities. They can further be implemented in
supergravity and therefore provide a promising avenue for future
research.

Galileon models \cite{Nicolis:2008in} arise naturally in the context
of massive gravity \cite{deRham:2010ik,deRham:2010kj}. These theories
and their generalizations
\cite{deRham:2010eu,Deffayet:2010qz,Kobayashi:2010cm,Deffayet:2010zh}
offer the intriguing option to start the cosmological evolution from a
nearly Minkowski space-time to a de Sitter-like expansion
\cite{Creminelli:2010ba,Creminelli:2012my}, thus alleviating the
initial value problem of inflationary cosmology.  We would like to
add, at this point, that the naturalness or unnaturalness of a set of
initial conditions, \eg starting with an empty flat universe, depends
on a particular researcher's viewpoint and is thus subjective.

Besides enabling inflationary models, a bounce can be induced via a
galileon field
\cite{Nicolis:2008in,Creminelli:2010ba,Creminelli:2012my} or its close
relative, a field with kinetic gravity braiding (KGB)
\cite{Deffayet:2010qz,Kobayashi:2010cm}.  These models make use of a
subclass of scalar field theories with higher order derivatives in the
action while maintaining second order equations of motion, as
classified by Horndeski \cite{Horndeski:1974wa}. Galileon Lagrangians
obey a symmetry under the Galilei transformation
\begin{equation}
\phi(x)\rightarrow \phi(x)+c+b_\mu x^\mu
\end{equation}
where $c$ and $b_{\mu}$ are constants\footnote{In the literature, the
  galileon is often denoted by $\pi$, a notation we shall not use here
  as the fields used in the literature on bouncing cosmology are
  commonly denoted $\phi$; here, we keep the latter notation.}. See
\cite{Deffayet:2013lga} for a recent review of the mathematical
properties and construction of galileon theories. The Lagrangian of
these types of fields can lead to NEC violation while avoiding
instabilities and ghosts.

Consequently, bouncing cosmological models have been put forward using
galileon fields, \eg the {{\it G-bounce}} in
\cite{Qiu:2011cy,Osipov:2013ssa} and \cite{Easson:2011zy} (a follow up
to KGB models) among others
\cite{Deffayet:2010qz,Pujolas:2011he,Qiu:2014nla}.  A common danger of
these models is the possibility of pressure/big rip singularities
\cite{Barrow:2004xh}, which are indeed present in the far past or
future of a G-bounce \cite{Easson:2011zy}.

In \cite{Qiu:2013eoa}, a nonsingular bounce in the framework of
galileon cosmology with an ekpyrotic phase was investigated, with the
addition of a curvaton instead of a matter phase to generate the
scale-invariant spectrum of perturbations. This work superseded that
of \cite{Li:2013hga} which set up the building blocks to obtain
scale-invariant entropy perturbations within the ekpyrotic scenario
via non-minimally coupled massless scalar fields. There, it was
suggested that the entropy perturbation could be converted into
curvature perturbations by means of a curvaton, as done in
\cite{Qiu:2013eoa}, or modulated (p)reheating
\cite{Battefeld:2004cd}. Non-Gaussianities for the model in
\cite{Qiu:2013eoa} where computed in \cite{Fertig:2013kwa}, see the
     {\it non-minimal entropic mechanism} in Table~\ref{table:}. This
     model does not entail intrinsic non-Gaussianities, but the ones
     arising from the conversion mechanism. This particular model is
     an example of a more general class of non-minimal ekpyrotic
     models studied in \cite{Ijjas:2014fja}.

A first attempt to implement galileons in supergravity turned out to
be problematic, since the bosonic sector of globally supersymmetric
extensions of the cubic Langrangian showed a reappearance of ghosts
\cite{Koehn:2013hk}; nevertheless, a follow-up study in
\cite{Koehn:2013upa} proved more successful: the necessary conditions
for a nonsingular, stable, cosmic bounce in $\mathcal N=1$
supergravity, and hence potentially allowed in string theory, are
derived in \cite{Koehn:2013upa}; this so-called {\it super bounce},
see Sec.~\ref{superbounce}, is based on supergravity versions of the
ghost condensate and cubic galileon scalar field theories that have
been used at the phenomenological level in the matter bounce scenario
\cite{Cai:2012va}. This bounce is free of most problems that hamper
many other nonsingular bounces, see Sec.~\ref{Fatal_effects} and table
\ref{table:}. It is therefore one of the most promising proposals.

\subsection{Massive gravity}

The idea of modifying gravity is not new. In 1939, Fierz and Pauli
raised the question of the existence of a consistent covariant theory
for massive gravity, whereby the graviton becomes massive, hence
leading to a modification of General Relativity \cite{Fierz:1939ix}.
However, the non-linear terms that curtail the discontinuity problem
\cite{vanDam:1970vg,Zakharov:1970cc}, give rise to the Boulware-Deser
(BD) ghost mode \cite{Boulware:1973my}.  The prevalence of ghosts made
the theory unstable and it was abandoned for decades until de Rham
\etal constructed a non linear extension
\cite{deRham:2010ik,deRham:2010kj}: the ghost could be removed in the
decoupling limit to all orders of perturbation theory through a
systematic construction of a covariant non-linear action.  It was soon
realized, however, that homogeneous and isotropic solutions in
non-linear massive gravity have a ghost
\cite{DeFelice:2012mx,DeFelice:2013awa}.  Extensions of non-linear
gravity models ensued \cite{D'Amico:2011jj,Gumrukcuoglu:2012aa} and
the graviton mass was allowed to vary by setting its mass via a scalar
field \cite{Huang:2012pe}. Motivated by this work, the cosmological
implications in flat and open universes were explored in
\cite{Saridakis:2012jy}; it was found that such an extension requires
a UV-modification of General Relativity, in addition to the one in the
IR. A pedagogical review of massive gravity can be found in
\cite{Hinterbichler:2011tt} (see also \cite{deRham:2014zqa}).

Nonsingular bouncing cosmologies have been constructed within massive
gravity.  An attempt to construct ghost and asymptotically free
modified gravity models that enable nonsingular bouncing solutions and
resemble General Relativity in the IR limit was made in
\cite{Biswas:2005qr,Biswas:2010zk}.  Using the results of
\cite{Saridakis:2012jy}, where the graviton was promoted to a function
of an extra degree of freedom, a nonsingular bounce and cyclic
cosmological evolutions at early times were studied in
\cite{Cai:2012ag}; in \cite{Langlois:2013cya}, bouncing cosmologies
were found to be generic in the context of massive gravity on de
Sitter; the bounce occurs while the cosmological matter satisfies the
strong energy condition. Other work include
\cite{Biswas:2011ar,Biswas:2012bp} and \cite{Koshelev:2013lfm}.

These models can provide a ghost-free bounce, but further implications
have not been explored.

\subsection{A nonsingular bounce in the multiverse?}
\label{multiverse}

\begin{figure}[tb]
\begin{center}
 \includegraphics[scale=0.77]{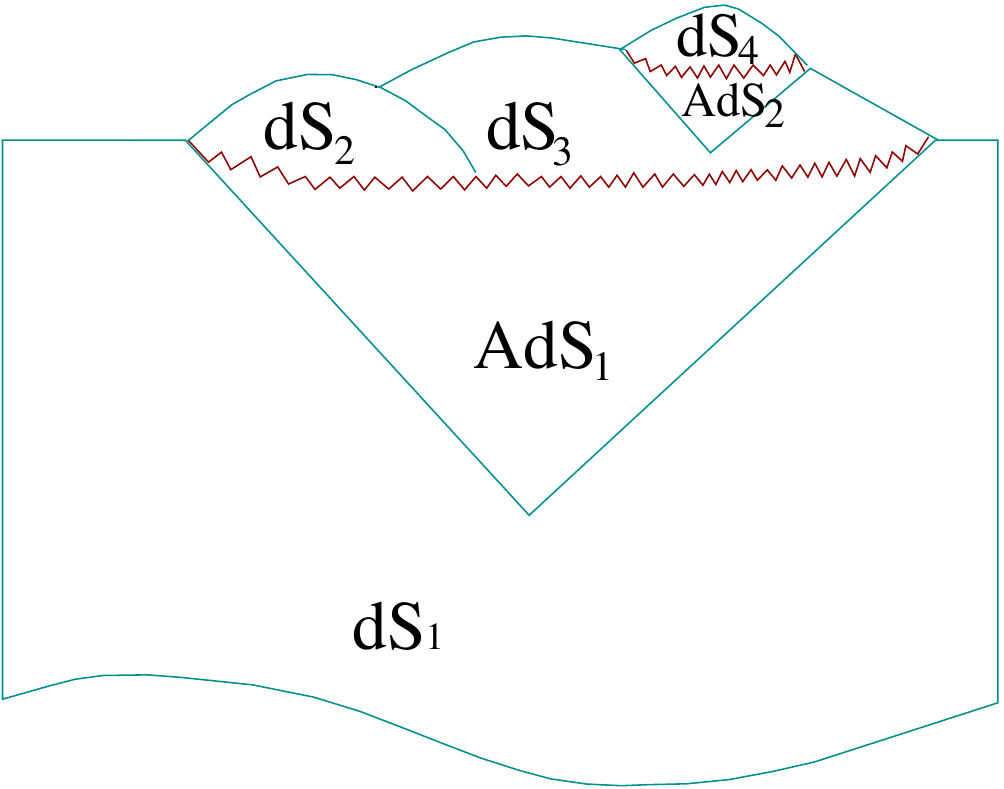}
   \caption{Causal diagram of transitions in the multiverse mediated
     by a phenomenological bounce \cite{Garriga:2013cix}. The AdS
     bubbles denote a contracting universe with a negative
     cosmological constant.   Red zig-zag lines indicate a bounce
     mediating a transition from AdS to dS or dS to AdS that could
     otherwise terminate in a big crunch.  \cite{Lehners:2012wz}
   \label{AdS}}
   \end{center}
\end{figure}

Attempts have been made to connect bouncing cosmologies to the
inflationary multiverse. The latter is made up of different space-time
regions populated by different meta-stable vacua. A transition from
one vacuum to the next may occur via quantum tunneling, generating a
daughter vacuum which expands within the parental one. Evolution after
the tunneling depends on whether the vacuum inside a bubble has
positive energy density or not. In the former case, the evolution is
asymptotically de Sitter (dS) and further nucleation occurs within the
bubble, the latter's AdS vacuum  (a contracting universe with
negative cosmological constant) eventually collapses into a big
crunch, developing curvature singularities where space-time
ends\footnote{ See \cite{Lehners:2012wz} for populating different
  vacua in eternal inflation and the possibility to encounter emergent
  or even cyclic universes in the nucleated bubbles. It should be
  noted that any quantification of such ideas is dependent on the
  measure. }.  Such bubbles are called terminal. It has been
speculated that the terminal singularity of the AdS vacuum is resolved
in a complete theory of quantum gravity, such as string theory -- see
Fig.~\ref{AdS} for a causal diagram. In the absence of such a
resolution, a phenomenological model yielding a nonsingular bounce
based on the introduction of a term $\propto-\rho^2$ into the
Friedmann equations, as in Sec.~\ref{LQG}, has been used in
\cite{Garriga:2013cix,Gupt:2013poa}. In this study and in related
works \cite{Piao:2004me,Piao:2009ku,Liu:2014uda}, the transition
between vacua during contraction and re-expansion was
computed. Putting aside the theoretical shortcomings of the model used
to replace the big crunch by a nonsingular bounce, the results of this
work are of interest: if the vacuum is AdS ($\mathcal{K}=-1$)
subsequent bounces take place until the field eventually emerges in a
de Sitter vacuum. During these transitions, the field usually jumps a
large distance of order $M_{\mbox {\tiny{Pl}}}$ in field space.
Hence, at least at the phenomenological level, the AdS bounces may
lead to transitions to remote parts of the landscape, reaching regions
otherwise inaccessible. However, tachyonic instability and parametric
resonance amplify scalar field fluctuations within the AdS bubble,
albeit less efficiently than in slow roll inflation.  If the
fluctuations remain small, the whole bubble transitions to a similarly
smooth vacuum; on the other hand, if fluctuations become large, the
bubble volume fragments into different final vacua after the bounce.
Transitions from one AdS vacuum to another one lead to further
amplification, enhancing the probability of bubble fragmentation. This
is reminiscent to models of eternal inflation discussed in
\cite{Johnson:2011aa}. Bubble wall fluctuations can give rise to
strong anisotropies in the contracting AdS bubble, leading to BKL
instabilities and Kasner periods, see Sec.~\ref{BKL}, with the
eventuality of further bubble fragmentation. In \cite{Liu:2014uda}, it
was found that bubbles fragment within two or three transitions based
on the enhancement of field perturbations induced by the amplification
of curvature perturbations. In a follow up study
\cite{vilenkin:2014yva} it was shown that even in the presence of AdS
bounces, space-time is still past-incomplete as in inflationary
cosmology. Thus the initial singularity is not resolved, but merely
pushed out of sight and hence, as in the corresponding inflationary
framework, physically inconsequential.

\subsection{Other models}  

The models presented above represent the mainstream ideas that have
been proposed to implement a bouncing alternative to inflation. We
conclude this general model presentation by identifying some
miscellaneous proposals \cite{Novello:2008ra}, which are generally
viewed as less fashionable and/or are hampered by conceptual problems.

 \subsubsection{Ho\v{r}ava-Lifshitz}
\label{Horava}
 
Ho\v{r}ava-Liftshitz (HL) gravity, first introduced in
\cite{Horava:2008ih}, is a power-counting renormalizable theory of
gravity with purportedly consistent UV-behavior and a fixed point in
the IR-limit \cite{Horava:2009uw,Horava:2009if}.  Therefore, as a
modification to General Relativity at high energies, this theory was
explored significantly within the context of cosmology: cosmological
solutions with matter and the possibility of a nonsingular bounce were
studied in
\cite{Calcagni:2009ar,Kiritsis:2009sh,Brandenberger:2009yt,Maier:2013yh}.
HL gravity was shown to have inconsistencies in \cite{Henneaux:2009zb}
and more recently, to be UV-incomplete \cite{Kimpton:2013zb}.  We
therefore do not dwell on these models further.

\subsubsection{Lee-Wick and Quintom}

Lee and Wick \cite{Lee:1969fy,Lee:1970iw} proposed, in the late
sixties, a finite version of QED; based upon this proposition,
Grinstein \etal constructed a modification to the standard model known
as the {\it Lee-Wick Standard Model} \cite{Grinstein:2007mp}; this
model aspires to provide an alternative to supersymmetry.  A feature
of these models is the presence of {\it phantom fields}
\cite{Caldwell:1999ew}.  Phantom fields have several conceptual
problems \cite{Cline:2003gs} since the equation of state parameter is
less than $-1$, but they might enable a bounce.  In addition, a future
singularity is present and the vacuum is unstable. Despite these
problems, Lee-Wick theory continues to be investigated and several
nonsingular bouncing models have been constructed within its
framework.  For instance, a nonsingular bounce caused by a Lee-Wick
type scalar field theory was studied in \cite{Cai:2008qw} providing a
realization of the matter bounce scenario; the authors found a
scale-invariant spectrum for both scalar perturbations and
gravitational waves, in agreement with \cite{Allen:2004vz}, see
Sec.~\ref{adiabatic_fluctuations_mb}. See \cite{Bhattacharya:2013ut}
for a follow up study.

Quintom models (two matter fields, one regular, the other with a wrong
sign kinetic term to violate the NEC \cite{Feng:2004ad}, see
\cite{Brandenberger:2012zb} for a review) alleviate the problem of a
future singularity, leading to a proposal of a quintom bounce
\cite{Feng:2004ad,Feng:2004ff,Guo:2004fq,Cai:2007zv,Cai:2007qw}.

Besides providing a nonsingular realization of a bounce, which is
hampered by the conceptual problems of Lee-Wick theory, these models
do not provide any additional desirable features; we shall therefore
not consider them any longer.  It should however be acknowledged that
the techniques developed for studying perturbations and extracting
observables in these toy models served as a stepping stone towards the
development of healthier scenarios such as galileon bounces in
supergravity, discussed in Sec.~\ref{galileon}.

\subsubsection{$F(R)$, $f(T)$ and Gauss-Bonnet gravity}

Most of the models developed above imply modifying gravity in one way
or another, yet they do not include the simplest such possibility,
namely that for which the Einstein-Hilbert Lagrangian is replaced by
an arbitrary function of the Ricci scalar
$F(R)$~\cite{DeFelice:2010aj}. Assuming a flat FLRW metric, one can
reconstruct the function $F$ that would be required to produce a given
bouncing behavior for the scale factor~\cite{Bamba:2013fha}. In this
context, dark energy models incorporating a bounce were constructed in
\cite{Astashenok:2013kka} and the interaction between dark energy and
dark matter was studied in \cite{Brevik:2014lpa}; it turns out that a
Gaussian bounce, with scale factor behaving as $a(t)\sim
\exp{(t/t_0)^2}$, occurs only in the unphysical region with $F'(R)<0$
and $F''(R)>0$, hence leading to an instability with respect to the
production of tensor perturbations. On the other hand, a power law
bounce $a\sim a_0 + (t/t_0)^q$ ($q\in 2\setN$) can take place with
simple monomial forms of $F$, in the stability region. In addition,
the latter models can be smoothly connected to more common contracting
and expanding phases.

A similar construction, leading to a bouncing phase connecting two
respectively contracting and expanding de Sitter phases, can also be
performed, yielding a possible oscillatory signal in the spectrum of
gravitational waves \cite{BouhmadiLopez:2012qp}. In this case, the
mass of the associated scalar field can become negative close to the
bounce, leading to instabilities, which manifest themselves in the
ensuing spectrum.

The same technique of Lagrangian reconstruction can be used assuming a
function of the Gauss-Bonnet invariant $\mathcal{G} \equiv R^2
-4R_{\mu\nu}R^{\mu\nu} +R_{\mu\nu\alpha\beta}R^{\mu\nu\alpha\beta}$
instead of one of the curvature Ricci scalar $R$.  Bouncing solutions
were explored in \cite{Bamba:2014mya}, some of which were found to be
stable.

Instead of choosing an arbitrary function of the Gauss-Bonnet term,
one can also consider a non local extension whereby an analytic but
otherwise arbitrary function of the D'Alembertian operator
$\square\equiv \nabla^\mu \nabla_\mu$ is inserted in between the
non-linear terms; for instance $R^2 \mapsto R f(\square) R$. This
procedure naturally introduces a new energy scale and the new terms
behave in the FLRW case as effectively negative energy fluids.
Because such terms are essentially non local, perturbations are
difficult to implement, but some arguments have been presented that
suggest these solutions to be stable \cite{Koshelev:2013ida}.

Instead of using the curvature scalar $R$ as the basic ingredient to
build the action by means of the torsionless Levi-Civit\`a connection,
one can also use the curvatureless Weitzenb\"ock connection
constructed from the {\sl vierbein} $e^{\scriptstyle A}_\mu$, the
metric $g_{\mu\nu}(x)= \eta_{\scriptstyle AB}e^{\scriptstyle A}_\mu
(x) e^{\scriptstyle B}_\nu (x)$, and the local Minkowski metric
$\eta_{\scriptstyle AB}$. This yields the so-called ``teleparallel"
Lagrangian \cite{Hayashi:1979qx}, which is nothing else but the
torsion scalar $T$, which provides a different, yet equivalent,
formulation of GR.  Adding an arbitrary function of the torsion
provides a natural extension, which has been investigated recently in
view of explaining the observed acceleration of the Universe
\cite{Linder:2010py}. An advantage of $f(T)$ models over $F(R)$ ones
is that their equations of motion remain second order, and therefore
reduce the risk of instabilities. The gravitational part of such
models can effectively violate the NEC, thus implying possible
bouncing solutions, even for vanishing spatial curvature
\cite{Cai:2011tc}. The procedure, however, requires special forms of
the otherwise arbitrary function $f(T)$.

The spectrum of perturbations predicted in the expanding phase for the
models detailed above, and hence their compatibility with the data, is
unknown.

\subsubsection{Brane worlds and extra-dimensions }
\label{brane_world}

String theory can be made mathematically self-consistent provide
spacetime has more than 4 dimensions, the extra dimensions being
usually assumed to be internal and small.  Branes are extended
objects in this framework, which can move in those internal
dimensions. The corresponding 4D-effective field theories can be
sufficiently rich to enable a bounce. For example, the ekpyrotic
scenario, as originally envisioned, can be seen as an example of a
brane-world set-up, see Sec.~\ref{ekcyclic}.

The Gauss-Bonnet action, although non-dynamical in 4 dimensions, can
also be obtained in the low energy limit of heterotic superstring
theory and studied in any other number of dimensions. Thus, such a
term is well suited to investigate brane-world scenarios: in
\cite{Maeda:2011px}, a 4 dimensional brane, on which a perfect fluid
with constant equation of state evolves, is embedded in a 5
dimensional Randall-Sundrum \cite{Randall:1999ee} like setting. The
conditions are derived under which the brane scale factor can
bounce. A branch singularity, actually a real curvature singularity,
exists at a finite physical radius in the bulk: when the brane
encounters this singularity, its scale factor instantaneously bounces
from a contracting to an expanding phase; there is no telling as to
what will happen with perturbations.

A bouncing solution was also found in the case in which the D3-brane
is the boundary of a 5 dimensional charged anti-de Sitter black hole
\cite{Mukherji:2002ft}: it is the charge $Q$ which provides the
negative energy regularizing term in the effective 4D Friedmann
equation, and thus permits the avoidance of the singularity through a
bouncing phase. On the 4D brane, this charge behaves as a stiff matter
fluid, $\rho\propto -Q^2 a^{-6}$.

 Relevant for the ekpyrotic scenario is the study in
\cite{Turok:2004gb}, which provides a semiclassical treatment of the
collision between two empty orbifold planes that approach each other
at constant speed. In this toy model, it is shown that the big
crunch/big bang transition appears smooth in the sense that certain
states can propagate smoothly across the transition. It is further
argued that interactions should be well-behaved since the string
coupling approaches zero during the transition. However, a realistic
transition remains an active field of research. 

Conceptually unrelated to brane-world scenarios in string theory, one
may entertain the idea of extra timelike dimensions. These might pose
conceptual questions, as it is not clear at the time of writing if
they are compatible with causality, if they predict tachyonic modes
and/or if they entail negative norm-states as is sometimes argued.
Ignoring these questions for the time being, it is possible to
construct bouncing cosmologies in this new framework. Considering a
Randall-Sundrum-like scenario with an extra timelike dimension instead
of a spacelike one, the effective energy momentum tensor contains a
term proportional to $-\rho^2$, which enables the transition from
contraction to expansion at high energies
\cite{Shtanov:2002mb,Brown:2004cs}. While these theories differ from
Randall-Sundrum models by merely a sign, they have never been
implemented in string theory.  Cosmological perturbations were studied
in \cite{Battefeld:2005cj,Geshnizjani:2005hc}, where it was found that
a scale-invariant spectrum can survive such a particular nonsingular
bounce, if it is generated in the preceding contracting phase. Since
the term quadratic in the energy momentum tensor does not add
additional degrees of freedom at the perturbed level, one cannot apply
the results derived in two field models, as in \cite{Bozza:2005xs}.

More recently, bouncing brane world cosmologies were considered in
\cite{Maier:2009zza,Maier:2013hr}.  Under certain restrictive
conditions, perturbations are found to be bounded in this category of
models, and sometimes sufficiently small to justify the use of
perturbation theory \cite{Maier:2013gua}.

\subsubsection{Non relativistic quantum gravity}

There are theoretical frameworks in which LI is not necessarily
fundamental but might instead arise as an emergent property of
space-time.  Such an approach permits theories in which this symmetry
is not implemented from the start, opening the possibility to quantize
the spatial degrees of freedom independently, and leading to a
non-relativistic quantization of gravity. Since Lorentz invariance is
an extremely well-tested symmetry of nature, see Sec.~\ref{LQG}, and
an integral part of high energy physics and string theory, it is often
challenging to reconcile such proposals with observations.  Ignoring
these conceptual pitfalls, one may try to construct bouncing
cosmologies under these conditions.

Effective field theory within such frameworks allows up to
6$^\mathrm{th}$ order spatial derivatives in the action, which can
contain all scalar combinations of the tensors $R_{ij}$ and $\nabla_i
R_{jk}$ (together with some matter contribution).  Restricting
attention to the FLRW metric, Cai \etal \cite{Cai:2009in}, find a
dark radiation term with negative energy density in the Friedmann
equations, provided that the spatial sections are not flat. Thus,
bouncing and even cyclic solutions, can be obtained. Incorporating a
matter-dominated contracting phase, perturbations have been found to
possibly induce a slightly red tilt in the spectrum, although the
cyclic continuation may generate backreaction problems.

\subsubsection{Mimetic matter}

The mimetic matter model \cite{Chamseddine:2013kea} mimics a phantom
field by introducing an auxiliary metric $\tilde g_{\alpha\beta}$ and
a scalar field $\phi$ in terms of which the actual metric reads
$g_{\mu\nu} = \tilde g^{\alpha\beta} \partial_\alpha\phi
\partial_\beta\phi \tilde g_{\mu\nu}$; this happens to be equivalent
to a dark matter component, the scalar field being not entirely
dynamical because of the normalization condition $g^{\mu\nu}
\partial_\mu\phi \partial_\nu\phi = 1$. The action is given by the
Ricci scalar of $g$ and contains just one extra longitudinal degree of
freedom. It can be supplemented by an arbitrary potential
$V(\phi)$. If the FLRW background metric is used, the scalar field
becomes a function of time, leaving the potential to behave in the
Friedmann equations as another arbitrary function of time.  Choosing
for instance $V\propto (1+\phi^2)^{-2}$ leads to bouncing solutions
\cite{Chamseddine:2014vna}. However, because of the non dynamical
nature of the scalar field involved, canonical quantization is not
always feasible and therefore setting initial conditions for
perturbations can be impossible.

\subsubsection{Nonlinear electromagnetic action}

Before inflation was conceived, Novello \etal \cite{Novello:1979ik}
proposed to implement a bounce in a cosmological framework, in which
the matter content is provided by a massless vector field
$A^\mu$. These models rely on two categories of modifications of
electromagnetism, \ie models with non-standard coupling to gravity,
using terms in the Lagrangian of the form $R A^2$, $R_{\mu\nu} A^\mu
A^\nu$, $R F_{\mu\nu}F^{\mu\nu}$, and models with scalar quantities
similarly built out of the curvature and the electromagnetic tensors.

Extending electromagnetism to include nonlinear terms such as those in
the Euler-Heisenberg corrections \cite{Heisenberg:1935qt} provides
another option to generate a bounce: the action becomes an arbitrary
function of the invariants $F_{\mu\nu}F^{\mu\nu}$ and
$\varepsilon_{\alpha\beta\mu\nu} F_{\alpha\beta}F^{\mu\nu}$. Such
terms are, however, usually only justified provided that the
electromagnetic fields vary slowly compared to the electron length
scale, but one can argue that similar terms should be obtained in the
more general situation one is concerned with in the early stages of
cosmological evolution.

By including a combination of such terms, the FLRW symmetry is kept by
demanding either that some averaging procedure is applied to the
electromagnetic field, or that the vector field has the special
timelike structure $A_\mu =A(t) \delta_\mu^0$. Bouncing solutions
arise because the extra terms contribute negative quantities of energy
density.  Similar ideas were revisited to provide more general
nonsingular solutions with either massless or massive vector fields
sourcing gravity \cite{Artymowski:2011tu}.

The models described in this section and others are discussed in
greater depth in \cite{Novello:2008ra}. We shall not consider them
further, since it is not clear whether they can actually be
constructed self-consistently, \ie without having insoluble intrinsic
difficulties (ghosts, violation of causality, shear/vector mode
overproduction, \etc), and because their cosmological relevant
consequences have not been established.

\subsubsection{Spinors and torsion}

A less known method of modifying gravity is provided by the
Einstein-Cartan-Sciama-Kibble extension
\cite{Kibble:1961ba,Sciama:1964wt}, in which the affine connection is
not necessarily symmetric, leaving the torsion tensor to behave as an
independent dynamical variable. These new degrees of freedom couple
only to spin densities and vanish outside material bodies, rendering
them useless in most contexts. Since there is no exterior in
cosmology, a fermionic field would induce an everywhere non-vanishing
spin density, whose coupling to the torsion behaves as a NEC violating
term in the Friedmann equations.  This can generate bouncing solutions
of a special kind, as the scale factor reaches a non-vanishing minimal
value at a cusp, \ie the Hubble rate is discontinuous at the bounce
point \cite{Brechet:2007cj,Brechet:2008tz,Poplawski:2011jz}. This
raises serious questions regarding stability and the fate of
perturbations.

A similar model, in which a topological sector is added to gravity,
was proposed in which the bounce behaves in a more regular way; by
adjusting the fermion number density and the mass, such a model can
reproduce a scale-invariant spectrum of perturbations
\cite{Alexander:2014eva}.

\section{Requirements for a successful  bounce}\label{requirements}

\begin{figure}[tb]
\begin{center}
\includegraphics[scale=0.95]{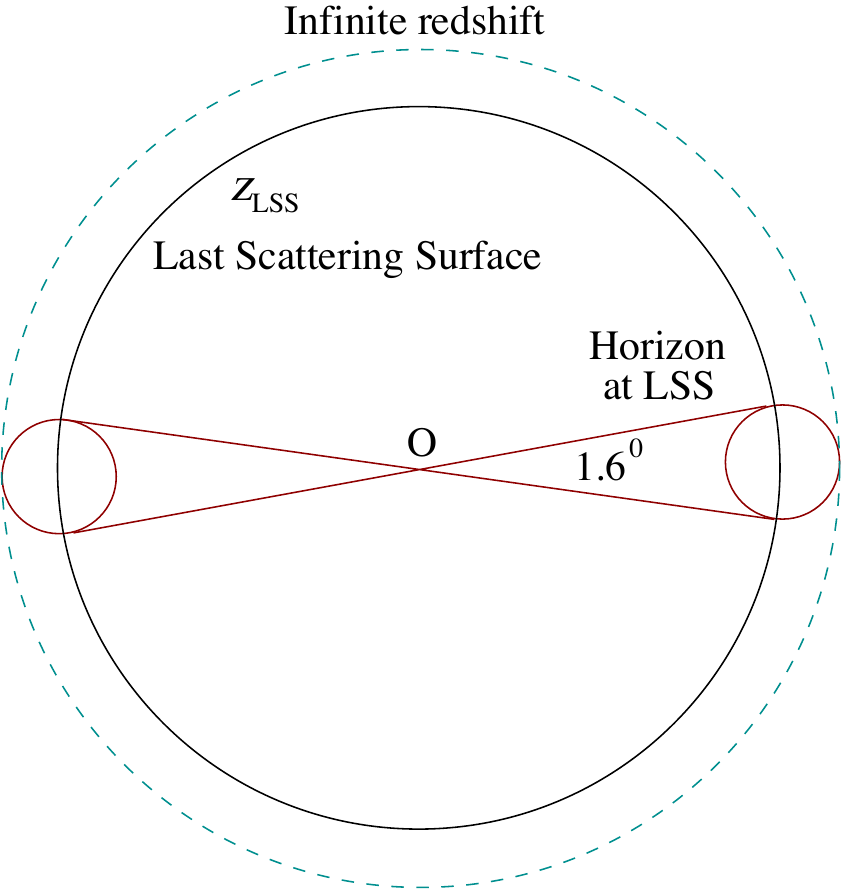}
\caption{Schematic view of the horizon problem: in the absence of
  inflation, causally connected patches of the surface of last
  scattering subtend small angles; yet, the temperature of the CMBR is
  isotropic to one part in $10^5$ over the entire sky and exhibit correlations
  at all observable scales, \ie on over angular scales much larger than
  the degree scale.
\label{LSS}}
\end{center}
\end{figure}

In order to provide a viable alternative to inflation, a model should,
at least, do as well as inflation in many respects. This implies that
such a model's cosmological predictions must not only be compatible
with the currently available data, but also have a sound theoretical
foundation. As we shall see, this is not an easy endeavor; before
moving to these difficulties, we discuss the common puzzles of the
standard hot big-bang model and their proposed solutions in bouncing
scenarios.

\subsection{Cosmological puzzles}

Inflation was proposed as a way out of three observational conundrums:
why is the universe isotropic on the largest accessible scales (the
horizon problem)? Why does the content of the universe sum up in the
exactly required fashion so as to make its spatial curvature
negligible (the flatness problem)? Why do we not observe an absurdly
large number of thermal relics, such as gravitinos in supersymmetric
theories, or topological defects from phase transitions, such as
primordial magnetic monopoles that should have been copiously produced
during a grand unification transition (the relic problem)? Before we
consider these questions, we would like to address a point that is
often ignored in the literature on inflationary cosmology,that of the
primordial singularity.

\subsubsection{A primordial singularity?}

Ever expanding cosmologies have been shown to be past incomplete
\cite{Borde:2001nh}, so that, as far as classical gravity is
concerned, the Universe began with an initial singularity. This
problem, if one sees it as one, is not directly addressed in the
inflationary framework: it is often postulated that the cosmological
evolution begins in an epoch during which the relevant physical
theories are well understood; previous phases are thought to be in the
realm of quantum gravity and it is assumed that they have limited
influence on the scales of observational relevance. Demanding that
inflation ``solves'' the singularity problem would be similar to
demanding that Big-Bang Nucleosynthesis (BBN) explains why the
Universe was homogeneous and radiation dominated at the time it took
place: it seems to us that one can make this hypothesis and assume
some other physics to provide the necessary explanation without
hampering the predictivity of BBN. Similarly, we think it is perfectly
reasonable to assume that the primordial singularity is somehow
resolved; an option would be to connect the currently expanding
Universe to a previously contracting phase through a bounce. In this
sense, studying bouncing solutions addresses an extremely relevant
question ignored, or perhaps overlooked, in the inflationary paradigm.

\subsubsection{Horizon problem}

\begin{figure}[tb]
\begin{center}
\includegraphics[scale=0.475]{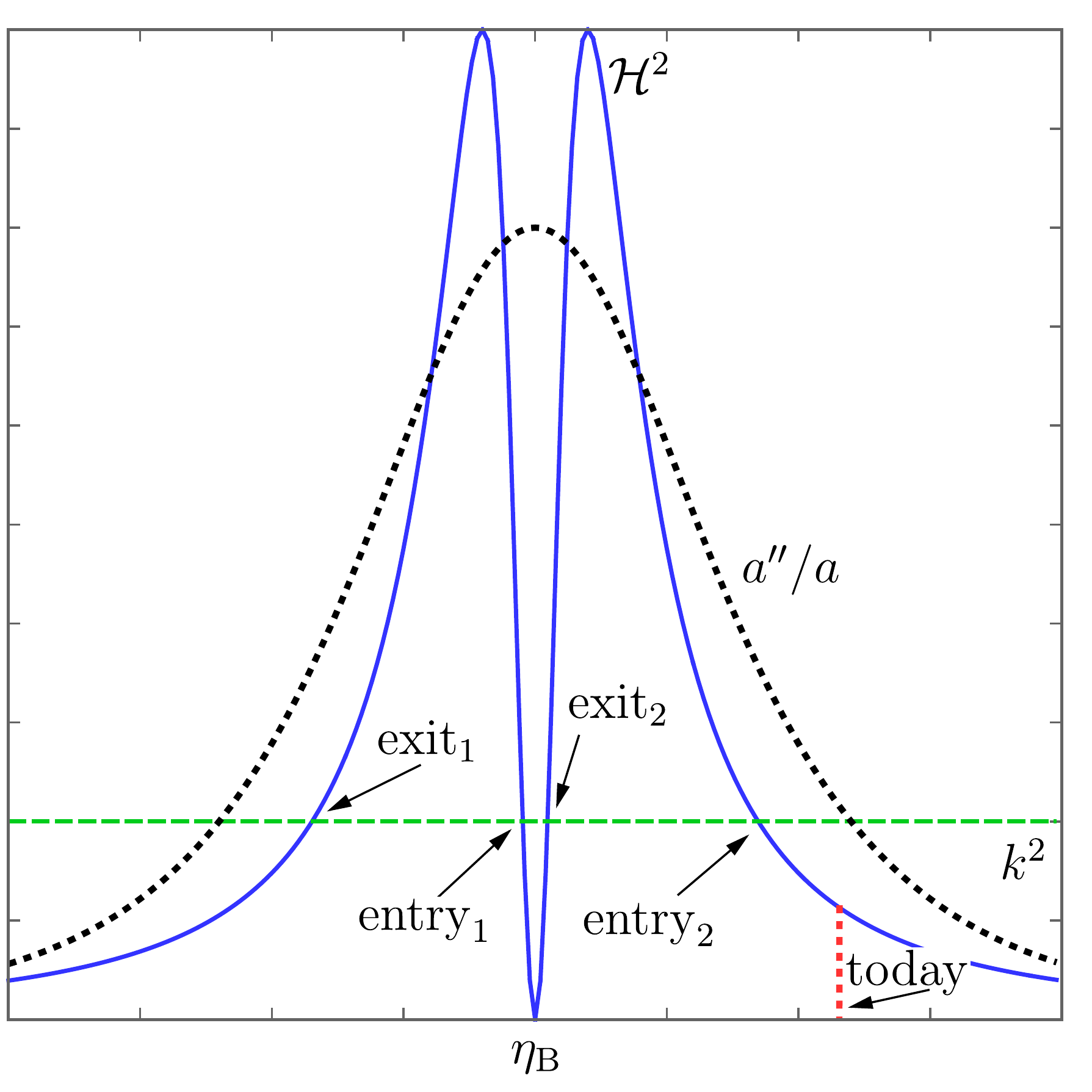}
\caption{Schematic of the Hubble crossing history of a mode with
  wavenumber $k$ (see also the time-line in Fig.~\ref{fig:4}): the
  mode first becomes larger than the Hubble scale at
  $t_\mathrm{out}^{(1)}=t^{(1)}_\mathrm{hc}$ in the pre-bounce phase,
  smaller at $t^{(1)}_\mathrm{in}=t_\mathrm{hc-entry}$, close to the
  actual bounce, and larger again for a second time shortly after the
  bounce at $t_\mathrm{out}^{(2)}=t^{(2)}_\mathrm{hc}$ before entering
  the Hubble radius later on, at $t^{(2)}_\mathrm{in}$. The plot is
  shown in terms of conformal time $\eta=\int a\dd t$ and the
  conformal Hubble factor is $\mathcal{H}\equiv
  a'/a$. 
  although correct in some models, this picture, contrary to its
  inflation counterpart, is not generically meaningful, as the
  potential entering in the perturbation equation is not necessarily
  proportional to the Hubble scale; this is illustrated with the
  example of the tensor mode potential $a''/a$ in (\ref{gralsoln})
  (dotted line) which clearly differs from $\Hu^2$ as, in particular,
  re-entry and exit of the Hubble sphere are seen to be absolutely
  irrelevant.
\label{spacetimeNSMB} 
}
\end{center}
\end{figure}

Big-bang cosmology provides a successful description of the evolution
of the Universe back to a fraction of a second after its birth; it is
consistent with the Hubble expansion, the cosmic background radiation
and the abundance of light elements.  Extrapolating to early times, we
encounter what is called the horizon problem.  The main assumption in
big bang cosmology is large-scale homogeneity and isotropy. This
assumption is in agreement with the cosmic microwave background
radiation, whose temperature, if measured in two different, opposing
patches of the sky, is the same to within at least one part in
$10^5$. Because the big bang contains an initial singularity, a causal
horizon exists beyond which one should not expect similar
thermodynamical properties.  Based upon this fact, opposite patches in
the sky could never have been in causal contact in the standard big
bang model, and yet they have the same CMBR temperature to one part in
$10^{5}$.

The horizon size $d_{_\mathrm{H}}\equiv a(t) \int_{t_\mathrm{i}}^t \dd
\tilde t/a\left(\tilde t\right)$, during a radiation and matter
dominated universe is of order $t$ when the origin of time is
$t_\mathrm{i}\ll t$ and the scale factor has power-law behavior for
all times.  At the time of last scattering the horizon size is
\cite{Peter:2009sa}
\begin{equation} 
d_{_\mathrm{H}} \approx \frac{1}{H(1+z_{_\mathrm{LSS}})^{3/2}}\,,
\end{equation}
where $z_{_\mathrm{LSS}}$ is the redshift of the last scattering
surface; the angular diameter distance to this surface is
\begin{equation}
d_{_\mathrm{A}}\approx \frac{1}{H(1+z_{_\mathrm{LSS}})}
\end{equation}
at the time of last scattering, so that the causal horizon size
subtends an angle of
\begin{equation}
\frac{d_{_\mathrm{H}}}{d_{_\mathrm{A}}}\approx\frac{1}{(1+z_{_\mathrm{LSS}})^{1/2}}\,
\end{equation}
radians. For a redshift to the surface of last scattering,
$z_{_\mathrm{LSS}}=1100$, we get
$d_{_\mathrm{H}}/d_{_\mathrm{A}}\approx 1.6^\circ$. Thus, patches of
the universe that were separated by more than this have no causal
reason to have the same temperature.  This is illustrated in
Fig.~\ref{LSS}.

In the big bang model, it is assumed that the universe was originally
highly homogeneous and isotropic on scales larger than the causal
horizon, indicating a high degree of fine-tuning. One might argue that
such initial conditions make no sense in the framework of GR.

Inflation solves this puzzle by adding a phase during which the scale
factor grows quasi-exponentially, in such a way that the causal
horizon grows larger than any other physically relevant scale. The
Hubble scale $H^{-1}\equiv a/\dot{a}$ remains more or less constant,
so the scale factor behaves roughly exponentially,
$a_\mathrm{inf}\propto \ex^{Ht}$, leading to an exponentially
increasing horizon, \ie $d^\mathrm{inf}_{_\mathrm{H}}\sim H^{-1}
\ex^{H\Delta t}$, with $\Delta T$ the duration of the inflationary
phase. It suffices that this duration be large enough, in practice
$H\Delta T\geq 60$, so that the resulting horizon scale is much larger
than the entire observable Universe today. Moreover, a given quantum
fluctuation of wavelength $\lambda$ sourced in the far past can start
out smaller than $H^{-1}$; due to its subsequent growth $\propto a$,
the wavelength becomes larger than $H^{-1}$, which remains roughly
constant. Nevertheless, it remains within the causal horizon, which
grows tremendously: no scale actually ever becomes
``super-horizon''. This is necessary for any consideration in GR,
including the setting of initial conditions, to make sense.

Bouncing models solve this puzzle in a completely different way. As
far as the background is concerned, consider a contracting phase
between $t_\mathrm{ini}<0$ and $t_\mathrm{end}<0$ dominated by a
perfect fluid with constant equation of state parameter $w$, so that
the scale factor behaves as $a_\mathrm{cont}\propto
(-t)^{2/[3(1+w)]}$; we assume the bounce to take place at $t=0$. The
contribution of this contracting phase to the horizon is (we correct a
misprint in \cite{Peter:2008qz} from which the argument is taken)
\begin{equation}
d^\mathrm{cont}_{_\mathrm{H}} = \frac{3(1+w)}{1+3w} t_\mathrm{end}
\left\{ 1 - \left( \frac{t_\mathrm{ini}}{t_\mathrm{end}}\right)^{(1+3w)/[3(1+w)]}
\right\},
\end{equation}
which can be made arbitrarily large for  $|t_\mathrm{ini}|\gg
|t_\mathrm{end}|$ provided that $w>-1/3$.

As for the perturbations, we consider that quantum fluctuations are
sourced in the far past, deep within the horizon and the Hubble
scale. The horizon itself grows at all times, and it is possible to
have it growing more rapidly than the scale factor, so that a
wavelength initially smaller than the horizon remains so at all
subsequent times. During a slow contraction, the wave modes stay
approximately constant, whereas the Hubble scale is rapidly shrinking
as the bounce is approached; thus, modes which are sourced by quantum
mechanical fluctuations inside the Hubble radius become super-Hubble
during the contraction\footnote{Colloquially, Hubble radius crossing
  of a particular mode is often referred to as ``horizon'' crossing,
  even though the Hubble radius is not a horizon in a contracting
  universe (and neither is it, as far as causality arguments are
  concerned, in an inflationary universe); we will not use this
  possibly confusing terminology \cite{Martin:2003bp}, particularly in
  view of the fact that both inflation and bouncing cosmologies are
  introduced in order to solve the horizon problem in such a way that
  the actual causal horizon becomes much larger than any relevant
  scale.}, but remain sub-horizon; thus a causal mechanism to seed the
observed structures on large scales is present. The horizon problem is
solved, because a lot more time is available to establish causal
contact, see Fig.~\ref{spacetimeNSMB}.

\subsubsection{Flatness problem}

The present density of the universe is close to the critical density,
$\Omega_{\mathrm{total}}\sim 1$, where
\begin{equation}
\Omega_\mathrm{total}=\Omega_{\Lambda}+\Omega_\mathrm{m}
+\Omega_\mathrm{r}+\Omega_\Ka,
\end{equation}
with $\Omega_{\Lambda}$, $\Omega_\mathrm{m}$ and
$\Omega_\mathrm{r}$ the relative energy densities of the cosmological
constant, matter and radiation, respectively, while the
dimensionless time-dependent curvature parameter is 
\begin{equation} 
|\Omega_\Ka|\equiv\frac{|\Ka|}{a^2H^2}\label{curvature}.
\end{equation} 
Deviations from $\Omega_\mathrm{total}=1$ grow in time in a
decelerating expanding universe. In order to have
$\Omega_\mathrm{total}\sim 1$ today, it must have been extremely close
to one in the early universe, indicating fine-tuning.  Data from the
CMBR and Type Ia supernovae indicate that $|\Omega_\Ka|\ll1$.  Since
\begin{equation}
\frac{\dd |\Omega_\Ka|}{\dd t} = -2 |\Ka |\frac{\ddot a}{\dot a^3},
\label{dOKdt}
\end{equation}
a non-accelerating expanding phase, such as one dominated by a
radiation or a dust-like fluid, always increases $|\Omega_\Ka|$;
hence, observing a small $|\Omega_\Ka|$ today requires fine-tuning of
its initial value.

Specifically, the temperature of the universe dropped from about
$10^{11}\,$K at $1\,$s to approximately $10^4\,$K at
$1.78\times10^8\,$s, the beginning of the matter dominated era; from
then on up until today, the scale factor has been increasing as
$a(t)\sim t^{2/3}$ (we ignore the cosmological constant), so that the
curvature parameter in (\ref{curvature}) has also been increasing as
$t^{2/3}\propto T^{-1}$. Thus, for $|\Omega_\Ka|<1$ today, it had to
be less than $10^{-4}$ at $T\approx10^4\,$K.  Furthermore, during the
radiation dominated era, we have $a(t)\sim t^{1/2}$, so that
$|\Omega_\Ka|\propto t\propto T^{-2}$, indicating that
$|\Omega_\Ka|<10^{-16}$ at $T\approx 10^{10}\,$K.  That the value of
this dimensionless parameter ought to be so small compared to unity in
the early Universe is called the flatness problem
\cite{Weinberg:2008zzc}.

\begin{figure}[tb]
\begin{center}
\includegraphics[scale=0.47]{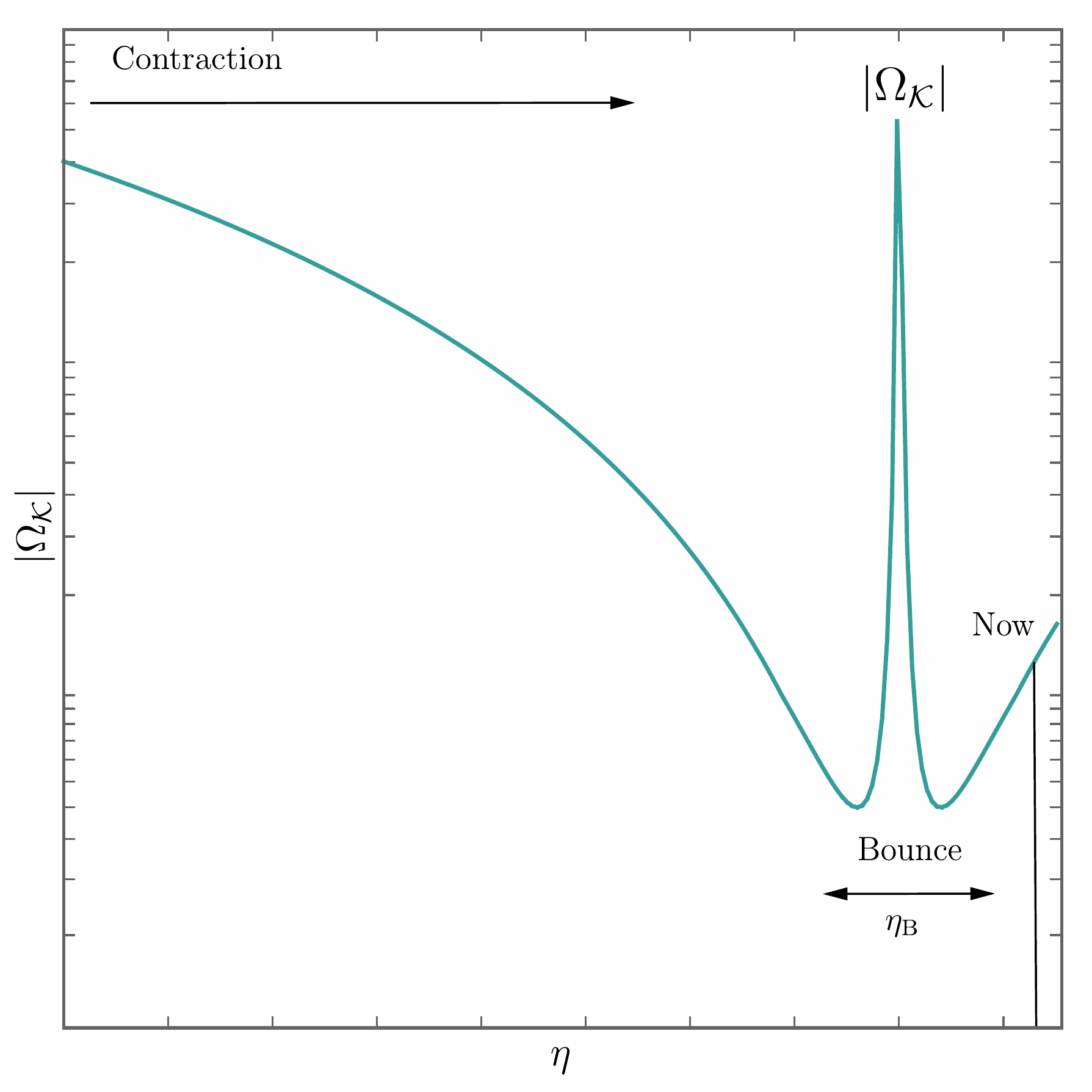}
\caption{How a long contracting phase solves the flatness problem:
  behavior of the relative curvature density during a bounce,
  $\Omega_\mathcal{K}$, as a function of conformal time $\eta$. During
  the contracting phase, the contribution of curvature to the total
  energy budget in the Friedmann equation decreases steadily. It then
  increases tremendously at the bounce (technically, it actually
  diverges when $\Hu\to 0$), but then returns almost to its pre-bounce
  negligible value, provided the bounce itself is sufficiently
  symmetric. If the elapsed time since the bounce to today is smaller
  than the time elapsed during the contracting phase, curvature still
  appears negligible today.}
\label{Omegak}
\end{center}
\end{figure}

Inflation solves this problem in a simple way: consider (\ref{dOKdt})
and add, for a sufficiently long period of time, a phase of
accelerated ($\ddot a >0$) expansion ($\dot a >0$). In this case, $\dd
|\Omega_\Ka|/\dd t <0$ and $|\Omega_\Ka|$ naturally evolves towards
small values. To ensure that $|\Omega_\Ka|\ll 1$ today, one needs
roughly $60$ e-folds of inflation if $|\Omega_\Ka|\sim 1$ initially.
At the end of a quasi-exponential phase with $a_\mathrm{inf} \propto
\ex^{Ht}$, (\ref{dOKdt}) indicates that $\Omega_\Ka =|\Ka|H^{-2}
\ex^{-2H\Delta t}$, again requiring $H\Delta T\sim 60$ in order for
the subsequent evolution of $\Omega_\Ka$ to be consistent with today's
upper bounds.

This problem is not solved in many alternative proposals to inflation:
in string gas cosmology fine-tuning is required and in models akin to
the original ekpyrotic proposal a vanishing $\Ka$ is selected
initially by symmetry arguments; for example, in the original
ekpyrotic proposal the brane that we live on is a BPS brane
\cite{Khoury:2001wf} (see however \cite{Kallosh:2001du}). In the
Ph\oe{}nix universe, the currently accelerated epoch is used as a
means of getting rid of the curvature term. In a way, the cyclic
scenario may be seen as yet another (very low energy) implementation
of the inflationary paradigm since in this model there are $60$
e-folds of cosmological constant domination before the contracting
phase, making everything flat \cite{Linde:2014nna}. In this sense, it
can be argued that it is not really an alternative to inflation.

However, in a bouncing scenario, the flatness problem can be solved in
an altogether different way: consider (\ref{dOKdt}) in a decelerating
($\ddot a <0$) and contracting ($\dot a^3 <0$) universe. The curvature
contribution can be made as small as desired during this contraction,
as shown in Fig.~\ref{Omegak}. However, close to a nonsingular bounce,
the curvature contribution grows again: consider the Friedmann
equation (\ref{FriedConf}) with the energy density $\rho$ of whatever
matter happens to contribute at that time
$$
1+\frac{\Ka}{a^2 H^2} = \frac{\rho}{3H^2},
$$
the r.h.s containing terms behaving like $a^{-3} H^{-2}$ (matter),
$a^{-4} H^{-2}$ (radiation) or even $a^{-6} H^{-2}$ (shear), all of
which dominate over the curvature term $Ka^{-2} H^{-2} $ whenever the
scale factor decreases ($a\to 0$). Therefore, if the bounce takes
place because of any term on the r.h.s., \ie in the energy density
$\rho$, it may be that the curvature remains negligible at the expense
of having a negative energy density source, thus potentially causing
new instabilities.

Subsequently, $\Omega_\Ka$ grows large at the bounce, although it has
to remain sufficiently small right after the bounce at the beginning
of the expansion phase, so that it can still be negligible today. The
amount of necessary ``fine-tuning'' is then transformed into a
requirement that the bounce be sufficiently symmetric: if the
curvature term was negligibly small at the end of the contracting
epoch, it stays roughly negligible at the beginning of the expansion
epoch. This fine tuning is exactly of the same nature as that found in
inflation, which requires the accelerated expansion phase to last at
least $60$ e-folds.  In many bouncing models, the relevant  amount
of contraction is in far excess of the amount of expansion after the
bounce until today.   As a result, the flatness problem is solved
as long as the bounce is sufficiently symmetric. This mechanism is
illustrated in Fig.~\ref{Omegak}.

\subsubsection{Avoidance of relics}
\label{relics}

Topological defects, exotic particles and even primordial black holes
(PBH) can be created during the early stages of the Universe. Since
estimates of the PBH production rate differ by many orders of
magnitude, they are commonly ignored by model builders. The other
kinds of relics must be considered carefully, since their production
rates are well understood, once a model is specified. See
\cite{Battefeld:2009sb} for a brief review by one of the authors of
this article and a collaborator, which we partly reproduce below.

Supersymmetric theories generically predict the existence of the
gravitino, the supersymmetric partner of the graviton. The gravitino
mass has its origins in spontaneous supersymmetry breaking and its
value ranges from GeV to TeV. Since the gravitino is long-lived, if
its dominant decay mode consists of a photon and its superpartner, it
provides a natural candidate for dark matter. However, even in the
absence of primordial gravitinos, they can be thermally produced
during the radiation dominated epoch: this is an example of a thermal
relic.  The presence of thermally produced relics such as gravitinos
imposes stringent constraints on the allowed maximal temperature in
the radiation epoch.  As a result, one finds an upper limit on the
reheating temperature of order $10^8\,$GeV. In any model of the early
universe, be it inflation or a bounce, this constraint must be
satisfied. Recall that supersymmetry is a key ingredient in string
theory, so these relics are natural in this context. This problem can
be alleviated by a second phase of reheating of a long-lived
oscillating scalar. Examples of such fields are the s-axion in
F-theory \cite{Heckman:2008jy} or moduli in G2-MSSM models arising
from M-theory compactifications \cite{Acharya:2008bk}, but a concrete
implementation in bouncing cosmologies has not been given.

A different example of heavy relics spoiling the subsequent evolution
of the Universe are super heavy magnetic monopoles as predicted in
theories entailing grand unification (GUT): since the photon is a
massless particle, we know for sure that at the current-day
temperature of the Universe, the gauge group of particle physics
contains a $U(1)$ subgroup.  Assuming a GUT based on a (semi)simple
gauge group, $\mathcal{G}$, the presence of a $U(1)$ factor in the
resulting low-energy symmetry group $\mathcal{X}$ implies that the
second homotopy group of the vacuum manifold $\mathcal{V} \sim
\mathcal{G}/\mathcal{X}$ is non trivial, \ie
$\pi_2(\mathcal{V})\not\sim \{\emptyset\}$. As a result, stable
solutions of the point-like kind (in 4D-spacetime) must form as
topological configuration in the symmetry-breaking Higgs and gauge
fields. With a symmetry breaking energy scale $E_{_\mathrm{GUT}}$ and
a unification coupling constant $q\ll 1$, this mechanism produces
objects whose mass can be estimated as $M_\mathrm{monopole} \sim
E_{_\mathrm{GUT}}/q$
\cite{Ringeval:2005kr,Peter:2013jj,Ade:2013xla,Huguet:1999bu}.

In the original hot big bang scenario, beginning with a singularity,
at least one such monopole per horizon is produced during the GUT
phase transition: this Kibble mechanism is due to causality, resulting
in an over abundance of order $\Omega_\mathrm{monopoles}\sim 10^{13}$
today, if the universe cooled down from the GUT temperature of
approximately $T_{_\mathrm{GUT}}\sim 10^{15}\,$ GeV.

The inflationary solution is again simple and natural
\cite{Guth:1980zm}: the accelerated expansion dilutes all prior relics
during inflation and as long as the reheating temperature is low
enough, no further relics are produced.  However, this problem
resurfaces acutely, for instance, in the S-brane bounce, since a
thermal component is present.  Since (p)reheating has not been studied
in bouncing cosmologies beyond simple estimates
\cite{Lehners:2011kr,Cai:2011ci}, it is unknown at the time of writing
whether or not thermal relics are formed\footnote{ Inflationary
  and bouncing solutions differ fundamentally; in the former, the
  reheating temperature is bounded from above by the energy scale of
  inflation, whereas, in the latter it is not a priory bounded. In
  \cite{Lehners:2008vx}, Lehners \etal envisioned reheating in the
  cylic scenario as occurring via the transfer of some kinetic energy
  between branes.  Given this reasoning, it is possible to argue that
  the temperature can be expected to be below the upper limits quoted.
  Yet, the uncertainty of reheating dynamics in the cyclic scenario,
  and other bouncing models, is considerable greater than in
  inflationary cosmology. This is pointed out in more detail in
  Sec. \ref{preheat}. }. Nevertheless, the defect question becomes
a crucial one, not on considerations of energy density and relative
contribution, but more fundamentally, because of the initial
conditions they demand: if many Higgs fields are originally present in
the large and cold universe, some of them must have vacuum expectation
values, which in turns means, for most of those, arbitrary phases. As
far as we know, there seems to be no natural and accepted way to set
up these phases; further, it is not even clear if such mechanisms
exist.\footnote{ Although the phase issue may still be an open one,
  the monopole question can possibly be solved in a string theory
  context since the effective field theory which emerges as its
  low-energy limit is in general not based on a simply connected GUT
  group, so that the relevant symmetry-breaking scheme may not include
  monopole formation
  \cite{Polchinski:1998rr,Braun:2005nv,Braun:2013wr}. }.

\subsubsection{Homogeneity and initial conditions: the Ph\oe{}nix
  universe as a case study}
\label{cyclic_ic}

Inflation initiates in a tiny region of space, assumed to be roughly
homogeneous; this region expands to a huge size, thus effectively
providing a mechanism to considerably alleviate (not solve), the
problem of having a homogeneous Universe. Given that we observe
homogeneity on sufficiently large scales, any alternative model should
also yield an explanation at least as satisfying as that provided by
inflation. In that respect, bouncing cosmology, with its contraction
phase, can be in trouble.

A simple possibility considered by one of the authors of this review
in \cite{Peter:2008qz} consists in arguing that a large universe
filled with diluted matter can be assumed to be initially roughly
homogeneous, as it is mostly empty (although one should impose some
extra constraints on the behavior of the Weyl tensor).  Provided the
contraction is sufficiently slow compared to the diffusion rate of the
particles present at such early stages, one expects any initial
inhomogeneity not only to remain small, but to be smoothed away
through diffusion processes, thus dynamically driving the universe
towards a homogeneous state of equilibrium. This method is, however,
not necessarily stable w.r.t.~the inclusion of a cosmological constant
\cite{Maier:2011yy}.

Another option is employed in the cyclic model
\cite{Steinhardt:2001st}, which traces back to Lema\^{\i}tre's closed,
oscillatory model of the universe undergoing repeated periods of big
bang, expansion, contraction and big crunch
\cite{Lemaitre:1933gd}. Contrary to Lema\^{\i}tre's model, the cyclic
universe has an added component, a phase of ekpyrotic contraction
\cite{Erickson:2003zm,Garfinkle:2008ei}, that smoothens and flattens
the universe. See Sec.~\ref{ekpyroticphase} and \cite{Lehners:2008vx}
for a review.  In accord with Lema\^{\i}tre's model, the underlying
idea of the original cyclic scenario is that the entire universe
partakes in the cycling.  Recycling the whole universe can be
problematic if the entropy density grows from cycle to cycle, since
our universe has low initial entropy; this problem can be avoided if
the universe increases sufficiently from one cycle to the next, so
that the entropy density does not grow. Nevertheless one still needs
to understand how the first cycle came into being.  Further,
initial conditions appear to be extraordinarily fine-tuned in any
model that uses the entropic mechanism to generate the scale-invariant
spectrum\footnote{See however the non-minimal entropic mechanism in
  \cite{Qiu:2013eoa,Li:2013hga,Fertig:2013kwa}, which has a stable
  direction in the entropy direction.}.

\begin{figure}[tb]
\begin{center}
\includegraphics[scale=0.59]{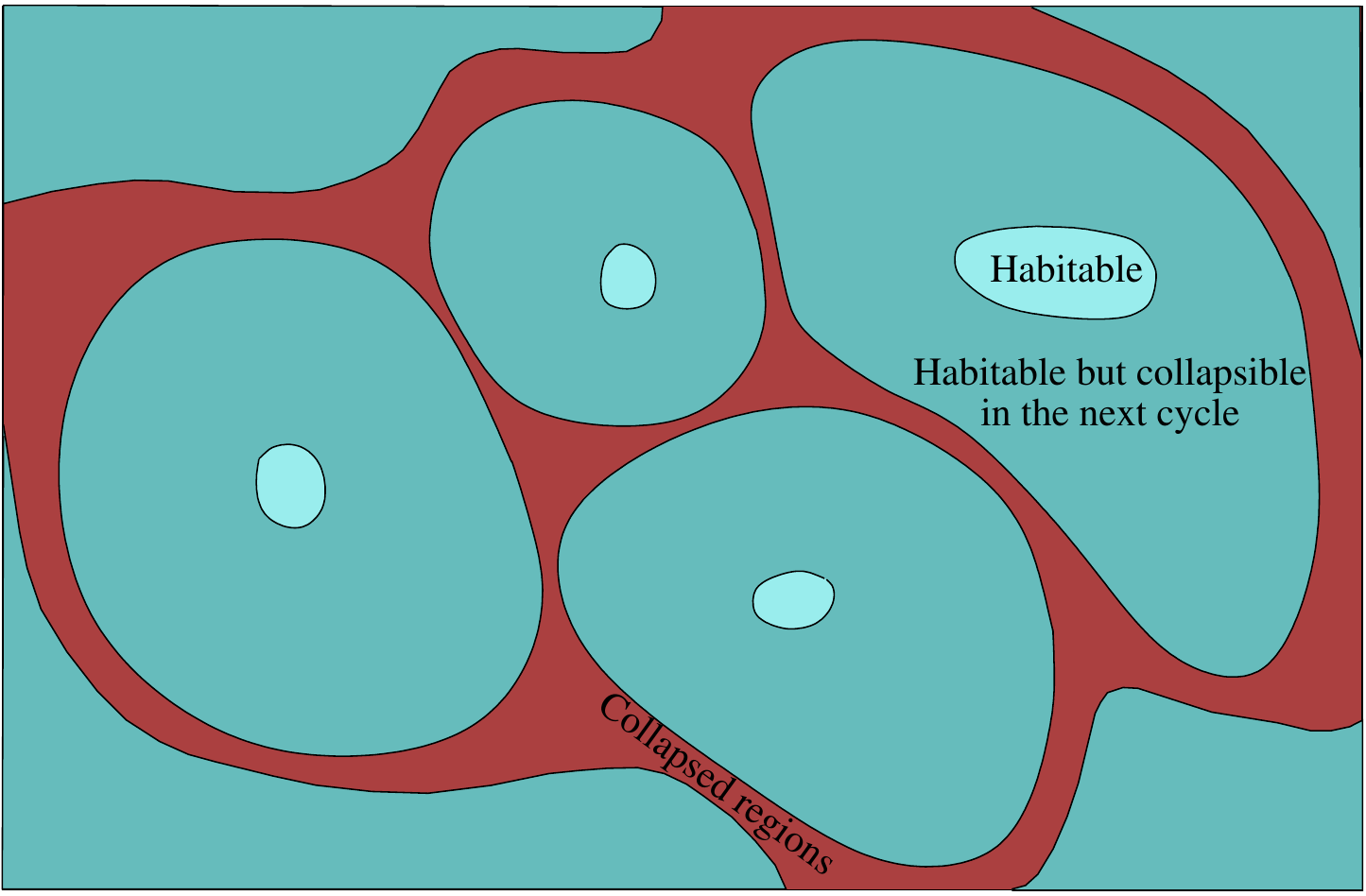}
\caption{ The Ph\oe{}nix Universe. Red regions have stopped cycling
  and are collapsed. Dark blue regions represent the smooth, flat
  regions (habitable) while the lighter blue areas are considerably
  empty and flattened during the dark energy phase to make it through
  the next bounce \cite{Lehners:2008qe}}.
\label{cyclic}
\end{center}
\end{figure}

\begin{figure*}[tb]
\begin{center}
\includegraphics[scale=0.8]{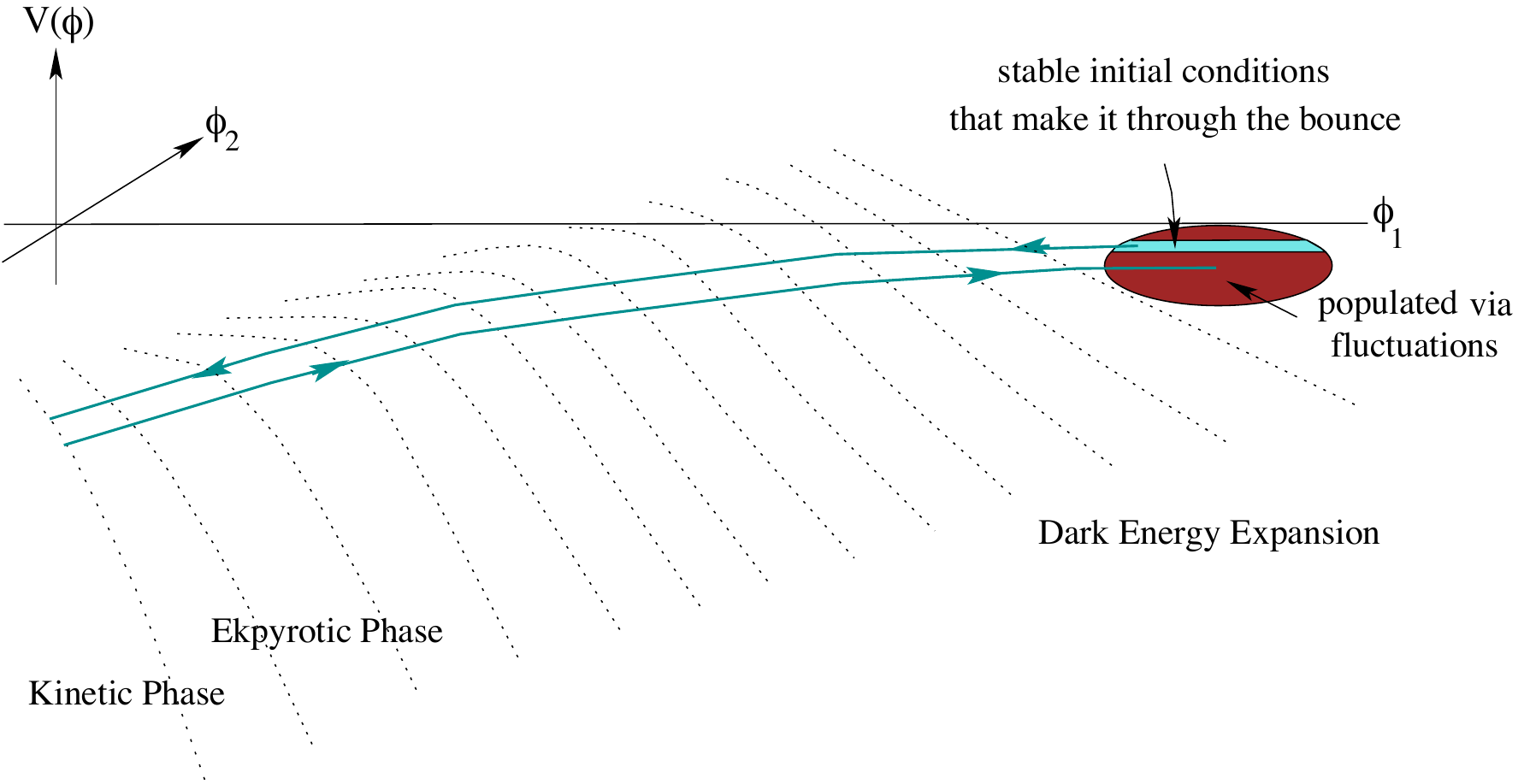}
\caption{Schematic of the potential for the Ph\oe{}nix universe
  \cite{Lehners:2008qe} during the dark energy domination. After the
  bounce, the trajectory goes back up the hill to the plateau, but,
  due to the presence of an unstable direction, it will be displaced
  from the initial trajectory. Initial values in the light blue region
  around the initial trajectory are sufficient to guarantee a
  successful next bounce. Other initial values, on the other hand,
  lead to a big crunch. During the dark energy domination, quantum
  fluctuations populate the red region. As long as the space-time
  region in which initial conditions lie within the blue patch grows
  from cycle to cycle, universes are reborn out of their ashes. This
  is guaranteed in the cyclic model if dark energy domination lasts
  for at least $60$ e-folds.}
\label{phoenix}
\end{center}
\end{figure*}

To alleviate these problems, the Ph\oe{}nix universe was proposed
\cite{Lehners:2008qe}: the universe is reborn from a surviving seed
found among its ashes, which goes hand in hand with generating
curvature perturbations by means of the entropic mechanism
\cite{Lehners:2007ac}. See Sec.~\ref{absenceNG}.  Namely, due to the
instability of the classical ekpyrotic trajectory along the potential
to transverse fluctuations, large portions of the universe are
converted into inhomogeneous remnants and black holes which are not
able to pass through the cycles.  Nevertheless, if a dark energy
expansion phase with at least $60$ e-folds is present before the
ekpyrotic contraction, a sufficiently large portion of space makes it
down the classical trajectory and through the big bang; this patch
grows from cycle to cycle. See Figs.~\ref{cyclic} and \ref{phoenix}.
The dark energy expansion makes space smooth and flat; thus, this
model seems to successfully address the question of flatness,
fine-tuning at the background level and initial conditions for
perturbations; in a way, such a model can be seen as a special
implementation of the low-energy inflationary paradigm with an added
curvaton mechanism to produce fluctuations, since the current-day
accelerating phase has a pre-big bang counterpart one could dub
inflationary.

An important consequence of the Ph\oe{}nix universe pertains to the
amplitude of primordial density fluctuations \cite{Lehners:2010ug}, as
parametrized by the so-called Mukhanov-Sasaki variable of the adiabatic mode
$Q_{\zeta}$, see (\ref{SM}) below. In contrast to inflation where
$Q_{\zeta}$ is fitted by hand to be in agreement with observations,
${Q_{\zeta}}_\mathrm{obs}\sim 10^{-5}$, in the cyclic model, patches
of the universe with the appropriate value of $Q_{\zeta}$ are
dynamically selected.  This model employs an entropic mechanism and a
change in the direction in which the scalar field moves after a turn
in field space to convert isocurvature modes to adiabatic ones, see
Sec.~\ref{absenceNG}. According to this mechanism, all patches that
make it through the bounce must have a value
\cite{Lehners:2011kr}\begin{equation} Q_{\zeta}\lesssim 10^{-4.5}.
\end{equation}

However appealing it might be, this model produces large amonts of
non-Gaussianities that stem from the entropic mechanism
\cite{Lehners:2011kr} together with hardly any gravitational waves,
see Sec.~\ref{rcyclic}. It is thus in tension with current
measurements \cite{Ade:2013uln,Ade:2013nlj,Ade:2013zuv}.

\begin{figure}[tb]
\begin{center}
\includegraphics[scale=0.7]{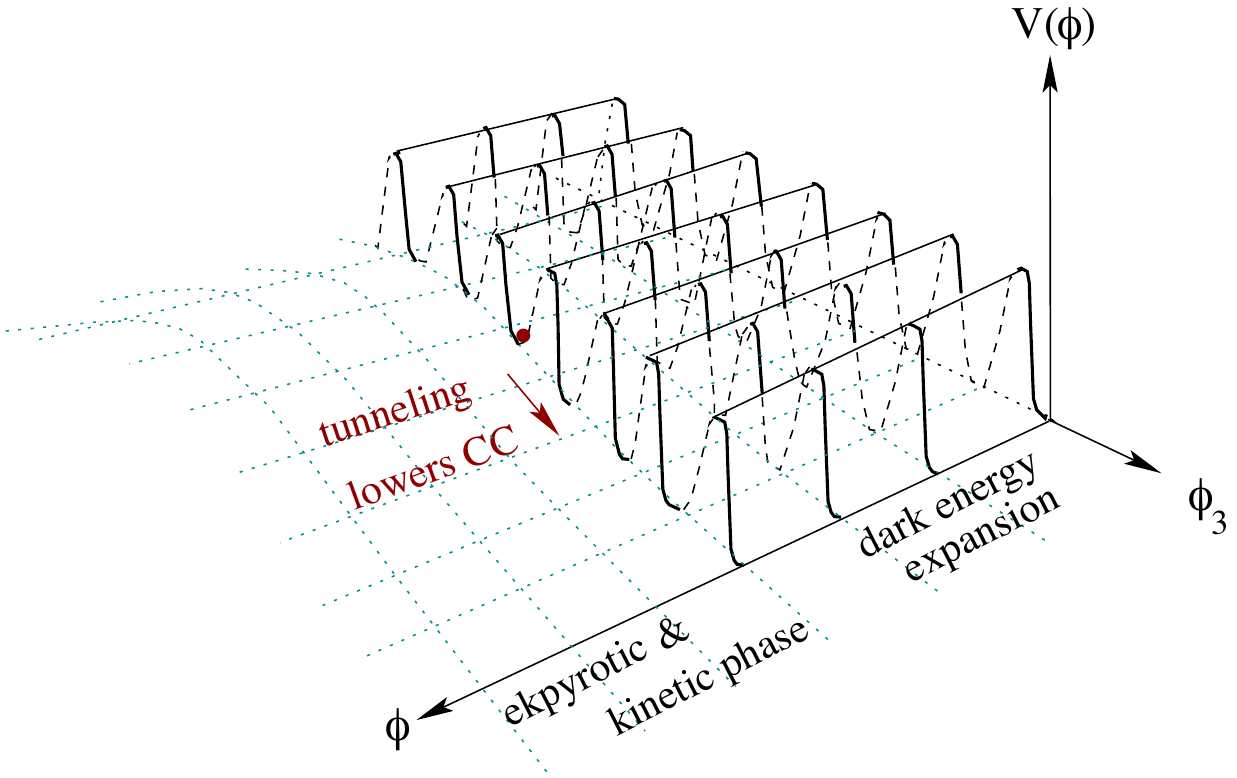}
\caption{Because the age of the universe is extended in cyclic models,
  one may invoke the proposal by Abbot \cite{Abbott:1988nx} to lower
  the value of the cosmological constant from cycle to cycle: a field
  $\phi_3$ orthogonal to the fields needed for the bounce is assumed
  to have a potential with many closely spaced minima on a shallow
  slope, with $\Delta V\sim V_{\Lambda}$. The occasional tunneling
  event reduces the value of the cosmological constant as the universe
  cycles, until the lowest lying positive vacuum is reached. Since
  tunneling to anti de Sitter is suppressed, habitable patches will
  eventually end up with a small cosmological constant. }
\label{cc}
\end{center}
\end{figure}

It is interesting at this stage to mention a possible renewal of the
solution to the cosmological constant problem proposed in
\cite{Linde:1986dq} (see also another discussion of the same idea in
\cite{Garriga:2003hj}). L. Abbott proposed a more involved version of
this idea \cite{Abbott:1988nx} which can be implemented in the cyclic
model \cite{Steinhardt:2006bf,Lehners:2009eg}.  The cosmological
constant, $\Lambda$, which is assumed to start out large and positive,
steadily decreases to lower values via tunneling along a long stream
of vacuum states, see Fig. \ref{cc} (this is not to be confused with
other classes of scenarios for which the cosmological constant starts
out large but is diminished by quantum gravity effects, thereby
inducing either a phase of inflation or the current acceleration
\cite{Romania:2012av}).  After each successive tunneling event, the
universe spends more and more time in the respective vacuum, because
the value of the cosmological constant is lower than before, reducing
the tunneling rate. The universe ends in a big crunch\footnote{It has
  been speculated that transitions from AdS to dS may be possible
  during a bounce, leading to an implementation of bouncing
  cosmologies in the inflationary multiverse \cite{Garriga:2013cix},
  see Sec.~\ref{multiverse}.} if the tunneling event leads to a
negative value of $\Lambda$.  The presence of many habitable patches
in the cyclic universe solves the cosmological constant problem {\sl
  \`a la} Weinberg \cite{Weinberg:1988cp}: after many cycles, any
habitable patch sits in the lowest vacuum with $V \geq 0$. The problem
of the original proposal by Abbott was that the tunneling events
required eons to commence, so that any initial matter in the universe
diluted, leading to an empty universe; the cyclic model circumvents
this problem because a large matter density is produced during each
cycle; hence the universe does not end up empty.
 
\subsection{New challenges}
\label{new_issues}

Just like having a phase of inflation generates new challenges, such
as the trans-Planckian problem \cite{Brandenberger:2012aj} or the
measure problem, implementing a bounce gives rise to new issues. These
belong to two categories: the first are cosmological in nature, for
instance, a phase of contraction can lead to an unwanted increase of
primordial shear; the second are of a theoretical kind, since in order
for a bounce to be able to take place in GR (with flat spatial
curvature), the underlying theory must be capable of violating the
Null Energy Condition (NEC) without introducing instabilities.

\subsubsection{The shear and the need for an ekpyrotic phase}
\label{ekpyroticphase}

In a contracting universe, anisotropies tend to grow, potentially
spoiling the isotropy seen in the CMBR.  Consider negligible initial
curvature and a metric of the Bianchi I type, \ie
\begin{equation}
\dd s^2 = -\dd t^2 + a^2(t) \left[ \ex^{2\theta_x (t)} \dd x^2 + \ex^{2\theta_y
(t)} \dd y^2
+\ex^{2\theta_z (t)} \dd z^2\right],
\end{equation}
with $\sum_i \theta_i \equiv \theta_x+\theta_y+\theta_z=0$. The
Einstein equations become
\begin{equation}
H^2 \equiv \left(\frac{\dot{a}}{a}\right)^2= \frac13\rho
+\frac16 \sum_i\dot\theta^2_i \equiv \frac13\left( \rho +
\rho_\theta\right),
\label{H2theta}
\end{equation}
where $i\in \{ x,y,z \}$, $\rho_\theta$ is the energy density
contained in the anisotropy stemming from the functions $\theta_i$
(the shear) and
\begin{equation}
\dot H = -\frac12 \left( \rho + P\right) -\frac12 \sum_i\dot\theta^2_i\,.
\label{dHtheta}
\end{equation}
Eqs.~(\ref{H2theta}) and (\ref{dHtheta}), together with the fact that
$\sum_i\theta_i=0$,
imply that $\ddot \theta_i + 3 H \dot \theta_i=0$, and
$\rho_\theta\propto a^{-6}$\,.

The FLRW limit, which is currently observed to be valid, must be
recovered. At first sight, this seems to be impossible in a
contracting phase: the continuity equation (\ref{DT0}) for the
constituents of the universe interacting exclusively via gravity with
energy densities $\rho_I$, pressure density $P_I$ and equation of
state parameter $w_I\equiv P_I/\rho_I$ is
\begin{equation}
\dot\rho_I+3H\rho_I\left(1+w_I\right)=0\,,
\end{equation}
and hence, each component, $I=$ radiation ($w_\mathrm{rad}=\frac13$),
non-relativistic matter ($w_\mathrm{dust}=0$), cosmological constant
$\Lambda$ ($w_\Lambda=-1$), curvature ($w_\Ka=-\frac13$), or shear
density ($w_\theta=1$), evolves as
\begin{equation}
\rho_I\propto a^{-3(1+w_I)}\,.
\end{equation}
The Friedmann equation (\ref{FriedCosm}), which provides the time
evolution of the Hubble parameter including the above components,
reads
\begin{equation}
H^2=\frac{1}{3}\left[-\frac{3\Ka}{a^2}+\frac{\rho_{\mathrm{m}0}}{a^3}+
\frac{\rho_{\mathrm{r}0}}{a^4}+
\frac{\rho_{\theta 0}}{a^6}+...+\frac{\rho_{\phi0}}{a^{3(1+w_{\phi})}}\right],
\end{equation}
where we have considered a last contribution from a yet-unknown
constituent labeled $\phi$ with equation of state parameter
$w_\phi$. In the absence of the latter constituent, it is clear that
when the universe contracts, \ie when $a\rightarrow 0$, the anisotropy
term, $\propto a^{-6}$, rapidly becomes dominant: if one starts with
even a slightly perturbed FLRW universe, one might end up with a
highly anisotropic Bianchi solution unless the primordial shear was
generated by quantum vacuum fluctuations; in this case, scalar and
vector perturbations, regardless of their magnitude
\cite{Vitenti:2011yc}, remain comparable \cite{Pinto-Neto:2013zya}:
the problem only arises in the presence of primordial classical shear
and it is absent in inflationary models because any pre-existing
anisotropy is diluted.  Fortunately, there is a simple mechanism to
solve the shear problem in a contracting universe: the incorporation
of an ekpyrotic phase.

\begin{figure}[tb]
\begin{center}
\includegraphics[scale=0.90]{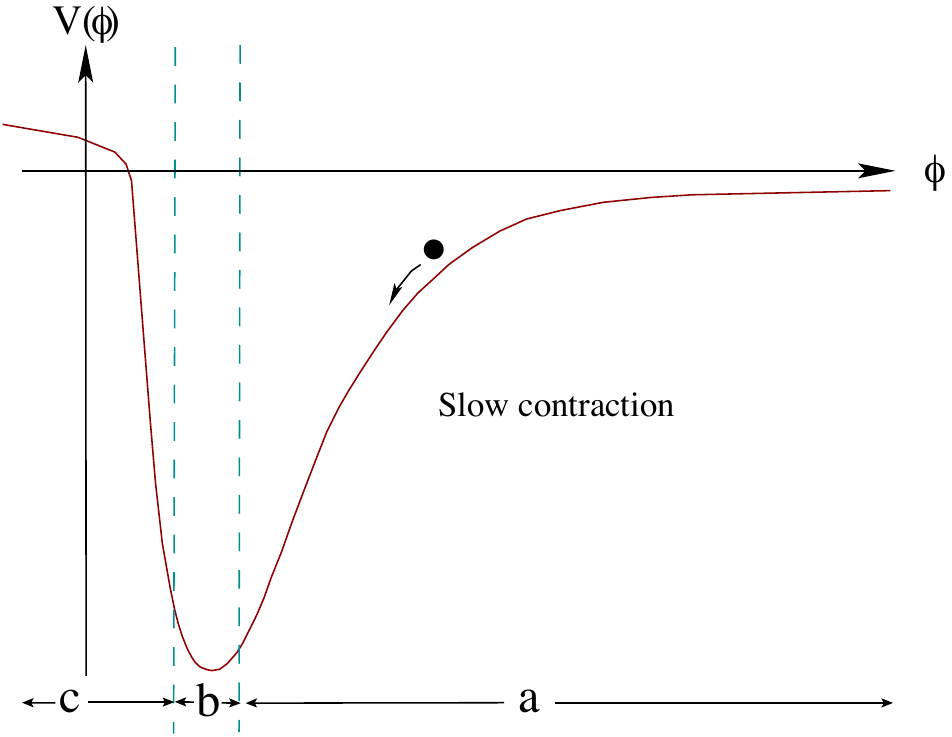}
\caption{Schematic of a generic ekpyrotic potential.  In region ${\bf
    a}$, the scalar field rolls down the steep, negative part of the
  potential, leading to a scaling solution (an attractor) leading to a
  slow contraction. In region ${\bf b}$, the scaling behavior ends and
  the potential is assumed to have a minimum that rises back to
  positive values to avoid a residual negative vacuum energy at the
  end of the ekpyrotic phase. In region ${\bf c}$, the universes
  reverses from contraction to expansion. The shape of this region is
  model-dependent.  The ekpyrotic phase alleviates the shear problem,
  see Sec.~\ref{ekpyroticphase}.}
\label{ekpyrotic}
\end{center}
\end{figure}

A generic ekpyrotic scenario requires a scalar field $\phi$, chosen to
have canonical kinetic energy without higher derivative interactions,
that is set-up to roll down a steep, negative potential $V(\phi)$; a
slow contraction ensues with an equation of state parameter $w_\phi\gg
1$, instead of an accelerated expansion which occurs in the slow-roll
potential of inflation.  Hence, the scalar field dominates at some
point and anisotropies become suppressed in
comparison. Fig.~\ref{ekpyrotic} depicts such a generic ekpyrotic
potential.

Let us illustrate this mechanism with a simple exponential potential
(which we will use for the calculation of correlation functions in
Sec.~\ref{Perturbations}) as in \cite{Khoury:2001wf},
\begin{equation}
V(\phi)\approx -V_0 \ex^{-c\phi},\label{Pekpyrotic}
\end{equation}
where\footnote{ To suppress anisotropies one needs $p>\rho$, that
  is $c^2>6$, which is identical to the requirement for having an
  atractor \cite{Heard:2002dr}. } $c\equiv\sqrt{2/p}\gg 1$, $p\ll
1$ and $V_0>0$; the energy density and pressure in the homogeneous
case are given by (\ref{rhoPscalar}).  As the field rolls down the
steep, negative region of the exponential potential, the scale factor
exhibits a power-law solution, similar to power-law inflation; this
solution, which causes a slow contraction of the universe, is an
attractor.  As discussed below, in order to meet the requirement of a
nearly scale-invariant spectrum, the potential must satisfy the fast
roll condition,
\begin{equation}
\epsilon\equiv\left(\frac{V}{V_{,\phi}}\right)^2\ll 1\,,
\label{epsFR}
\end{equation}
the notation $V_{,\phi}$ denoting a derivative of $V$ with respect to
$\phi$.  Condition (\ref{epsFR}) is satisfied in a steep, nearly
exponential potential, see region ${\bf a}$ in Fig \ref{ekpyrotic},
but other potentials can be used.  Using the FLRW metric (\ref{FLRW}),
the equations of motion are given by the Klein Gordon equation
\begin{equation}
\ddot\phi+3H\dot\phi=-\frac{\dd V}{\dd\phi}\,,
\end{equation}  
along with the Friedmann equations,
\begin{eqnarray}
3H^2&=& V+\frac{1}{2}\dot\phi^2\,,\\
\dot{H}&=&-\frac{1}{2}\dot\phi^2\,.
\end{eqnarray}
With the potential (\ref{Pekpyrotic}) these equations are solved by
\begin{eqnarray}
a(t)&\propto&(-t)^p\,,\label{scaling_solution}\\
H&=&\frac{p}{t}\,,\\
\phi(t)&=&\sqrt{2p}\ln\left[-\sqrt{\frac{V_0}{p(1-3p)}}t\right]\,,
\end{eqnarray}
where $t$ is negative and increases towards zero.  The equation of
state parameter $w_\phi=P/\rho$ during the scaling solution on the
exponential potential is that of a stiff fluid, $w_\phi=-1+2/(3p) \sim 2/(3p) \gg 1$.
Hence, the ratio of curvature $\propto a^{-2}$ and anisotropy $\propto
a^{-6}$ to the scalar field's energy density $\rho_{\phi}=a^{-3(1+w_\phi)}$
scales as
\begin{equation}
  \frac{a^{-2,-6}}{a^{-3(1+w_\phi)}}\propto
  \frac{a^{-2,-6}}{a^{-2/p}}\underset{p\ll1}{\longrightarrow}
  a^{2/p}\sim\frac{1}{H^2}\,,
\end{equation}
where we used $p\ll 1$ in the last steps, see \cite{BKthesis} for a
comprehensive discussion.  Thus, any pre-existing curvature and
anisotropy ends up being sub-dominant close to the bounce. If the
ekpyrotic phase lasts $N$ e-folds, curvature and anisotropy diminish
by a factor of $\ex^{-2N}$, and the universe is homogeneous and
isotropic prior to the bounce.  See Sec.~\ref{BKL} below
for a discussion on
how such a super-stiff equation of state can avoid the BKL
instability.

When the scaling behavior ends, the potential must rise back from its
negative minimum to positive values, in order to avoid a lingering
large negative vacuum energy, region $\bf{b}$ in
Fig.~\ref{ekpyrotic}. The ekpyrotic phase may be used in conjunction
with different bounce mechanisms such as a galileon bounce, see
Sec.~\ref{galileon}. Thus the potential in region $\bf{c}$ is model
dependent.

\subsubsection{BKL instability}
\label{BKL}

A bounce can be disrupted when an instability to the growth of
anisotropy is present, see for instance
Sec.~\ref{quatuminducedanisotropy}, so that the energy corresponding
to anisotropy $\propto a^{-6}$ dominates. In 1970, Belinsky,
Khalatnikov and Lifshitz (BKL) showed that any initial anisotropy
grows unstable as the universe contracts towards a big crunch
\cite{Belinsky:1970ew}.  It can then be amplified sufficiently in a
contracting phase to spoil a bounce if $P<\rho$
\cite{Belinsky:1970ew,Lifshitz:1963ps}; this BKL instability can be
avoided by means of an ekpyrotic phase, for which $P\gg
\rho$. However, if spatial curvature is amplified and dominates, a
{\it mixmaster} phenomenon takes place, in which space contracts
and expands in different directions.  The result is an
inhomogeneous, anisotropic, singular crunch \cite{Misner:1969hg}.

It has been argued that such a mixmaster behavior can also occur in
models with ultra-stiff matter, $w\gg 1$, \cite{Barrow:2010rx}:
assuming anisotropic pressures in such models, Barrow \etal argued
that the isotropic FLRW universe would cease being an attractor. Thus,
distortions and anisotropies are expected to be strongly amplified
during contraction in cyclic or ekpyrotic cosmologies.  However, prior
analytic \cite{Erickson:2003zm} and numerical studies
\cite{Garfinkle:2008ei} of an ekpyrotic contraction showed that while
anisotropies and inhomogeneities indeed grow in some regions, leading
to an effective equation of state parameter $w=1$, other regions
remain smooth and isotropic. The proper volume ratio of the latter to
the former grows exponentially along time slices of constant mean
curvature. In this sense, the ekpyrotic smoothing mechanism is indeed
effective in avoiding the mixmaster behavior in most, if not all,
regions. \\

As we show below, such an ekpyrotic phase discussed in
Sec. \ref{ekpyroticphase} could generate, just as in inflation, an
almost scale-invariant curvature perturbation spectrum
\cite{Battarra:2013cha}.  Whether or not this can be made to agree
with the observational data is a different topic, as one then needs to
transfer these perturbations through the bouncing phase, see
Sec.~\ref{modemixing}.

\subsubsection{NEC violation}
\label{NEC_violation}

The success of a bounce depends on several factors. First and
foremost, the Hubble parameter must change its sign from negative to
positive at the bounce, commonly requiring the violation of the {\it
  Null Energy Condition} (NEC) in the framework of GR,
\begin{equation}
T_{\mu\nu}n^{\mu}n^{\nu}\geq 0\label{nec},
\end{equation}
where $n^{\mu}$ is an arbitrary null vector ($g_{\mu\nu} n^\mu
n^\nu=0$); for an ideal fluid with stress-energy tensor given by
(\ref{Tmn}), Eq.~(\ref{nec}) is equivalent to $(\rho + P) (u\cdot
n)^2$, which implies, since $u\cdot n\not= 0$ ($u$ being timelike and
$n$ lightlike),
\begin{equation}
\rho+P\geq 0\label{eq}.
\end{equation}
Indeed, in the absence of curvature, the Einstein equations imply
\begin{equation}
\dot H=-\frac{1}{2}(\rho+P)\label{dotH},
\end{equation}
which, along with (\ref{eq}), prohibits a nonsingular bounce: in order
to have a Hubble rate always increasing from negative to positive
values, the null energy condition must be violated for a finite amount
of time $\Delta t$, \ie $\rho + P<0$ for $\Delta t >0$ around the
bounce time.  Hence, the violation of the NEC is crucial for models
with vanishing (or negative) spatial curvature described by GR that
weave together a contraction and an expansion phase.

The violation of the NEC usually leads to ghosts, indicating dangerous
instabilities at the classical and/or quantum level and
superluminality \cite{Dubovsky:2005xd}. See \cite{Rubakov:2014jja} for
a recent review.  Ghosts can be avoided in certain set-ups by
employing, for example, a ghost condensate (higher derivatives)
\cite{ArkaniHamed:2003uy,Creminelli:2006xe}, conformal galileons
\cite{Nicolis:2008in} (conformally invariant scalar field theories
with particular higher-derivative interactions)
\cite{Creminelli:2012my} or DBI conformal galileons (a 3-brane moving
in an anti-de Sitter bulk, AdS$_5$), used, for example, in the
inflationary models of DBI Genesis \cite{Hinterbichler:2012yn}.

Desirable properties for a healthy bounce according to
\cite{Hinterbichler:2012yn,Elder:2013gya} include,
\begin{enumerate}
\item A stable Poincar\'e-invariant vacuum.
\item The $2\rightarrow 2$ scattering amplitude about this vacuum
  obeys standard analyticity conditions.
\item A time-dependent, homogeneous, isotropic solution which allows
  for a stable violation of the NEC.
\item A subluminal propagation of perturbations around the
  NEC-violating background.
\item A stable solution against radiative corrections.
\end{enumerate}

Many proposed bouncing models have problems satisfying these
conditions, as shown in table \ref{interpolation} taken from
\cite{Elder:2013gya} for a limited selection of models, highlighting
the improvements of some iterations within the framework of galileon
theories.  \cite{Elder:2013gya} provides an extended review of what is
presented here.  The models in this table are comprised of a ghost
condensate \cite{ArkaniHamed:2003uy}, Galileon genesis
\cite{Nicolis:2009qm,Nicolis:2008in,Creminelli:2010ba,Creminelli:2012my}
based on conformal galileons, DBI genesis \cite{deRham:2010eu}, and a
recent proposal by Elder, Joyce and Khoury (EJK model)
\cite{Elder:2013gya}.

A common stumbling block is superluminality, which only the most
recent iterations of galileon models were able to avoid. It was
further argued in \cite{Creminelli:2013fxa,deRham:2013hsa} that
Galileon theories plagued with superluminality can be mapped into
their dual analogues, which are free of superluminality; this idea has
not yet been employed elsewhere.

As argued in \cite{Elder:2013gya}, any theory needs to admit a smooth
interpolation between a Poincar\'e-invariant vacuum and the
NEC-violating region. Such a smooth interpolation is impossible in a
single-field dilation invariant theory \cite{Rubakov:2013kaa} (see
also \cite{Abramo:2005be} in the different but related context of
phantom dark energy).  By relaxing the assumption of dilation
invariance, the EJK model in \cite{Elder:2013gya} admits a solution
that allows not only a healthy interpolation between a NEC-satisfying
phase at early times and a NEC-violating phase at late ones, but is
also subluminal.

Since many recent proposals of bouncing cosmologies employ ghost
condensates and/or galileons, see table \ref{table:}, we provide in
the next section a brief introduction to the galileon theories
outlined above, even though they are not bouncing models on their
own. The cosmological super-bounce \cite{Koehn:2013upa}, based on the
incorporation of a galileon into supergravity, avoids the problem of
superluminality and is one of the more promising bouncing models to
date.

\subsection{Frameworks and partial answers}
\label{ghosts}

Implementing a bounce is not an easy task since the underlying theory
must have many new properties. In order to be capable of having a
bouncing solution, the theory must accommodate the violation of the
NEC without initiating instabilities, which would spoil the very
existence of the bounce. For instance, whatever the value of the
spatial curvature, a bouncing phase initiated by a perfect fluid with
energy momentum tensor (\ref{Tmn}), even including entropy modes, will
always be absolutely unstable in the framework of GR
\cite{Peter:2001fy}.

Viewing General Relativity as a low energy effective theory, there are
two basic options: to modify gravity or to consider matter content
with special properties. A third alternative is to consider
non-perturbative results in string theory.  After a brief excursion on
modified gravity models we introduce galileon theories, since this
special class of exotic matter exhibits desirable properties for
building bouncing cosmologies. For instructive purposes we follow with
one concrete example, the matter bounce, which employs galileon
fields. We conclude with two models based on T-duality in string
theory, the S-brane bounce and a Hagedorn phase in string gas
cosmology; the bounce in the latter set-ups is not describable within
General Relativity or dilaton gravity, but is based on
non-perturbative symmetries.

\begin{table*}[hbt]
\centering
\setlength{\tabcolsep}{0.7pc}
\begin{tabular}{ c  c  c  c  c }
	\hline\hline
	&{  Ghost condensate}        & {    Galilean Genesis}   
	& {    DBI Genesis}         & {EJK Theory} \cite{Elder:2013gya} \\ \hline
	{\bf\cancel{NEC} vacuum}       &  {\color{green}\cmark}  & {\color{green}\cmark} 
    &  {\color{green}\cmark} &  {\color{green}\cmark}\\		
	 No ghosts                     &  {\color{green}\cmark}  &  {\color{green}\cmark}
    &  {\color{green}\cmark} &  {\color{green}\cmark} \\
	  Sub-luminality               &  {\color{green}\cmark}  &  {\color{green}\cmark}
    &  {\color{green}\cmark} &  {\color{green}\cmark} \\
	{\bf Poincar\'e vacuum}        & {\color{red}\xmark}    & {\color{red}\xmark}  
    & {\color{green}\cmark} & {\color{green}\cmark} \\
	No ghosts                      &  --                    &  --                  
    &{\color{green}\cmark} & {\color{green}\cmark}\\		
	S-Matrix analyticity ($2\to2$) &   --                   &  --                  
    & {\color{green}\cmark}&  {\bf {\color{green} \cmark}}\\
	Sub-luminality                 & --                     &--                    
    & {\color{red}\xmark} & {\color{green}\cmark}\\		
	{\bf Interpolating solution}   &  --                    &  --                  
    & {\color{red}\xmark} & {\color{green}\cmark}  \\ 
	{\bf Radiative stability}      & {\color{green}\cmark}   & {\color{green}\cmark} 
    & {\color{green}\cmark} & {\color{red}\xmark}  
	 \\ \hline\hline
\end{tabular}
\caption{\small Successes and drawbacks some of models that violate
  the NEC, taken from \cite{Elder:2013gya}.}
\label{interpolation}
\end{table*}
\subsubsection{Modified gravity}

There are many ways to modify gravity. Historically, the first models
included a scalar mode in addition to the metric (a tensor) of General
Relativity; since these scalar tensor theories are equivalent to
ordinary GR with a modified matter content, we do not consider them
separately. More complicated theories have been suggested, where terms
are added to render the cosmological evolution explicitly
singularity-free \cite{Brandenberger:1993ef}. Such models, which are
appropriate to describe a bouncing phase, can be expressed as
\begin{equation}
\mathcal{S}=\frac12\int\dd ^4x\,\sqrt{-g}\, \left[ R+\sum_i{\phi_i}{I^{(i)}} -
V(\phi_{1},\phi_2,\dots)\right],
\end{equation}
where the $I^{(i)}$ are undetermined functions of curvature
invariants, related to the Lagrange multipliers through the
Euler-Lagrange equation $I^{(i)}=\dd V/\dd {\phi_i}$. Demanding that
$V\sim \sum_i\phi_i^2 + \cdots$ for $\sum_i\phi_i^2 \to 0$ and $V\to
2\Lambda$ for $\sum_i\phi_i^2\to \infty$, with $\Lambda$ a constant,
the theory reproduces GR for low curvatures and yields a de Sitter
solution for high curvature. Bouncing solutions can be implemented in
this framework \cite{Abramo:2009qk}; these solutions are stable and
connect a contracting de Sitter phase to an expanding one.

Massive gravity can also lead to bouncing solutions.  Defining gravity
with respect to a de Sitter metric, bouncing solutions exist
\cite{Langlois:2013cya}, if positive spatial curvature is present,
even for a fluid satisfying the strong energy condition $\rho + 3
P>0$. However, these solutions are either doomed by a future
singularity (curvature singularity or one that is unique to massive
gravity) or tend to an asymptotic de Sitter regime.

Generically, modifications of gravity are expected to arise after
quantization of gravity.
Methods employing high energy physics have been proposed, most notably
string theory, in which case one may often switch from the point of
view of modified gravity to new components of the energy momentum
tensor, as discussed in Sec.~\ref{stringy}. Similar attempts have been
made in Loop Quantum Gravity (see Sec.~\ref{LQG}).

As we discussed in Sec.~\ref{sec:QC}, a complementary approach one may
contemplate consists in an effective low-energy modification, as in
the Wheeler-De Witt approach, whereby quantization is attempted in a
super-space consisting of the set of all possible 3-metrics
$\{h_{ij}\}$.  Such an approach is usually unfeasible unless an
additional mini-superspace approximation is performed, in which the
infinite number of degrees of freedom is reduced to a few, for
instance by considering only homogeneous and isotropic metrics. In
\cite{AcaciodeBarros:1997gy}, it was shown that in the presence of a
simple perfect fluid, the singularity can always be avoided,
independent of the fluid's equation of state and the spatial
curvature.  Moreover, perturbations in such models can be treated
self-consistently.  In \cite{Peter:2006id,Pinho:2006ym}, a model
dominated by a dust-like perfect fluid was shown to produce a
scale-invariant spectrum of perturbations \cite{Peter:2008qz},
otherwise commonly obtained in the so-called matter bounce (see
Sec.~\ref{workingmodel}). As this review aims at concentrating on
classical GR or modifications thereof, we shall now move on to the
second option to implement a bounce, namely that of changing the
constituent behavior acting as a source in the Einstein equations.

\subsubsection{Modified matter content \label{ghost}}

{\it $\bullet$ Ghost fields\\}

\label{GhostField}
To violate the NEC by brute force at the phenomenological level, one
may include a ghost field with the Lagrangian
\begin{equation}
\mathcal L=\frac{1}{2}(\partial\phi)^2-V(\phi),
\end{equation}
where $\phi$ has a canonical kinetic term with the wrong sign. In the
absence of a potential ($V\to 0$), the energy density and pressure are
equal ($\rho=P=-\frac12 \dot\phi^2$), so that $\rho+P=-\dot\phi^2<0$
and the NEC is violated.  Such an inclusion produces large amounts of
adiabatic perturbations \cite{Peter:2001fy}, because intrinsic entropy
modes are absent. To counter this instability and render the total
energy density positive, one may couple the negative energy fluid to
an ordinary perfect fluid with positive energy, such as radiation. The
resulting entropy modes, which were previously absent, render the
model phenomenologically viable: using a free scalar field to mimic a
stiff fluid (\ie one with an equation of state parameter $w=1$), such
a model was shown to be stable.  The conditions to produce a
scale-invariant spectrum of perturbations were calculated in
\cite{Peter:2002cn}. This model,  however intrinsically problematic
because relying on an unstable negative energy field, may be seen
as a precursor of a ghost condensate: the temporary effective
violation of the NEC merely permits the bounce to occur, but the
negative energy component is otherwise subdominant for most of the
universe history.  No implementation was suggested in practice 
(see also \cite{Chimento:2005ua} for a bounce introduced by K-essence,
where the conditions for a regular bounce, as well as anisotropies,
were discussed): A more general phase space analysis of models with
generalized kinetic terms can be found in \cite{DeSantiago:2012nk},
where conditions under which a bounce is possible in the NEC violating
regime are derived.

The same idea was applied to the special situation in which, instead
of radiation, the positive energy fluid is a scalar field whose
dynamics is driven by an exponential potential \cite{Allen:2004vz},
providing a model for the matter bounce. Contrary to many other
bouncing models, the amplitude of tensor modes can be large, of order
$r\sim\mathcal{O} (30)$ \cite{Allen:2004vz,Cai:2008qw}, with a
scale-invariant spectrum, in excess of current measurements, see
Sec.~\ref{rcyclic}.

Xue \etal studied in \cite{Xue:2013bva} the evolution of fluctuations
(in the same framework) non-perturbatively by means of simulations.
They found that some regions of space could undergo a regular bouncing
epoch, provided initial conditions were sufficiently close to being
homogeneous and isotropic, while other regions would collapse, see
Sec.~\ref{viabilityofptb} for details. Similar to a multiverse version
of inflation, in which causally disconnected pieces of the Universe
behave in different ways, our observable part is but a small patch of
the full cycling multiverse.

Evidently, while simple ghost fields are unrealistic, the resulting
simplicity of bouncing models enables one to address otherwise
inaccessible questions.\\
 
{\it $\bullet$ Ghost-condensate\\}

As summarized in \cite{Elder:2013gya}, the violation of the NEC in
theories that involve one derivative per field, $\mathcal
L(\phi^I,\partial\phi^I)$, implies the existence of either ghosts,
gradient instabilities or superluminal propagation of perturbations
\cite{Dubovsky:2005xd,Buniy:2006xf}.  In the previous paragraph, we
discussed a few models which assumed such ghost fields for a finite
duration. This can be implemented explicitly in the ghost-condensate
technique according to which a field manages to violate the NEC only
for some time, being dynamically driven first into this regime, and
subsequently out of it.

\begin{figure}[tb]
\begin{center}
\includegraphics[scale=1.1]{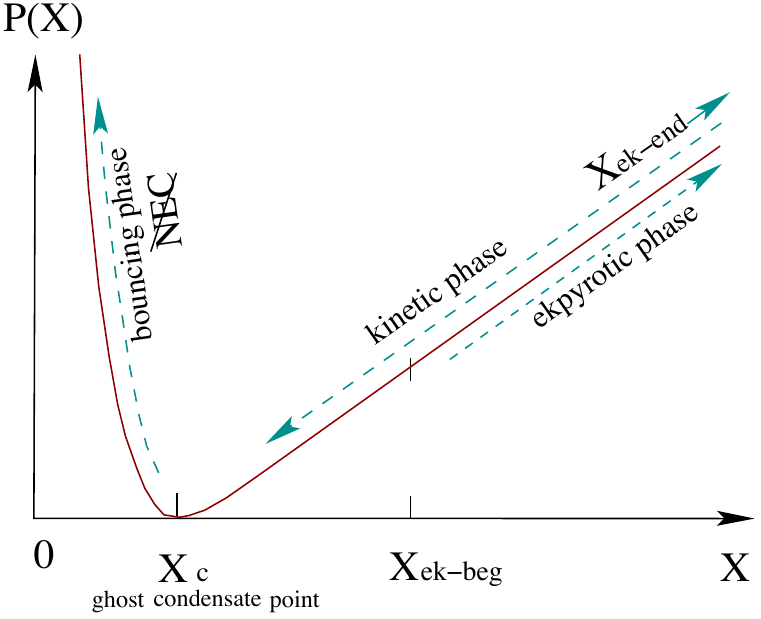}
\caption{Schematic of $P(X)$ in the vicinity of the ghost condensate
  point at $X_\mathrm{c}$ \cite{Xue:2011nw}. To the right of
  $X_\mathrm{c}$ the kinetic term approaches the canonical one which
  is valid during the ekpyrotic and kinetic phases. To the left of
  $X_\mathrm{c}$ the kinetic term has the wrong sign enabling the
  violation of the NEC and the occurrence of a bounce. Labels are
  consistent with Fig.~{\ref{fig:4}}.}
\label{BPEktransition}
\end{center}
\end{figure}

A ghost condensate \cite{ArkaniHamed:2003uy} can arise from a
higher-derivative theory containing the Lagrangian density
\begin{equation}
\mathcal{L}=P(X),\label{lagrangian}
\end{equation}
where the pressure function $P(X)$ is an arbitrary differentiable
function of the standard kinetic energy term
\begin{equation}
X\equiv-\frac{1}{2}\partial_{\mu}\phi\partial^{\mu}\phi,
\end{equation}
and which, in the cosmological context in which the scalar field
depends only on time, reads $X=\frac12\dot{\phi}^2$; the canonical
kinetic term of (\ref{canLagphi}) is obtained in the limit where $P\to
X$.  Using the flat FLRW metric (\ref{FLRW}) with $\Ka=0$, the scalar
equation of motion becomes (we follow \cite{Koehn:2012te} in this
section)
\begin{equation}
\frac{\dd}{\dd t}(a^3P_{,X}\dot\phi)=0\,.\label{scalareom}
\end{equation}
If $\phi$ is a constant, (\ref{scalareom}) is trivially
satisfied. However, if $X$ is a constant and $P_{,X}=0$ at
$X=X_\mathrm{c}$, the equation of motion allows for the
ghost-condensate solution
\begin{equation}
\phi=\sqrt{2X_\mathrm{c}}t\,.
\end{equation}
The energy density is given by
\begin{equation}
\rho=2XP_{,X}-P\,,
\end{equation}
where the pressure is identified with the Lagrangian function $P$.
Since, by definition, $X>0$, a violation of the NEC can only result
provided there exists a region for which $P_{,X}<0$. Therefore, the
shape of the function $P(X)$ should have a local minimum, denoted by
$X_\mathrm{c}$ in Fig.~\ref{BPEktransition} which shows schematically
the typical shape of $P(X)$ enabling a temporary NEC violating
phase.  The theory is well-behaved near the ghost condensate
point, if seen as an effective description of only the bounce -- the
model is self-consistent and does not contain a ghost. However, if
seen as a fundamental theory, the model is problematic since it does
not exhibit a Lorentz-invariant vacuum  \cite{Elder:2013gya}: to
avoid ghosts, $P_{,XX}(X_\mathrm{c})>0$ has to hold. On the other
hand, we need $\lim_{X\to0}P_{,X}>0$ for a Lorentz-invariant vacuum,
which cannot be reached from the minimum of $P(X)$ at $X_\mathrm{c}$
without inducing pathologies.
 
 In the validity range of the ghost condensate regime, the
Einstein's equations (\ref{FriedCosm}) imply
\begin{equation}
\dot H=\frac{\Ka}{a^2}-\frac{1}{2}(\rho+P)\underset{\Ka\to0}{\longrightarrow}
-\frac12 (\rho+P),
\end{equation}
so that a nonsingular bounce is possible in this regime. This
mechanism was employed in \cite{Creminelli:2006xe} and in the new
Ekpyrotic scenario
\cite{Buchbinder:2007ad,Creminelli:2007aq,Buchbinder:2007tw},
which, however, suffers from several instabilities
\cite{Xue:2010ux}; see Sec.~\ref{Fatal_effects}. It was also
implemented as the bounce mechanism for the matter bounce in
\cite{Lin:2010pf}.

In the next subsections we investigate galileon theories which can
improve upon the ghost condensate.\\

{\it $\bullet$ Galileon genesis\label{galileon_genesis}\\}

An alternative to violate the NEC without introducing gradient
instabilities is given by conformal galileons
\cite{Buniy:2006xf,Nicolis:2009qm} (see also Kinetic Gravity Braiding
\cite{Deffayet:2010qz} used in the G-bounce \cite{Easson:2011zy}).
The simplest conformally-invariant galileon Lagrangian in
\cite{Creminelli:2010ba} takes the form (we follow again
\cite{Elder:2013gya}),
\begin{equation}
\mathcal
L=f^2\ex^{2\phi}(\partial\phi)^2+\frac{f^3}{\Lambda^3}
(\partial\phi)^2\Box\phi+\frac{f^3}{2\Lambda^3}(\partial\phi)^4
\label{galileonlagrangian1},
\end{equation}
where $\phi$ is the galileon field and $f$ and $\Lambda$ are
constants.  This theory has a time-dependent solution, $\phi=-\ln
(-H_0 t)$ with $H_0^2=2\Lambda^3/(3f)$ for $ -\infty<t<0$. To remain
within the confines of EFT we need $f\gg \Lambda$ so that $H_0\ll
\Lambda$.  Since\footnote{We corrected a typo in
  \cite{Elder:2013gya}.}
\cite{Nicolis:2009qm,Creminelli:2012my,Creminelli:2010ba}
\begin{equation}
\rho+P=-\frac{2f^2}{H_0^2t^4},
\end{equation}
the NEC is violated for this solution. Furthermore, perturbations are
stable and propagate luminally, but small deformations of the
solution, which break homogeneity/isotropy \cite{Easson:2013bda}, lead
to superluminality.

To avoid superluminality, \cite{Creminelli:2012my} reduced the
symmetry by considering a deformation of the original Lagrangian
(\ref{galileonlagrangian1}) to
 \begin{equation}
\mathcal
L=f^2\ex^{2\phi}(\partial\phi)^2+\frac{f^3}{\Lambda^3}(\partial\phi)^2
\Box\phi+\frac{f^3}{2\Lambda^3}(1+\alpha)(\partial\phi)^4
\label{galileonlagrangiandeformed},
\end{equation}
where $\alpha$ is a dimensionless parameter of order unity. The case
$\alpha\neq 0$ breaks conformal invariance, while preserving dilation
invariance. The solution for the galileon is unchanged, while
$H_0^2=2\Lambda^3/[(1+\alpha)3f]$ is simply rescaled.  The
time-dependent solution violates the NEC if
\begin{equation}
\rho+P=-\frac{2f^2}{H_0^2t^4}\frac{3+\alpha}{3(1+\alpha)}
\end{equation}
is negative, requiring
\begin{equation}
\alpha >-1 \quad \mbox{or}\quad \alpha<-3\,. 
\end{equation}
Expanding the Lagrangian around this solution to second order, one can
read off the speed at which perturbations propagate, namely
\begin{equation}
c_{_\mathrm{S}}^2=\frac{3-\alpha}{3(1+\alpha)}\,.
\end{equation}
In order to avoid instabilities we need $c_{_\mathrm{S}}^2>0$ and in
order to avoid superluminality we need $c_{_\mathrm{S}}^2\leq1$. Thus
the NEC can be violated while retaining stability and subluminality
for
\begin{equation}
0<\alpha< 3\,.
\end{equation}

Similar to the ghost-condensate model, no Lorentz-invariant vacuum is
present, even if the set-up were extended to include higher-order
conformal galileons \cite{Creminelli:2012my}. However, perturbations
are stable on all scales, opposite to ghost-condensate models, which
are unstable on large scales during the NEC-violating phase.

Lagrangians of the above type are used in the galileon genesis
scenario
\cite{Nicolis:2009qm,Nicolis:2008in,Creminelli:2010ba,Creminelli:2012my},
among other proposals \cite{Deffayet:2010qz,Kobayashi:2010cm}.\\

{\it $\bullet$ DBI genesis via conformal galileons\\}

 Even though the aforementioned models lack a Lorentz-invariant
vacuum, either one of them may describe our Universe as an effective
theory during the bounce; nevertheless, none of them can be seen as a
fundamental theory. To improve upon this shortcoming, even more
complicated galileon theories have been considered, while still
avoiding superluminality. An example is to consider Dirac-Born-Infeld
(DBI) conformal galileons \cite{Hinterbichler:2012yn}, as summarized
in \cite{Elder:2013gya}, which we follow below. These theories
describe the motion of a 3-brane by means of an effective scalar field
$\phi$ in an $\mathrm{AdS}_5$ geometry.  The resulting 4D Lagrangian
for $\phi$ is given by \cite{Hinterbichler:2012yn}
 \begin{equation}
 \mathcal L=c_1\mathcal L_1+c_2\mathcal L_2+c_3\mathcal L_3+c_4\mathcal
L_4+c_5\mathcal L_5\,.
 \end{equation}
 where
 \begin{eqnarray}
 \mathcal{L}_1&=&-\frac{1}{4}\phi^4,\\
 \mathcal{L}_2&=&-\frac{\phi^4}{\gamma},\\
\mathcal{L}_3&=&-6\phi^4+\phi[{\Phi}]+
\frac{\gamma^2}{\phi^3}\left(-[\phi^3]+2\phi^7\right),\\
 \mathcal{L}_4&=&12\frac{\phi^4}{\gamma} +
\frac{\gamma}{\phi^2}\left\{
[{\Phi}^2]-\left([{\Phi}]-6\phi^3\right)
 \left([\Phi]-4\phi^3\right)\right\}\\
 &&+2\frac{\gamma^3}{\phi^6}\left\{
-[{\phi}^4]+[{\phi}^3]\left([{\Phi}]-5\phi^3\right)-2[{\Phi}]\phi^7
+6\phi^{10}\right\},\nonumber\\
\nonumber
\mathcal{L}_5&=&54\phi^4-9\phi[{\Phi}]+\frac{\gamma^2}{\phi^5}
\left(9[{\phi}^3]\phi^2
+2[{\Phi}^3]-3[{\Phi}^2][{\Phi}]\right.\\
\nonumber && +\left.12[{\Phi}^2]\phi^3
+[\Phi]^3-12[\Phi]^2\phi^3+42[{\Phi}]\phi^6
  -78\phi^4\right)\\
\nonumber &
&+3\frac{\gamma^4}{\phi^9}\Big\{-2[\phi^5]+2[\phi^4]
\left([\Phi]-4\phi^3\right)\\
 \nonumber &&+[\phi^3]\left([\Phi^2]-[\Phi]^2+8[\Phi]
\phi^3-14\phi^6\right)\\
 &&+2\phi^7\left([\Phi]^2-[\Phi^2]\right)-8[\Phi]\phi^{10}+12\phi^{13}\Big\},
\end{eqnarray}
with $5$ free coefficients $c_{I=1...5}$, and
\begin{equation}
\gamma\equiv \left[1+\frac{(\partial\phi)^2}{\phi^4} \right]^{-1/2}
\end{equation}
is the Lorentz factor for the brane motion; $\Phi$ denotes the matrix
of second order derivatives $\partial_{\mu}\partial_{\nu}\phi$,
$[\Phi^n]\equiv \mathrm{Tr} (\Phi^n)$, and $[\phi^n]\equiv\partial\phi
\cdot\Phi^{n-2}\cdot\partial\phi$, with indices raised by the
Minkowski metric $\eta^{\mu\nu}$ \cite{Nicolis:2008in}.  The
$\mathcal{L}_I$ are Lovelock invariants and guarantee second order
differential equations \cite{deRham:2010eu}.
 
Suitable choices of the free coefficients, $c_{I}$'s, allow for a
solution $\ex^{\phi}\propto 1/t$, leading to the stable violation of
the NEC \cite{Hinterbichler:2012yn}. A nice feature of this model is
that the speed of sound of fluctuations is subluminal while preserving
conformal invariance; further, the solution is stable against
radiative corrections and a stable Poincar\'e-invariant vacuum is
present. However, weak-field deformations of the vacuum may again lead
to superluminality and it is not clear if this is a mere pathology
\cite{Creminelli:2013fxa,deRham:2013hsa} and how it should be cured.
Nevertheless, this is the first model in the literature that allows
the coexistence of a NEC-violating solution and a stable
Poincar\'e-invariant vacuum, although no interpolating solution
between them is given in \cite{Hinterbichler:2012yn}. \\

{\it $\bullet$ Elder, Joyce, Khoury (EJK) model \\}

A smooth transition from a Poincar\'e-invariant vacuum to a
NEC-violating phase is necessary, if such a model is to be used to
generate a bounce. Rubakov showed in \cite{Rubakov:2013kaa} that such
an interpolation is impossible in single-scalar field theories that
obey dilation-invariance.  Violating the latter, a time-dependent,
smooth interpolation between such solutions is presented in
\cite{Elder:2013gya} (we refer to this work as the {\it EJK
  model}). There, the galileon Lagrangian (\ref{galileonlagrangian1})
is generalized to
\begin{equation}
\mathcal L=\mathcal{Z}(\phi)\ex^{2\phi}(\partial\phi)^2+
\frac{f_0^3}{\Lambda^3}(\partial\phi)^2\Box\phi+
\frac{1}{\mathcal{I}(\phi)}\frac{f_0^3}{2\Lambda^3}
(\partial\phi)^4
\label{galileonlagrangian2},
\end{equation}
where, $\mathcal Z(\phi)$ and $\mathcal{I}(\phi)$ break scale
invariance and are chosen to interpolate between a NEC satisfying
solution at early times and a violating one at late times. In
\cite{Elder:2013gya}, these functions are taken to be
\begin{eqnarray}
\mathcal
I(t)\!\!&=&\!\!\frac{\mathcal{I}_0}{\left(1+
  \displaystyle\frac{t}{t_*}\right)^8}\,,\\
\mathcal
Z(\phi)\!\!&=&\!\!\frac{f^2_0}{\left(\ex^{\phi-\phi_{\infty}}-1\right)^4}\!
\left[\ex^{4(\phi-\phi_{\infty})}-
  \!\left(1+\frac{f^2_{\infty}}{f_0^2}\right)\right],
\end{eqnarray}
where $f_0,f_{\infty},\Lambda,\mathcal{I}_0$ and $\phi_{\infty}$ are
constants.  For suitable choices of these parameters, it was shown in
\cite{Elder:2013gya} that this model contains most of the necessary
ingredients for a stable bounce: a Poincar\'e-invariant vacuum,
stable, subluminal perturbations and an interpolation to a
NEC-violating phase. See table \ref{interpolation} for a reproduction
of the results presented in \cite{Elder:2013gya}.
 
To summarize, particular galileon theories appear to be promising
candidates to construct nonsingular bouncing models that avoid
pathological problems associated with the violation of the NEC such as
ghosts, instabilities and/or superluminality.

\subsubsection{Example: a nonsingular matter bounce}\label{workingmodel}
\begin{figure*}[tb]
\begin{center}
\includegraphics[scale=0.99]{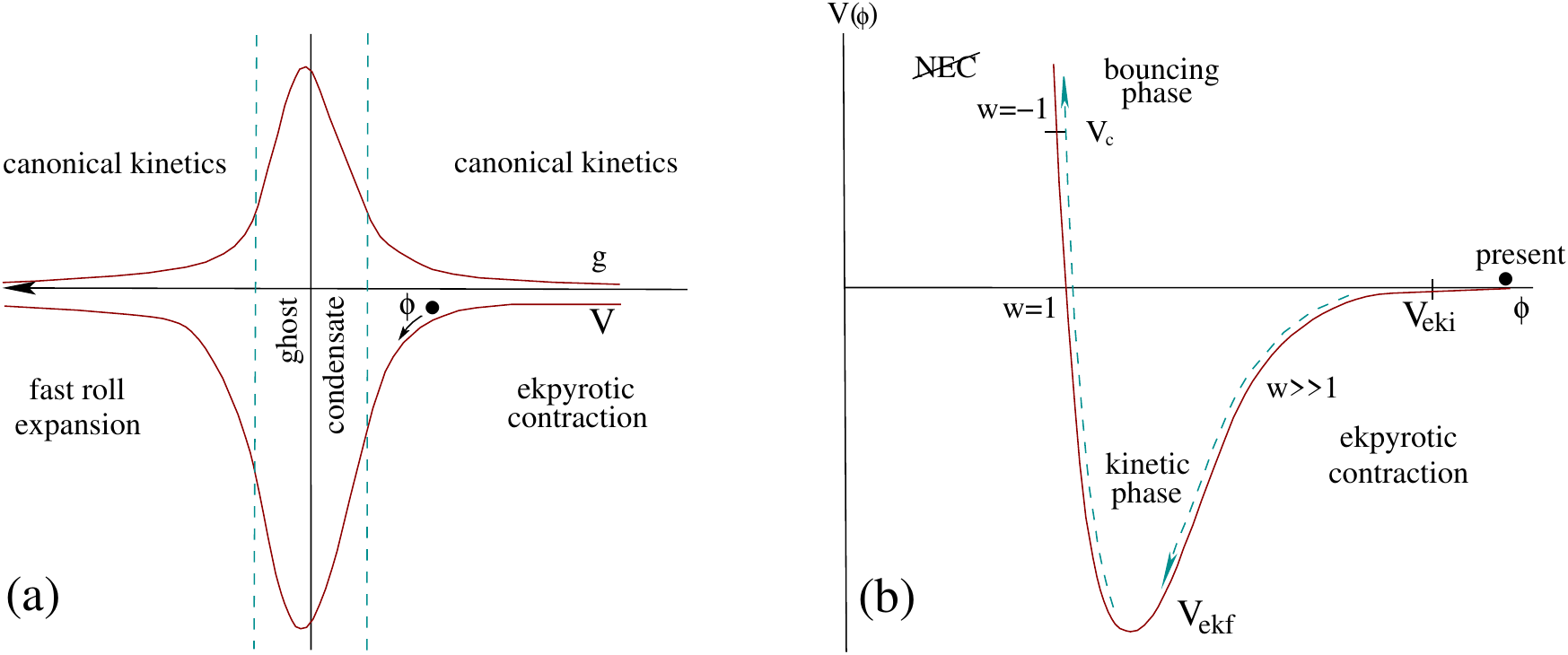}
\caption{Schematic of the potential in the model proposed in (a)
  \cite{Cai:2012va} and (b) in \cite{Xue:2011nw}. In (a), the ghost
  condensate is triggered at a certain value of $\phi$ via the
  function $g(\phi)$ of the galileon; $\phi$ does not necessarily slow
  down during the ghost condensate phase, enabling a fast bounce after
  the ekpyrotic contraction in order to avoid the regrowth of the
  initial anisotropy, instability C, see Sec.~\ref{Fatal_effects}.  In
  (b) the potential $V(\phi)$ for the scalar field $\phi$ rises in
  order to slow down the field; the ghost condensate is triggered at a
  small value of the kinetic term, $X_\mathrm{c}$, as set by the
  kinetic term $P(X)$. Such a ghost condensate bounce takes
  considerable time after the ekpyrotic phase and suffers from a
  regrowth of anisotropies as the field slows down, see
  Sec.~\ref{initialanisotropy}.}
\label{comparison}
\end{center}
\end{figure*}

Nonsingular bouncing models are attractive since four dimensional
General Relativity remains applicable throughout.  The price one pays
is that a component to the energy content of the universe must violate
the NEC, often leading to dangerous instabilities, see
Sec.~\ref{NEC_violation} and Sec.~\ref{Fatal_effects}. This was
emphasized for instance in \cite{Xue:2011nw} in which the action,
besides gravity, is that a of a scalar field $\phi$ with Lagrangian,
\begin{equation}
\mathcal{L}= P(X) -V(\phi)\,,
\end{equation}
where the function $P(X)$ is depicted in Fig.~\ref{BPEktransition} and
the potential is illustrated on the right hand side of
Fig.~\ref{comparison}. In such an approach, the ekpyrotic phase must
last a sufficiently long time in order to drive the shear contribution
$\rho_\theta$ in (\ref{H2theta}) to vanishingly small values. At this
stage, the universe enters a rapid phase of kinetic domination
(labeled ``kinetic phase'' in Fig.~\ref{BPEktransition}) driving the
field to the minimum of its potential and hence, to the
ghost-condensate point. At this stage, the Hubble rate is large and
negative, so that the ensuing phase must also last a long time in
order to lower $|H|$. Once the field has passed the ghost-condensate
value, the NEC is violated, allowing for a bounce to eventually take
place; this long period of NEC violating contraction tends to increase
the shear exponentially, thereby ruining the benefit of the ekpyrotic
phase.

\begin{figure*}[tb]
\begin{center}
\includegraphics[scale=0.65]{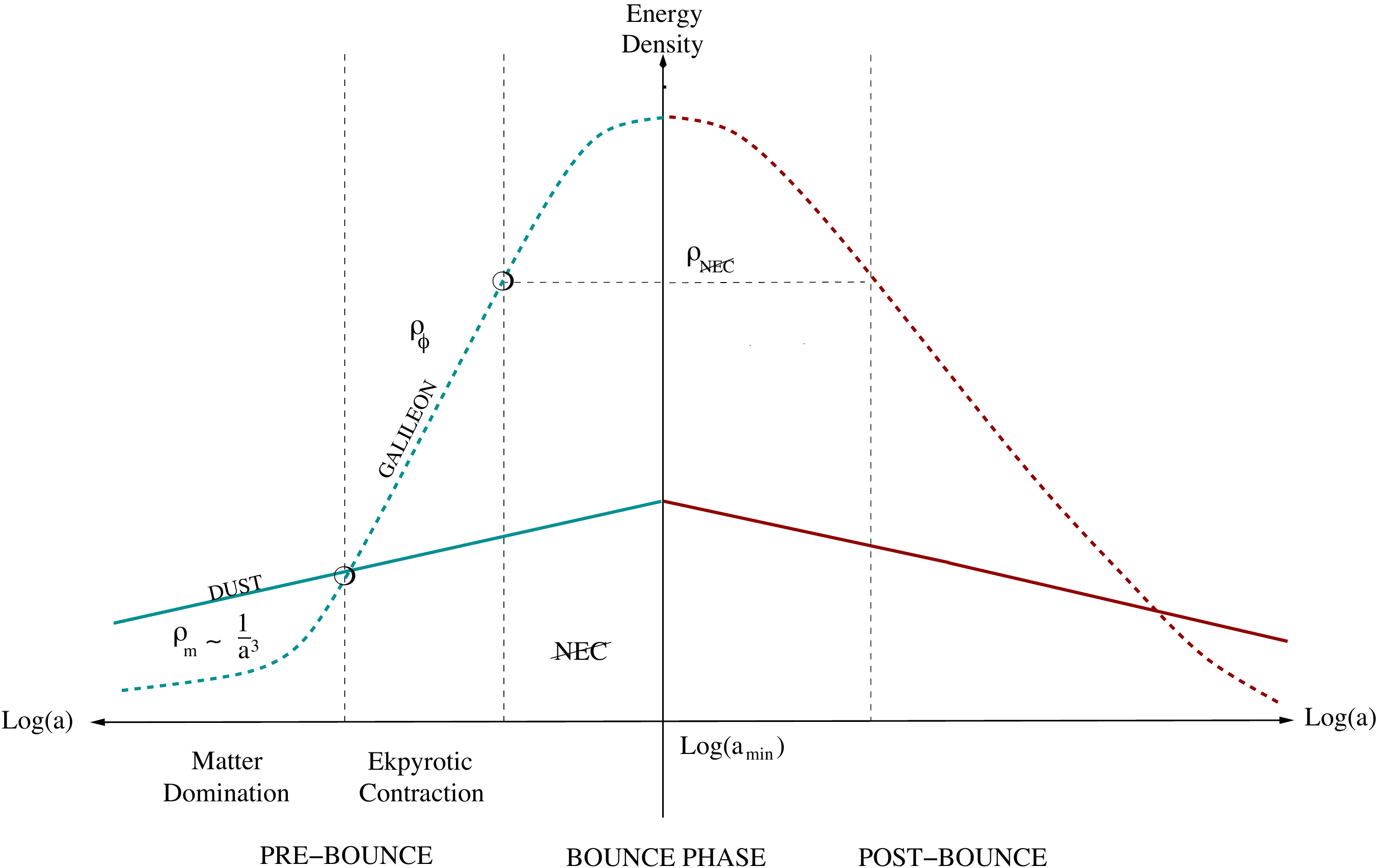}
\caption{Schematic of the nonsingular model proposed in
  \cite{Cai:2012va}, see also fig \ref{comparison} (a). The initial
  conditions are chosen so that the model starts in the matter
  dominated, contracting phase. The ekpyrotic phase of contraction is
  reached once $\phi$ takes over: the energy density of the ekpyrotic
  scalar field grows faster than that of regular matter or anisotropy.
  The end of the ekpyrotic phase signals the beginning of the bouncing
  phase, ghost condensation, which is followed by a fast-roll
  expansion and a final transition to the expansion in a standard big
  bang cosmology.}
\label{MB}
\end{center}
\end{figure*}

A proposal that appears to avoid this instability is given in
\cite{Cai:2012va}. In this model, the curvature fluctuations are
expected to be scale invariant due to the presence of a matter phase
and the bounce is nonsingular, bypassing the initial big bang
singularity\footnote{ The nomenclature to call this bounce a
  ``matter bounce'' is unfortunate as the model requires no prior matter
  dominated contraction phase; it is, nevertheless, used as such in
  the literature.}. The universe undergoes a contraction, stops and
reverses to expansion at a finite value of the scale factor while
General Relativity remains valid.  The NEC is violated via a ghost
condensate and a galileon field.  Adopting the form of the Kinetic
Gravity Brading (KBG) model of \cite{Easson:2011zy}, the Lagrangian is
\begin{equation}
\mathcal L=K(\phi,X)+G(\phi,X)\box\phi\label{LMP},
\end{equation}
where $K$ and $G$ are functions of a dimensionless scalar field $\phi$
\begin{equation}
K(\phi,X)=[1-g(\phi)]X+\beta X^2-V(\phi)\,.
\end{equation}
The value of $\beta$ is chosen so that the kinetic term is bounded
from below at high energy scales and the dimensionless function
$g(\phi)$ is chosen so that a phase of ghost condensation occurs
briefly when $\phi$ approaches $\phi=0$, see Fig.~\ref{comparison}(a);
$G$ is a galileon type operator,
\begin{equation}
G(X)=\Upsilon X\Box \phi\,,
\end{equation}
with $\Upsilon$ a positive-definite number and
\begin{equation}
\Box\phi\equiv g^{\mu\nu}\nabla_{\mu}\nabla_{\nu}\phi\,.
\end{equation}
This term is introduced to stabilize the gradient term of
perturbations.  The potential is shown in Fig.~\ref{comparison}
together with the function $g(\phi)$. This function has a maximum
which is only slightly larger than unity. The dynamics differs
considerably from that of the previous model: right after the
ekpyrotic contraction that suppresses the shear to negligible values,
the NEC is only violated briefly, leading to a rapid bounce
immediately followed by a fast-roll expanding phase. The shear growth
is present, but has no time to develop efficiently, leading to
logarithmic growth only. When expansion takes over, the shear
decreases again: it is easy to set up initial conditions such that
this shear never dominates the overall dynamics\footnote{ Note
  that in this model, a large, flat, empty universe is assumed. These
  initial conditions may be seen as unnatural and hence highly
  fine-tuned.}. The schematic of such a bounce is illustrated in
Fig.~\ref{MB}.

Specifically, using the matter action (\ref{LMP}) leads to the
modified Einstein equations
\begin{eqnarray}
R_{\mu\nu}-\frac{1}{2}g_{\mu\nu} R&=&
\left(-K+2XG_{,\phi}+G_{,X}\nabla_{\sigma}X
\nabla^{\sigma}\right) g_{\mu\nu}\nonumber\\
&&+(K_{,X}+G_{,X}\Box\phi-2G_{,\phi})\nabla_{\mu}\phi\nabla_{\nu}\phi\nonumber\\
&&-G_{,X}(\nabla_{\mu}X\nabla_{\nu}\phi+\nabla_{\nu}X\nabla_{\mu}\phi),
\end{eqnarray}
where $F_{,\phi}$ and $F_{,X}$ denote derivatives of $F$ with respect
to $\phi$ and $X$ respectively.

To achieve a nonsingular homogeneous bouncing solution with an
ekpyrotic phase of contraction, the authors of \cite{Cai:2012va}
choose the potential,
\begin{equation}
V(\phi)=-\frac{2V_0}{\ex^{-\sqrt{\frac{2}{q}}\phi}+\ex^{b_V\sqrt{\frac{2}{q}}\phi}}\,,
\end{equation}
where $V_0$ is a positive constant with mass dimension and $q$, $b_V$
are free parameters (see Fig.~\ref{comparison}a for a schematic). The
choice of $g(\phi)$ is crucial for the success of a nonsingular
bounce, since $g$ should dominate the kinetic term for $|\phi|\ll 1$
to violate the NEC. The function
\begin{equation}
g(\phi)=\frac{2g_0}{{\ex^{-\sqrt{\frac{2}{p}}\phi}+\ex^{b_g\sqrt{\frac{2}{p}}\phi}}}\,,
\end{equation}
with  free parameters $p$, $g_0$ and $b_g$, was chosen in \cite{Cai:2012va}. 

To summarize, the model entails the choice of two free functions
$V(\phi)$ and $g(\phi)$ and 2 constants, which in \cite{Cai:2012va}
reduces to the specification of $8$ parameters, namely $\{V_0, q, g_0,
p, \Upsilon,\beta,b_V,b_g \}$; a set of numerical values was then
chosen, among many others, merely to provide a proof of concept to
illustrate a bounce that avoids the BKL instability: the calculation
was done in the Bianchi I case in which a flat, homogeneous but
anisotropic Universe contracts under the domination of a dust-like
fluid.   As mentioned above, the resulting curvature perturbations
are scale invariant as shown in \cite{Wands:1998yp}. The
instabilities discussed in Sec.~\ref{Fatal_effects} are not present in
this context because the bounce is rapid and immediately followed by a
rapid expansion that counterbalances any prior increase of the shear,
which is produced marginally (logarithmically) during the bounce. The
duration of the bounce, which is short in comparison to the Hubble
time just before it, is however long enough to ignore quantum gravity
corrections, which would spoil the predictability of the model.

\subsubsection{The cosmological super-bounce}
\label{superbounce}

\begin{figure*}[tb]
\begin{center}
\includegraphics[scale=0.65]{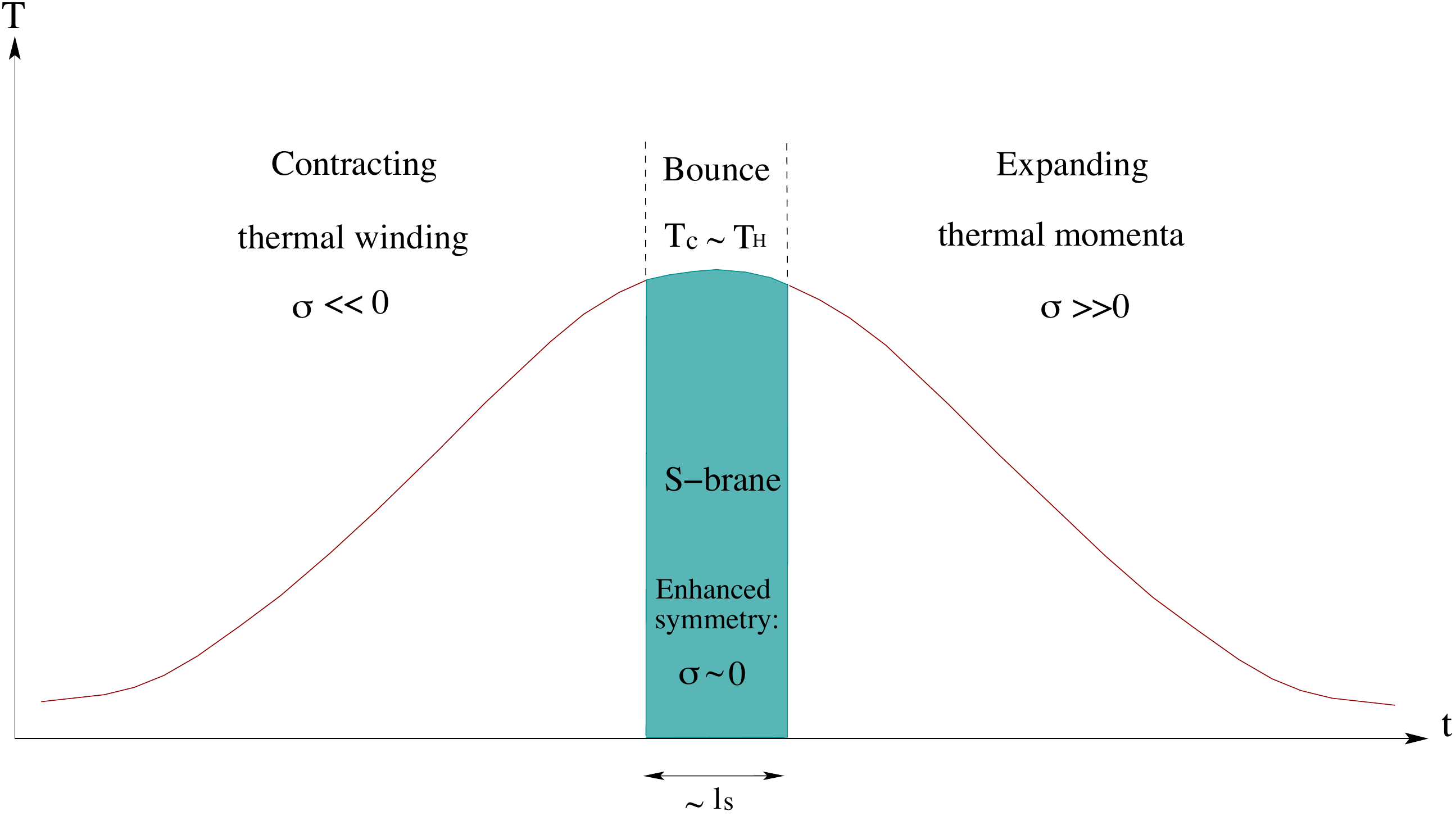}
\caption{A schematic of the temperature evolution in the S-brane
  bounce model: two dual regimes with a thermal string gas, the
  expanding and contracting phases, are connected via an S-brane,
  which enables the violation of the NEC for a brief period of time
  without introducing fatal instabilitites. The S-brane arises as a
  consequence of an extended symmetry once the temperature approaches
  the critical (Hagedorn) temperature $T_\mathrm{c}$
  \cite{Kounnas:2011fk,Kounnas:2013yda}.}
\label{sbrane}
\end{center}
\end{figure*}

Based on prior work in \cite{Cai:2012va}, Lehners \etal constructed a
{\it {super bounce}} in \cite{Koehn:2013upa} based on a galileon
Lagrangian given by
\begin{equation}
\mathcal L=-\frac{R}{2}+P(X,\phi)+g(\phi)X\Box\phi\,,
\end{equation}
where
\begin{equation}
P(X,\phi)=k(\phi)X+t(\phi)X^2-V(\phi)\,,
\end{equation}
$V(\phi)$ is the usual ekpyrotic potential, as depicted in
Fig.~\ref{comparison}(a), and
\begin{eqnarray}
k(\phi)&=&1-\frac{2}{(1+2\kappa\phi^2)^2}\,,\\
t(\phi)&=&\frac{\bar t}{(1+2\kappa\phi^2)^2}\,,\\
g(\phi)&=&\frac{\bar g}{(1+2\kappa\phi^2)^2}\,,
\end{eqnarray}
while $\bar t$, $\bar k$ and $\bar g$ are constants. These functions
are chosen such that the effective ghost condensate $(\mathcal L\sim
-X+X^2)$ and the galileon term violate the NEC briefly, leading to a
bounce.  The functions $t$ and $g$ are non-zero as $k$ passes through
zero to prevent a singularity.  Opposite to \cite{Cai:2012va}, where
$t=\bar t$ and $g=\bar g$, the higher derivative terms turn off for
large values of $\phi$, particularly during the ekpyrotic phase, to
simplify an implementation into supergravity. The functional form of
$t$, $k$ and $g$ entails considerable freedom of choice.

As mentioned in Sec.~\ref{galileon}, the above Lagrangian can be
implemented in supergravity after a computational tour de force, see
\cite{Koehn:2013upa} for details. The resulting bounce is devoid of
the many pitfalls reviewed in Sec.~\ref{Fatal_effects}, and thus, at
the time of writing, one of the most successful nonsingular bouncing
models, see Table~\ref{table:}.

Observables, such as the scalar spectral index and non-Gaussianities,
are consistent with the known results in ekpyrotic scenarios, see
Sec.~\ref{Perturbations}, since curvature perturbations stay frozen on
super Hubble scales during the bounce, as shown in
\cite{Battarra:2014tga}: here, the computation was performed in the
harmonic gauge with a detailed discussion of the validity of linear
perturbation theory, see Sec.~\ref{viabilityofptb}.
\subsubsection{Models based on T-duality in string theory}

{\it $\bullet$ S-brane models\\}\label{stringy}

String theory is a UV complete theory of quantum gravity and as such,
it provides a consistent framework for model building. Hence, a
cosmological model free of singularities should be feasible via the
incorporation of fundamental duality symmetries and stringy degrees of
freedom in a time-dependent setting, see
\cite{Florakis:2010is,Kounnas:2011fk,Kounnas:2011gz,Kounnas:2013yda}
and \cite{Kiritsis:1994np,Cornalba:2002fi,Cornalba:2003kd} for early
work.

An example in type II superstring theory, $\mathcal N=(4,0)$,
compactified to four dimensions entails a weakly coupled string gas as
a thermal component to enable a bounce
\cite{Florakis:2010is,Kounnas:2011fk,Kounnas:2011gz}. Such a
nonsingular cosmological scenario in dilaton gravity is usually
hindered by the appearance of Hagedorn instabilities, which occur at a
critical temperature close to the string scale, $T_\mathrm{c}\sim
1/\ell_\mathrm{s}\sim \mathcal{O}(10^{15})$ GeV (the Hagedorn
Temperature), and signal non-trivial phase transitions
\cite{Atick:1988si,Antoniadis:1991kh,Barbon:2001di,Barbon:2004dd}.
The Hagedorn singularity as well as the big bang singularity can be
resolved \cite{Kounnas:2011fk}; the nonsingular\footnote{ The
  S-brane's delta function source is a delta function in temperature
  and thus, in time. Therefore, it does not constitute a singularity
  with respect to curvature invariants, which we take as the
  definition of a singular bounce. In essence, this delta function
  source is conceptually closer to the presence of D-branes and indeed
  cosmological perturbations can be matched across the S-brane
  unambiguously. The string theoretical description of the S-brane
  bounce goes beyond the scope of this review and we refer the
  interested reader to the original literature
  \cite{Florakis:2010is,Kounnas:2011fk,Kounnas:2011gz}. }
cosmological evolution is governed by a phase transition, described in
terms of a spacelike brane (S-brane\footnote{The S-brane may be viewed
  as a defect that interpolates between the two distinct geometrical
  phases of the underlying conformal field theory. At the level of the
  low energy effective description, the condensate of massless states
  that appears at $T_\mathrm{c}$ forms a short-lived, space-filling
  object, whose lifetime is comparable to the temporal resolution of
  the EFT and is therefore modeled as a delta function. At the CFT
  level, an operator becomes marginal at $T_\mathrm{c}$ which converts
  directly winding states into momentum ones with a coefficient
  determining the efficiency of this interaction proportional to
  $\alpha'$.}), between two dual thermal phases.

This setup realizes temperature duality (T-duality) of string theory
\cite{Florakis:2010is,Kounnas:2011fk,Kounnas:2011gz,Kounnas:2013yda}. The
crucial component in this scenario is a thermal string gas, see
\cite{Brandenberger:1988aj,Tseytlin:1991xk} and
\cite{KalyanaRama:1997xt,Bassett:2003ck,Easther:2004sd,Battefeld:2005av,Kaloper:2007pw,Brandenberger:2008nx,Greene:2008hf}
for reviews on string gas cosmology and early work: light thermal
momentum modes and light thermal winding modes exchange roles via
T-duality. Near the critical temperature, new light degrees of freedom
appear due to an enhanced symmetry, giving rise to the S-brane which
serves as a glue between the two dual thermal phases. The S-brane has
localized (in time) negative pressure, but no energy density. It thus
violates the NEC in a controlled manner; in the low energy
description, the S-brane, which has a thickness set by the string
scale, can be treated as a $\delta$-function source in the Einstein
equations. Since the violation of the NEC is extremely brief, the
instabilities hampering a ghost-condensate bounce, see
Sec.~\ref{ghost}, are absent.
 
The action of the physical system in the string frame, denoted by a
tilde, reads
\begin{eqnarray}
S&=&\int \dd^4x\sqrt{-\tilde g}\ex^{-2\phi}\left(\frac{\tilde
R}{2}+2\tilde\nabla_{\mu}\phi\tilde\nabla^{\mu}\phi\right)\label{stringaction}\\
&&+\int \dd^4x\sqrt{-\tilde g}\tilde P-\int \dd\tilde\beta
\dd^3\xi\sqrt{\tilde\gamma}\ex^{-2\phi}\kappa\delta
\left(\tilde\beta-\tilde\beta_\mathrm{c}\right)\nonumber\,,
\end{eqnarray}
where $\phi$ is the dilaton, $\tilde g$ is the determinant of the
string frame metric, $\tilde R$ is the corresponding Ricci scalar,
$\tilde P\propto \tilde T^4$ the pressure of the thermal string gas,
$\tilde \gamma$ is the determinant of the induced metric at the
location of the S-brane, $\kappa$ is the brane tension and
$\tilde\beta$ is the inverse of the temperature. For more details see
\cite{Kounnas:2011fk,Kounnas:2011gz,Kounnas:2013yda,Brandenberger:2013zea}.
Since the thermal string gas has a constant equation of state, it can
be modeled by a derivatively coupled scalar field via
\begin{equation}
\int \dd^4x\sqrt{-\tilde g}\tilde P=\int \dd^4x\sqrt{-\tilde
g}n^*\sigma_\mathrm{r}(-\partial_{\mu}\psi\partial^{\mu}\psi)^2\,,
\end{equation}
where $n^*$ is the number of massless degrees of freedom and
$\sigma_\mathrm{r}$ is Boltzmann's radiation constant.  Hence, the
temperature is identified with
\begin{equation}
\tilde T^2=-\tilde g^{\mu\nu}\partial_{\mu}\psi\partial_{\nu}\psi\,,
\end{equation}
a relation which continues to hold at the perturbed level.

The dilaton and the scale factor in the Einstein frame reverse at the
critical temperature $T_\mathrm{c}$, with the latter's bounce
corresponding to a smooth reversal of contraction to expansion so that
the big bang curvature singularity is absent; at the phase transition
the dilaton attains its maximal value, determined by the brane
tension, the critical temperature $T_\mathrm{c}$, and the number of
thermally excited massless degrees of freedom; as a consequence, the
transition can take place entirely in the weakly coupled regime.
 
Each of the three phases, contraction, bounce and expansion, admits a
local effective field theory description \cite{Kounnas:2011fk}: the
temperature in the contracting regime can be identified with the
inverse period of the Euclidean time cycle, $T=1/(2\pi R)$. The
expanding phase corresponds to the dual of the contracting one with
$\tilde{R}=R_\mathrm{c}^2/R$ and temperature $T=1/(2\pi
\tilde{R})$. Defining the thermal modulus
$\sigma=\ln(R/R_\mathrm{c})$, we have
\begin{enumerate}
\item Contraction: light thermal windings regime, $R_\mathrm{c}/R\gg
  1$, ($\sigma<0 $ and $|\sigma|\gg 1$, \ie large and negative).
\item Bounce: enhanced symmetry regime -- thermal states become
  massless, $|R/R_\mathrm{c} -R_\mathrm{c}/R|\ll 1$, ($\sigma\sim 0$).
\item Expansion: light thermal momenta regime, $R/R_\mathrm{c}\gg 1$,
  ($\sigma>0 $ and $\sigma\gg 1$, \ie large and positive).
\end{enumerate}
Fig.~\ref{sbrane} shows a schematic representation of these phases.

To study fluctuations, one needs to perturb the metric and the dilaton
together with the thermal component. Modeling the latter via a
derivatively coupled scalar field
\cite{Mukhanov:2005sc,Dubovsky:2005xd,Boubekeur:2008kn}, the
transition of fluctuations between contracting and expanding phases is
studied in \cite{Brandenberger:2013zea}. Applying the Israel junction
conditions \cite{Israel:1966rt}, Sec.~\ref{sbounceptbs}, across the
S-brane, the curvature fluctuation $\zeta$ in the expanding phase can
be computed from the perturbations generated via quantum fluctuations
in the degrees of freedom during the contracting phase. Incorporating
a matter dominated phase during contraction \cite{Wands:1998yp}, a
scale-invariant spectrum of curvature perturbations can be generated
at late times. A tilt of the power-spectrum may result from changes to
the speed of sound caused by the dilaton admixture to the matter
fluid, but a computation of this effect has not been performed yet.

This implementation of a nonsingular bounce in string theory is a
promising proof of concept. However, the current proposal suffers from
several short-comings: for instance, since no ekpyrotic phase is
present, the BKL instability commonly arises \cite{Belinsky:1970ew},
see Sec.~\ref{BKL}. It may be problematic to incorporate an ekpyrotic
field, see Sec.~\ref{ekpyroticphase}, since its fast growing energy
density might dominate over the thermal component.  It was argued in
\cite{Brandenberger:2013zea} that the presence of kinetic energy in
the dilaton, which also redshifts as $a^{-6}$, would alleviate the
problem to some degree. However, it is crucial that the thermal
component dominates during the bounce, hence, fine-tuning appears
unavoidable. Lastly, since the temperature is usually very high,
$T_\mathrm{c}\sim 1/\ell_\mathrm{s}\sim \mathcal{O}(10^{15})$ GeV,
thermal relics, such as gravitinos, and potentially topological
defects from phase transitions, such as magnetic monopoles, may be
produced \cite{Battefeld:2009sb} (see Sec.~\ref{relics}), but this is
model-dependent. In addition, large non-Gaussianities may be produced
\cite{Gao:2014hea,Gao:2014eaa}.\\

{\it $\bullet$ A Hagedorn phase in string-gas cosmology \\}
\label{sgc}

\begin{figure}[tb]
\begin{center}
\includegraphics[scale=0.4]{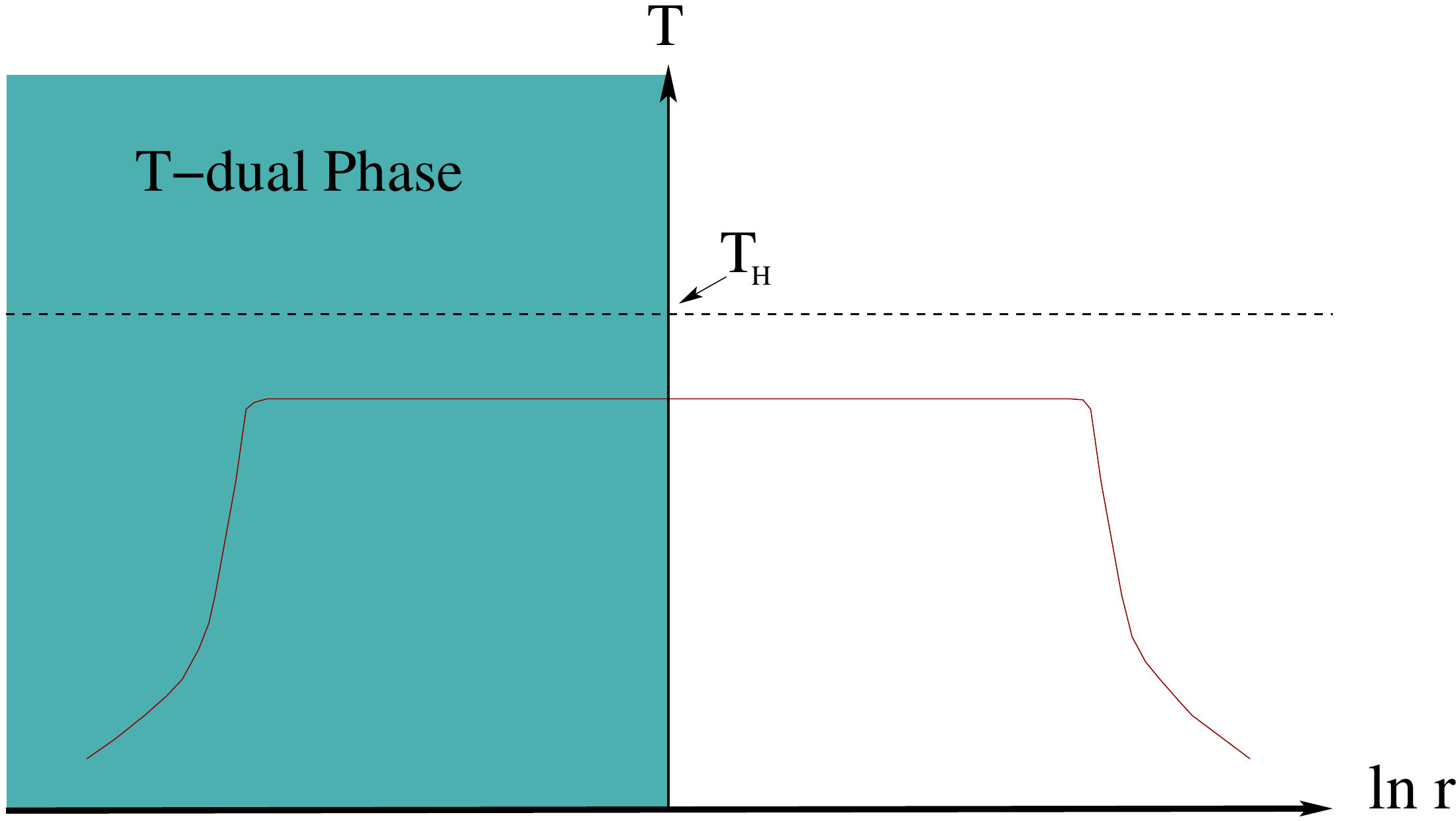}
\caption{Schematic evolution of the temperature, red curve, plotted
  over a dimensionless radius, $r$, in a toroidal compactification of
  heterotic string theory in string-gas cosmology. Perturbations are
  generated in the quasi-static Hagedorn phase $T\sim T_\mathrm{H}$,
  see \cite{Brandenberger:2011et} for a review. The transition from a
  Hagedorn phase to a contracting expanding FLRW universe cannot be
  computed within the realm of General Relativity or dilaton
  gravity. T-duality, a non-perturbative symmetry of string theory,
  indicates that this Hagedorn phase may be incorporated into a
  bouncing universe via non-perturbative techniques. In the T-dual
  description, string winding and momentum modes exchange roles and
  the scale factor is replaced by its inverse.}
\label{hagedorn}
\end{center}
\end{figure}

In a speculative, yet interesting proposal motivating the above
S-brane bounce, Brandenberger, Vafa and Nayeri
\cite{Nayeri:2005ck,Brandenberger:2006pr} (see
\cite{Brandenberger:2011et,Battefeld:2005av} for reviews and
\cite{Kaloper:2007pw,Kaloper:2006xw} for criticisms) attempted to
generate a scale-invariant spectrum of density fluctuations via
temperature fluctuations during a quasi-static Hagedorn phase of a
string gas comprised of closed heterotic strings\footnote{The S-brane
  bounce has its roots in string-gas cosmology, but operates in Type
  II string theory, opposite to the models of this section in
  heterotic string theory. Note that the role of T-duality also
  differs, as it involves the Euclidian time cycle in the S-brane
  bounce, and the scale factor of a toroidal compactification in
  string-gas cosmology.}. If the dilaton and all other moduli fields
are assumed to be stabilized at these high temperatures, the resulting
spectrum is indeed scale invariant.  A tilt is introduced through the
transition from the Hagedorn phase to a radiation dominated expanding
FLRW universe. This model may be viewed as a bounce if a pre-Hagedorn
phase is introduced via T-duality, see Fig.~\ref{hagedorn}. A concrete
realization (that is free of ghosts) invokes an infinite number of
higher derivative interactions, see
\cite{Biswas:2006bs}. Alternatively, this proposal can be seen as an
emergent universe without a bounce.

This model also predicts an observably large gravitational wave
spectrum, similar to inflationary slow roll models, but with a slight
blue tilt
\cite{Brandenberger:2014faa,Brandenberger:2006xi,Brandenberger:2006vv,Nayeri:2006uy},
opposite to the inflationary red one. Thus, this model may be
compatible with a high level of gravitational waves such as claimed by
B{\footnotesize ICEP2} \cite{Ade:2014xna,Ade:2014gua}, see
Eq.~(\ref{BICEP2}) below.

The string gas proposal has several points that need be addressed:
\begin{itemize}

\item the flatness problem is not solved, but requires fine-tuned
  initial conditions,

\item a quantitative understanding of the Hagedorn phase is
  incomplete, so that the transition from the Hagedorn phase to a
  radiation-dominated FLRW universe cannot yet be computed.  Due to
  the high temperature, $T_{\mathrm{H}}\sim T_{\mathrm{GUT}}$, the use
  of General Relativity or dilaton gravity may not necessarily be
  justified. A phenomenological proposal for a nonsingular, ghost free
  bounce induced by an infinite series of higher derivative
  corrections was given in \cite{Biswas:2006bs}. If the Hagedorn phase
  is directly connected to the expanding FLRW phase, one should expect
  the over-production of thermal relics and topological defects (if
  present in the spectrum of the theory) from phase transitions,
  because the temperature is so high, as pointed out by one of the
  authors of this review in \cite{Battefeld:2009sb},
  
\item keeping the dilaton dynamical, as one might expect at these high
  energies, destroys the scale invariance of this spectrum
  \cite{Kaloper:2006xw}. While dilaton gravity is not expected to be
  applicable during the Hagedorn phase, it is uncertain if all similar
  scalar quantities can be frozen during this phase, see footnote
  \ref{stabilization} in Sec.~\ref{implementationST} for the related
  challenges of moduli stabilization.
\end{itemize} 

\section{Cosmological Perturbations}
\label{Perturbations}

In addition to providing a solution to problems of the big bang, a
bounce should produce almost Gaussian (\ie with a low level of
non-Gaussianities) perturbations whose spectrum is nearly scale
invariant and slightly red in order to agree with current observations
as revealed by the P{\footnotesize LANCK} data
\cite{Ade:2013uln,Ade:2013nlj,Ade:2013zuv}. Further, if the
B{\footnotesize{ICEP2}} interpretation is confirmed, gravitational
waves at the level of $r\sim \mathcal{O}(0.1)$ ought to be present at
these scales \cite{Ade:2014xna}.  See however,
\cite{Flauger:2014qra,Mortonson:2014bja} for a study highlighting that
the polarization observed by {B{\footnotesize{ICEP2}} might be caused
  by dust instead of primordial gravitational waves. 

Given that a bounce is feasible, the aforementioned conditions impose
severe constraints on the overall scenario, particularly during the
contracting phase. If only a single degree of freedom is present in
the contracting phase, so that the perturbations are adiabatic, the
equation of state parameter must be $w\sim 0$ in order to generate a
nearly scale-invariant power spectrum\footnote{A duality between
  contracting and inflating universes exists at the level of the power
  spectrum, see \cite{Wands:1998yp}.} \cite{Finelli:2001sr}. Models
with $w\approx 0$ are called ``matter bounces'' because their equation
of state mimics that of ordinary dust; it does not imply the presence
of actual matter, since such an equation of state can also be achieved
by a scalar field oscillating in a quadratic potential.
Alternatively, a scale-invariant power spectrum can be generated by an
entropic mechanism, similar to the curvaton scenario, see
Sec.~\ref{absenceNG}. However, anisotropies/shear often arise, as
discussed in \cite{Notari:2002yc,Xue:2010ux,Xue:2011nw}, which can
spoil the bounce\footnote{The generating mechanism of a
  scale-invariant power spectrum is essentially decoupled from the
  physics of the actual bounce.} and/or the spectrum, see
Sec.~\ref{Fatal_effects}. Entropic mechanisms also generate
non-Gaussianities via the conversion mechanism, which are in tension
with current observational limits, see Sec.~\ref{absenceNG}.

An equation of state parameter larger than unity during the
contraction (ekpyrotic phase) can prevent the growth of
anisotropy/shear. Models whereby the bounce is realized via galileons,
in which the Lagrangian contains higher order derivatives, avoid the
appearance of ghosts while the extra additional degrees of freedom
violate the NEC. Promising models which address the anisotropy problem
in Galileon cosmology and might generate a scale-invariant spectrum
include \cite{Chimento:2005ua,Cai:2012va,Cai:2013vm,Koehn:2013upa},
see Sec.~\ref{galileon}.

Models with an ekpyrotic phase do not generate measurable
gravitational waves, whereas matter bounce models generically predict
a tensor-to-scalar ratio in excess of observational limits, see
Sec.~\ref{rcyclic}.

In the subsequent sections we discuss cosmological perturbations
(scalar and tensor) and non-Gaussianities in a few concrete models. We
do not aim to provide a complete overview of all proposals, but
deliberately focus on some promising ones to explain the crucial steps
of the computations and highlight important results with broad
applicability.

\subsection{The viability of perturbation theory in a contracting
  universe}
\label{viabilityofptb}
 
Is a perturbative analysis possible in nonsingular models?  For
singular models, such as the original ekpyrotic scenario, cosmological
scalar perturbations diverge as $a\rightarrow 0$ \cite{Lyth:2001nv},
casting doubt on the perturbative treatment. In such scenarios, it is
understood that at some point the 4D effective theory breaks down, and
one has to resort to the full description in string theory. One might
hope that the singularity and thus the divergence of perturbations is
absent. However, merely going to the higher dimensional setting, for
example, in the ekpyrotic universe \cite{Enqvist:2001zk}, and properly
incorporating metric perturbations does not necessarily alleviate this
problem, as shown in \cite{Battefeld:2004mn}.  A similar problem is
present for vector perturbations in a contracting universe
\cite{Battefeld:2004cd}.

For a nonsingular bounce, perturbations are not necessarily divergent,
but still strongly growing in the contracting phase.  The current lore
in the literature is that linear perturbation theory fails in some
gauges, such as the longitudinal \cite{Lyth:2001nv,Martin:2001ue} and
comoving ones \cite{Kodama:1996jh}, while it remains valid in others,
for example, in the uniform $\chi$ field gauge \cite{Allen:2004vz}.
As shown in \cite{Xue:2010ux,Xue:2011nw}, the curvature perturbation
and anisotropy grow rapidly during contraction, casting doubt on the
viability of perturbation theory, see Sec.~\ref{Fatal_effects}.

One of the first occurrences where this problem surfaced was the
pre-big bang scenario, see \cite{Gasperini:2002bn} for a review. In
\cite{Gasperini:2004ss} it was argued that one could go to a gauge
where scalar perturbations are at most logarithmically growing, but it
should be noted that the usual gauge invariant variables still obey a
power-law growth.

The question regarding the validity of the perturbative linear regime
during the bounce is particularly evident in the longitudinal gauge
due to the Newtonian potential's rapidly growing mode
\cite{Creminelli:2004jg} -- the Newtonian potential corresponds to the
metric perturbation function $A$ in Eq.~(\ref{FLRWmetric}) below and
equals the Bardeen Potential $\Phi$ in this gauge
\cite{Bardeen:1980kt}.  However, this mode becomes a decaying one
during the expansion phase, hinting that this growth may be a gauge
artifact and/or that $\Phi$ is not a good tracer to check the validity
of perturbation theory (one may define non-divergent, gauge-invariant
variables by multiplying with appropriate background functions). It is
possible to find other gauges in which this growing contribution is
absent \cite{Allen:2004vz}.  A generalization of these ideas to a
large class of models was attempted in \cite{Vitenti:2011yc}, where a
set of conditions for linearity is obtained that allows the
perturbative series to be valid. The spectrum of modes considered in
\cite{Vitenti:2011yc} became frozen during a matter dominated
contracting phase, but the actual bounce was kept general.  The
conditions in \cite{Vitenti:2011yc} arise by demanding that all terms
in the perturbed Einstein equations remain small. One then obtains a
set of conditions by requiring the smallness of the perturbed volume
expansion rate as well as that of the second spatial derivative (the
covariant Laplacian with respect to the spatial metric) acting on the
two Newtonian potentials $A$ and $\psi$ in Eq.~(\ref{FLRWmetric}), as
well as the shear, which is set by the off-diagonal metric
perturbations. Of all the gauges considered in \cite{Vitenti:2011yc},
the synchronous gauge ensures that perturbations near the bounce stay
finite and small at all times \cite{Vitenti:2011yc}.

 This advantage of the synchronous gauge, as well as the presence
of a dynamic attractor in the ekpyrotic scenario, was first pointed
out in \cite{Creminelli:2004jg}. In the presence of the aforementioned
attractor, the cosmological solution becomes locally homogeneous,
isotropic and flat. Thus, as long as the bounce is not sensitive to
exponentially small corrections, the bounce appears identical to every
observer. As a consequence, unambiguous predictions can be made with
respect to the spectrum of fluctuations. Based on this method,
Creminelli \etal \cite{Creminelli:2004jg} showed that the original
single field ekpyrotic model does not possess a scale-invariant
spectrum.  It is further shown why the Bardeen potential generically
diverges, see Sec.~\ref{failureN}. 

Linearity conditions can also be defined covariantly, as in
\cite{Kumar:2012tr} for a radiation and dust-like single-fluid FLRW
background. It has been claimed in \cite{Kumar:2012tr} that these
conditions are violated as the scale factor shrinks and the bounce is
approached.  However, as shown in \cite{Pinto-Neto:2013zya}, the
conditions of \cite{Kumar:2012tr} reduce to the ones in
\cite{Vitenti:2011yc} and a violation of linear perturbations is not
necessarily present.  

 Going beyond linear perturbation theory, a recent, interesting
and encouraging study of adiabatic perturbations generated was
performed in \cite{Xue:2013bva}. A simple model of the bounce was used
to follow perturbations through a nonsingular matter-like contraction
akin to \cite{Peter:2002cn,Allen:2004vz,Cai:2007zv}, see
Sec.~\ref{GhostField}, the actual bounce and into an expansion
phase. The main goal of this study was to find a well-behaved gauge in
which a non-perturbative numerical analysis can be performed. A
pressing problem preventing the use of constant mean curvature slices,
as commonly used in standard perturbative analysis, is that these
slices stop being space-like during the transitions to the bouncing
phase in the presence of inhomogeneities. A possible way out is
provided by the use of the harmonic gauge, where coordinates satisfy
the condition
\begin{equation}
\nabla_{\alpha}\nabla^{\alpha}x^{\mu}=0\,,
\end{equation} 
which entails 
\begin{equation}
g^{\alpha\beta}\Gamma^{\gamma}_{\alpha\beta}=0\,.
\end{equation}
In this gauge, coordinate singularities are absent during the bouncing
phase when $a$ and $H$ are non-monotonic. Further, the equations of
motion for metric components are wave-like and thus easy to solve. In
\cite{Xue:2013bva} results in the harmonic gauge were also translated
to the commonly used non-linear generalization of the curvature
perturbation in the covariant formalism, $\zeta$.  At the time of
writing, the harmonic gauge appears to be the gauge of choice to
compute unambiguously the evolution of perturbations in a bouncing
universe.

The non-perturbative, numerical solutions in \cite{Xue:2013bva} show
that the universe does bounce in regions where the universe is
homogeneous and isotropic enough. More precisely, the universe
undergoes a nonsingular bounce in regions where the ratio between
energy density in the anisotropy $\rho_\theta$ [see
  Eq.~(\ref{H2theta})] and that of the field $\rho_{\chi}$ satisfies,
\begin{equation}
\left|\frac{\rho_\theta}{\rho_{\chi}}\right|<1.
\end{equation}
On the other hand, in  regions where 
\begin{equation}
\left|\frac{\rho_\theta}{\rho_{\chi}}\right|>1,
\end{equation}
the universe crunches: since energy in the ghost field, $\rho_{\chi}$,
drives the nonsingular bounce, and because it can never catch up with
$\rho_\theta$, the universe in this particular region 
collapses into a singularity.

If a nonsingular bounce takes place close to the critical ratio above,
adiabatic modes are generically strongly coupled during the bounce,
which affects the power spectrum and may generate large
non-Gaussianities. However, if the perturbations are of the same order
as the observed primordial ones, nonlinearities are unimportant, the
bounce is unscathed, and the strong coupling problem for super-Hubble
modes does not arise.  Thus, in this case, the nonsingular bounce is
not expected to lead to large non-Gaussianities \cite{Xue:2013bva}.

For more details on the use of the harmonic gauge see
\cite{Battarra:2014tga}, which makes contact to other commonly used
gauges and provides another application to the superbounce, see
Sec.~\ref{superbounce}: the validity of linear perturbation theory was
tested and it was confirmed that the equations of motion for certain
variables, such as the curvature perturbation $\mathcal R$ in
(\ref{R}), contain singularities. However, in the harmonic gauge, the
relevant equations are nonsingular and linear perturbation theory
remains valid in this particular bounce model. Furthermore, it was
found that long wavelength perturbations are unaffected by the bounce;
hence, the pre-bounce power spectrum passes through the bounce
unaltered.

In the following, we focus on the generation mechanism of a
scale-invariant spectrum in the pre-bounce phase, employing the
curvature perturbation $\zeta$. It should be understood that the
harmonic gauge should be used to follow the perturbations through the
bounce, as mentioned above, to check whether or not a spectrum
survives in a particular bouncing model of the universe.

\subsection{Generating a nearly scale-invariant power spectrum in
  a contracting  universe}
\label{sis}

Measurements of the cosmic microwave background radiation
\cite{Ade:2013uln,Ade:2013nlj,Ade:2013zuv} reveal adiabatic, highly
Gaussian temperature fluctuations with a nearly scale-invariant power
spectrum.  These fluctuations trace back to curvature fluctuations
that must have been generated during a preceding inflationary or, as
in our case, a contracting phase of the early universe
\cite{Lehners:2011kr}.  These fluctuations need to show correlations
on super-Hubble scales at the time of decoupling, which can be
achieved either during a rapid accelerated expansion or a slow
contraction, see \cite{Wands:1998yp,Khoury:2001zk} for a duality at
the level of the power spectrum between these two options.  For
example, during an ekpyrotic phase, quantum fluctuations cross the
Hubble scales and are converted into classical, local density
perturbations. The scaling solution (\ref{scaling_solution}) shows
that the scale factor is almost constant, meaning that a mode's
wavelength stays also constant. The Hubble length, ${\mathcal
  H}^{-1}\sim t$ (assuming the bounce to take place at $t=0$), see
Fig.~\ref{spacetimeNSMB}, shrinks as $t\rightarrow 0$, so that any
mode eventually becomes super-Hubble close to the bounce.

We first provide a brief overview of cosmological perturbation theory
\cite{Peter:2009sa,Mukhanov:1990me}, following the review
\cite{Bassett:2005xm} in order to set our notation and then proceed by
providing the computation of the power spectrum in a two-field
ekpyrotic contracting universe as a concrete example of a bouncing
universe. This computation highlights the failure of single field
models, since they carry a deep blue spectral index, and explains the
current preference of two-field models, whereby scale-invariant
perturbations in an entropic field are subsequently converted to the
adiabatic mode. We also introduce the $\delta N$ formalism for our
discussion of non-Gaussianities in Sec.~\ref{absenceNG}. In addition,
we show how a scale-invariant spectrum can arise in the adiabatic mode
in a matter-dominated contracting universe. We briefly explain the
Deruelle-Mukhanov matching conditions, which are invoked whenever
distinct phases are attached to each other.

\subsubsection{Basics of cosmological perturbation theory}
\label{BasicsPT}

Our goal is the computation of the gauge invariant, comoving curvature
fluctuation $\zeta$, which is commonly used to impose observational
bounds.  Consider the perturbed FLRW metric
\cite{Mukhanov:1990me,PeterUzan2009}, including only scalar degrees of
freedom,

\begin{eqnarray}
  \dd s^2&=&a^2\left\{ -(1+2A)\dd\eta^2+2B_{,i}\dd\eta \dd x^i\right. \nonumber\\
  & & +\left.
  \left[\left(1-2\psi\right)\delta_{ij}+2E_{,ij}\right]\dd x^i \dd
x^j\right\},
\label{FLRWmetric}
\end{eqnarray}
in conformal time.  The four variables in the metric entail two
degrees of freedom and two gauge modes. A popular choice is the
Newtonian gauge where $B=E=0$ so that the gauge-invariant Bardeen
potentials read ${\Phi}=A$ and ${\Psi}=\psi$. The curvature
perturbation on hypersurfaces orthogonal to comoving worldlines is
defined as
\begin{equation}
\mathcal{R}\equiv \psi+\frac{H}{\dot\phi}\delta\phi\label{R}\,,
\end{equation}
where $\delta\phi$ is the scalar field perturbation.
This gauge-invariant variable coincides with $-\zeta$ on large scales,
see Eq.~(\ref{zeta}) below. In Fourier space, one may define the
so-called Mukhanov-Sasaki variable
\begin{equation} v_k\equiv z\delta\phi_k
\end{equation}
 with 
 \begin{equation}
 z\equiv a\sqrt{2\displaystyle\frac{\dot H}{H^2}},
 \end{equation}
and $k$ the wavenumber (see Sec.~\ref{gravitationalinstability} for
the inclusion of a general speed of sound $c_{_\mathrm{S}}$). In the
following, we provide expressions in the presence of many fields,
which we denote with a latin subscript $I=1\dots\mathcal{N}$; to avoid
clutter, we often skip the Fourier index $k$ on variables. The
curvature fluctuation $\zeta$ is easily expanded by means of the
$\delta N$ formalism
\cite{Starobinsky:1986fxa,Sasaki:1995aw,Sasaki:1998ug,vernizzi:2006ve,Rigopoulos:2005ae,Lyth:2005fi}
as,
\begin{eqnarray}
\zeta&\simeq&\sum_IN^I\delta\phi_{I}+\frac{1}{2}
\sum_{IJ}N^{IJ}\delta\phi_{I}\delta\phi_{J} \nonumber\\
& & +\frac{1}{3!}\sum_{IJK}N^{IJK}\delta\phi_{I}\delta\phi_{J}
\delta\phi_{K}+\cdots,
\label{curvature_perturbation}
\end{eqnarray}
where $\delta\phi_{I}\equiv\delta\phi_{I}|_{\psi=0}$, $N$ is the
number of e-folds
\begin{equation}
N\equiv\int_t^{t_{\mathrm{end}}}H\dd t,
\end{equation}
using an initially flat hypersurface as well as a final uniform
density hypersurface, $N_I=\partial N/\partial\phi_I$,
$N_{IJ}=\partial^2N/\partial\phi_I\partial\phi_J$, \etc and
$N^I=\delta^{IJ}N_J$.

It is useful to introduce the gauge-invariant Mukhanov-Sasaki
variables
\begin{equation}
Q_I\equiv\delta\phi_I+\frac{\dot\phi_I}{H}\psi\,\label{SM},
\end{equation}
that satisfy the equations of motion in Fourier space
\cite{Gordon:2000hv},
\begin{equation}
\ddot Q_I+3H\dot Q_I+\frac{k^2}{a^2}Q_I
+\sum_J\left[V_{IJ}-\frac{1}{a^3}
\left(\frac{a^3}{H}\dot\phi_I\dot\phi_J\right)^.\right]Q_J=0.
\label{SMEOM}
\end{equation}
Sometimes, rescaled Mukhanov-Sasaki variables are defined in the
$\psi=0$ gauge as $u_I\equiv a\delta\phi_I$. Note that in models with
only one scalar field and $z\propto a$, the two rescaled
Mukhanov-Sasaki variables are proportional to each other $v_{k}\propto
u_k$.  These variables are defined such that the friction term in
their respective equations of motion vanishes,
\begin{equation}
v''_k+\left(k^2 -\frac{z''}{z}\right) v_k=0\label{eomvk},
\end{equation}
where a prime denotes a derivative with respect to conformal time.

Useful relationships between these variables are
\begin{equation}
\mathcal{R}\equiv\sum_I\left(\frac{\dot\phi_I}{\sum_J\dot\phi_J^2}\right)Q_I,
\end{equation}
and 
\begin{eqnarray}
\mathcal{R}&\equiv&\Psi-\frac{H}{\dot H}\left(\dot\Psi+H{{\Phi}}\right),\\
-\zeta&=&\mathcal{R}+\frac{2\rho}{3(\rho+P)}\left(\frac{k}{aH}
\right)^2{{\Psi}}\label{zeta}.
\end{eqnarray}
We freely switch between $\mathcal{R}$ and $-\zeta$ in what follows,
since we are only interested in the limit of large scales $k\ll aH$.
 
The power spectrum of curvature fluctuations is defined as the Fourier
transform of the 2-point function,
\begin{equation}
\langle\mathcal{R}(k_1)\mathcal{R}(k_2)\rangle \equiv (2\pi )^3
\delta(\bm{k}_1+\bm{k}_2)P_{\mathcal{R}} (k_1).
\end{equation}
One often defines a rescaled dimensionless power spectrum by
\begin{equation}
\mathcal{P}_{\mathcal{R}}\equiv\frac{k^3}{2\pi^2}P_{\mathcal{R}}(k)
=\frac{4\pi k^3}{(2\pi)^3}|\mathcal R|^2,
\label{Powerspectrum}
\end{equation}
whose amplitude satisfies the COBE normalization $\mathcal P_{\mathcal
  R}=2.41\times 10^{-9}$ \cite{Bunn:1996py} and we introduced the
notation $|\mathcal{R}|^2\equiv P_{\mathcal{R}}$. Using the $\delta N$
formalism \cite{Dvali:2003em}, one may also compute the power spectrum
via
\begin{equation}
\mathcal P_{\zeta}=\sum_I(\delta N_I)^2\mathcal P_{\delta\phi_I|\psi=0},
\end{equation}
where \begin{equation}
\mathcal P_{\delta\phi_I|\psi=0}\equiv\frac{4\pi
k^3}{(2\pi)^3}\left|\frac{u_I}{a}\right|^2.\label{Powerspectrumfield}
\end{equation}
The scalar spectral index $n_\mathrm{s}$ is defined through
\begin{equation}
 n_\mathrm{s}-1\equiv\frac{\dd\ln\mathcal{P_{\zeta}}}{\dd\ln
k},\label{scalarspectralindex}
 \end{equation}
 which has to equal the P{\footnotesize LANCK} measurement
 $n_\mathrm{s}=0.9603\pm 0.0073$ \cite{Ade:2013zuv}. Thus, a blue
 power spectrum ($n_\mathrm{s}>1$), such as that obtained in the
 simplest matter bounce, is ruled out.
 
 Similarly, a tensor power spectrum can be defined, see
 Sec.~\ref{rcyclic}, whose amplitude is usually constrained via the
 tensor to scalar ratio
 \begin{equation}
 r\equiv\frac{\mathcal P_{_\mathrm{T}}}{\mathcal P_{\mathcal R}},
 \end{equation}
 which has to satisfy $r< 0.11$, according to P{\footnotesize LANCK}
 \cite{Ade:2013zuv} and $r=0.2^{+0.07}_{-0.05}$ if the B{\footnotesize
   ICEP2} interpretation of the data is correct
 \cite{Ade:2014xna}.  See however,
 \cite{Flauger:2014qra,Mortonson:2014bja}

\subsubsection{Two-field ekpyrosis}
\label{isocurvature}

As we shall see, it is impossible to generate a scale-invariant
spectrum in the adiabatic mode during an ekpyrotic phase; this led to
the investigation of two-field models \cite{Notari:2002yc}. Such
scenarios are often natural in concrete settings in string theory and
supergravity, see \cite{Koehn:2013upa} for a recent example. The
presence of at least two scalar fields gives rise to isocurvature
(entropy) perturbations, which, in turn, can source curvature
perturbations. Hence, if the isocurvature perturbations acquire a
nearly scale-invariant spectrum, the curvature perturbations may
inherit this spectrum via curvaton-like mechanisms, see
Sec.~\ref{absenceNG} \cite{Notari:2002yc,Lehners:2007ac}.

In what follows, we review the origin of scale-invariant perturbations
in an isocurvature mode during an ekpyrotic contracting phase.  As a
concrete example, consider two scalar fields $\phi_I$, $I=1,2,$ with
canonical kinetic terms and an uncoupled, exponential potential
\begin{equation}
V=-\sum^2_{I=1}V_I\ex^{-c_I\phi_I},\label{potential}
\end{equation}
where $c_I\equiv \sqrt{2/p_I}>0$ and $V_I>0$.
The equations of motion are
\begin{equation}
\ddot\phi_I+3H\dot\phi_I=-\frac{\dd V}{\dd \phi_I},\label{FE1}
\end{equation}
and the Friedmann equations read,
\begin{eqnarray}
3H^2&=&V+\sum^2_{I=1}\frac{1}{2}\dot\phi^2_I,\label{FE2}\\
\dot H&=&-\sum^2_{I=1}\frac{1}{2}\dot\phi^2_I.\label{FE3}
\end{eqnarray}
Eqs.~(\ref{FE1}--\ref{FE3}) with potential (\ref{potential}) possess
two attractor solutions, each corresponding to a single field
ekpyrotic solution, whereby either $\phi_1$ or $\phi_2$ dominate.  In
addition to these two attractor solutions, there is a third
alternative where the ratio of the kinetic energy to potential energy
is constant; the {\it adiabatic field} along this trajectory is not an
attractor.

Following the review \cite{Bassett:2005xm}, let us define,
\begin{equation}
\phi\equiv\frac{c_2\phi_1+c_1\phi_2}{\sqrt{c_1^2+c_2^2}},
\end{equation}
and an isocurvature field perpendicular to it
\begin{equation}
\xi\equiv\frac{c_1\phi_1-c_2\phi_2}{\sqrt{c_1^2+c_2^2}}.
\end{equation}
The potential becomes
\begin{equation}
V=-U(\xi)\ex^{-c\phi},
\end{equation}
where $c^{-2}\equiv c_1^{-2}+c_2^{-2}$ and
\begin{equation}
U(\xi)=V_1\ex^{-(c_1/c_2)c\xi}+V_2\ex^{(c_2/c_1)c\xi}\label{U},
\end{equation}
which has a maximum at
\begin{equation}
\xi_0=\frac{1}{\sqrt{c_1^2+c_2^2}}\ln\Big(\frac{c_1^2V_1}{c_2^2V_2}\Big).
\end{equation}
The unstable solution corresponds to $\xi=\xi_0$, while $\phi$ rolls
down the exponential potential in (\ref{potential}). The resulting
equation of state parameter satisfies $w\gg 1$.  For this solution,
$\phi$ is identical to the adiabatic field along the trajectory,
\begin{eqnarray}
\phi&\equiv&\int\sum_I\hat\phi_I\dot\phi_I \,\dd t,\\
\hat\phi_I&\equiv&\frac{\dot\phi_I}{\sqrt{\sum_J\dot\phi^2_J}},
\end{eqnarray}
whereas $\xi$ is the perpendicular isocurvature field.

The tachyonic instability in $\xi$ gives rise to a scale-invariant
spectrum of perturbations in $\xi$ in a contracting
universe\footnote{Setting initial conditions to this solution may
  appear fine-tuned, but can be natural in the framework of the
  Ph\oe{}nix universe, see Sec.~\ref{cyclic_ic}}.  We make the
assumption that $\xi$ stays close to $\xi_0$ throughout so that
$\dot\xi\approx 0$ and $U(\xi)\approx U(\xi_0)=$const. The Friedmann
equations leads to a power law contraction $a\propto (-t)^p$ with
$p\equiv 2/c^2\ll 1$.  From here onwards, we follow
\cite{Battefeld:2007st} (see also \cite{Tolley:2007nq}) and switch
back to conformal time $\eta=\int a^{-1} \dd t$ so that
\begin{equation} 
a\propto (-\eta)^{p/(1-p)} \propto
(-\eta)^{1/(\bar\epsilon-1)},
\label{scale_factor_epsilon}
\end{equation}
with $\bar\epsilon$ defined below through Eq.~(\ref{bar_epsilon}).

The Mukhanov-Sasaki variable for multiple fields is defined in
(\ref{SM}) and its equation of motion is given by (\ref{SMEOM}). In
the $\psi=0$ gauge, $Q_{\phi}=\delta\phi$ and $Q_{\xi}=\delta\xi$. In
terms of these variables, the comoving curvature perturbation is
$\zeta=\sum_I\dot\phi_IQ_I/\sum_J\dot\phi^2_J=Q_{\phi}/\dot\phi$.

In the absence of a curved trajectory in field space, there is no
conversion of isocurvature modes into adiabatic ones and the equations
decouple: using $\dot\xi=0$ and $V_{,\xi}=0$ in Eq.~(\ref{SMEOM}), we
obtain
\begin{eqnarray}
0&=&\delta\ddot\xi+3H\delta\dot\xi+\left(\frac{k^2}{a^2}+
V_{,\xi\xi}\right)\delta\xi,\\\label{eqq}
0&=&\delta\ddot\phi+3H\delta\dot\phi+\left[\frac{k^2}{a^2}+
V_{,\phi\phi}+\frac{1}{a^3}\left(\frac{a^3}{H}\dot\phi^2\right)^.\right]
\delta\phi\nonumber.\\
\end{eqnarray}  
Working in conformal time and making use of $u_{\xi}\equiv a\delta\xi$
and $u_{\phi}\equiv a\delta\phi$, the equations above reduce to
\begin{equation}
u''_I+\left(k^2-\frac{\beta_I}{\eta^2}\right)u_I=0\label{eomuI},
\end{equation}
with $I=\chi,\phi$ and
\begin{eqnarray}
\beta_{\xi}&\equiv&\frac{2\bar\epsilon^2
-7\bar\epsilon+2}{(\bar\epsilon-1)^2},\\\label{beta1}
\beta_{\phi}&\equiv&\frac{2-\bar\epsilon}{(\bar\epsilon-1)^2},
\end{eqnarray}
where we introduced
\begin{equation}
\bar\epsilon\equiv\frac{3}{2}(1+w)=\frac{1}{p}=\frac{1}{2\epsilon}\gg1,
\label{bar_epsilon}
\end{equation}
with the fast roll parameter defined in (\ref{epsFR}).  Imposing the
Bunch-Davies vacuum initial conditions well inside the Hubble radius
$k^2\eta^2\gg\beta_I$ so that $\beta_I$ can be neglected, we have
\begin{equation}
u_I=\frac{1}{\sqrt 2k}\left[b\left( \bm{k}\right)\ex^{-ik\eta}+
b^\dagger\left(\bm{k}\right)\ex^{ik\eta}\right],\label{ptbu1}
\end{equation}
where $b$ satisfies $\langle b(\bm{k}) b^\dagger(-\bm{\tilde
  k})\rangle =\delta^3(\bm{k}-\tilde{\bm{k}})$. Keeping the $\beta_I$
term in the equation of motion we obtain
\begin{equation}
u_I=\sqrt{-k\eta}\left[C(\bm{k})H^{(1)}_{\nu_I}(-k\eta)+C^*(-\bm{k})
H^{(2)}_{\nu_I}(-k\eta)\right],\label{ptbu2}
\end{equation}
where $H^{(1,2)}_{\nu_I}$ are Hankel functions of the first and second
kind respectively, with index
\begin{equation}
\nu_I=\frac{1}{2}\sqrt{1+4\beta_I}.\label{nu}
\end{equation}
Expanding the Hankel functions for large arguments we can match $u_I$
from (\ref{ptbu1}) and (\ref{ptbu2}) to determine $C$ as
\begin{equation}
C(\bm{k})=\sqrt\frac{\pi}{4k}\ex^{\frac{i\pi}{4}(2\nu_I+1)}b(\bm{k}).
\end{equation}
After Hubble crossing we can use the small argument limit of the
Hankel functions to arrive at
\begin{equation}
u_I=\frac{1}{\sqrt 4\pi
k}2^{\nu_I}\Gamma(\nu_I)(-k\eta)^{\frac{1}{2}-\nu_I}\tilde b(\bm{k}),
\end{equation}
where 
\begin{equation}
\tilde b(\bm{k})\equiv\frac{1}{i}\left[\ex^{\frac{i\pi}{4}(2\nu_I+1)}b(\bm{k})
-\ex^{-\frac{i\pi}{4}(2\nu_I+1)}b^+(-\bm{k})\right].
\end{equation}
The power spectrum in (\ref{Powerspectrumfield}) becomes
\begin{equation}
\mathcal P_I=\frac{4\pi k^3}{(2\pi)^3}\frac{1}{a^2}|u_I|^2,
\end{equation}
so that we can read off the scalar spectral indices 
\begin{equation}
n_I-1 \equiv \frac{\dd\ln\mathcal{P}_I}{\dd\ln k}
= 3-2|\nu_I|\label{scalarspectral}.
\end{equation}
Further expanding $\nu_I$ from (\ref{nu}) and (\ref{beta1}) in terms
of the fast roll parameter we obtain the spectral indices
\begin{eqnarray}
n_{\phi}-1&\simeq& 2+4\epsilon,\\
n_{\xi}-1&\simeq& 4\epsilon\label{nxi}.
\end{eqnarray}
The adiabatic mode carries a deep blue spectrum as in single field
ekpyrosis\footnote{Since these modes are not amplified and thus their
  wavefunction is not squeezed, it has been argued that they cannot be
  described classically, as pointed out in \cite{Tseng:2012qd}. As a
  result, the application of matching conditions, as delineated in
  Sec.~\ref{sbounceptbs}, to the ekpyrotic scenario, would be
  flawed. In the entropic mechanism, perturbations become classical
  and the conversion process causes super-Hubble perturbations to
  fully decohere \cite{Battarra:2013cha}; the lack of amplification,
  and thus squeezing, of gravitational waves in ekpyrotic scenarios
  casts doubts on treating the latter classically, as done in
  Sec.~\ref{rcyclic} and \cite{Boyle:2003km}. A study for
  gravitational waves similar to \cite{Battarra:2013cha} has not yet
  been performed. \label{footnotedecoherence}} \cite{Lyth:2001nq} and
the isocurvature mode carries a small blue tilt, respectively.  By
choosing a slightly more complicated potential than the
choice made in (\ref{potential})
one may also generate a slightly red spectrum for the isocurvature
field \cite{Fertig:2013kwa}, see also
\cite{Lehners:2007ac,Buchbinder:2007ad}.  To reiterate, the
isocurvature fluctuations still need to be converted into adiabatic
ones in realistic models (Sec.~\ref{absenceNG}), and to be passed
through the bouncing phase to the expanding one.

\subsubsection{The growing mode in the Newtonian gauge}
\label{failureN}
  
As explained in Sec.~\ref{viabilityofptb},
the Bardeen potential $\Phi$ commonly grows very large or even
diverges as the bounce is approached, which is easy to see. For
instance, using the same scaling
solution as in the previous section, valid for the ekpyrotic scenario,
(\ref{scaling_solution}), the equation of motion for $\Phi$ becomes
\cite{Khoury:2001zk}
\begin{equation}
\ddot\Phi+\frac{2+p}{t}\dot\Phi+\frac{k^2}{{(t/t_0)}^{2p}}\Phi=0.
\end{equation}
The properly normalized solution is 
\begin{equation}
  \Phi(\tau)=\frac{p \sqrt\pi}{2 a(1-p)\sqrt{2k}} 
  \left[\frac{J_{\frac{1+p}{2-2p}}(-k\tau)}{\sqrt{-k\tau}}
   +i\frac{Y_{\frac{1+p}{2-2p}}(-k\tau)}{\sqrt{-k\tau}}\right],
\end{equation}
where we used similar steps as in the previous section; hence, at late times
\begin{equation}
\Phi\sim k^{(-3/2-p)}t^{1-p}+k^{(-1/2+p)}t^p_0\,.
\end{equation}
Evidently, the first term diverges as the bounce is approached,
indicating that the Newtonian gauge should not be used close to the
bounce. As we saw in Sec.~\ref{viabilityofptb}, one should use the
harmonic gauge to unambiguously follow perturbations through the
bounce. Nevertheless, familiar gauges, such as the Newtonian one, can
be used in the pre-bounce phase.

\subsubsection{Adiabatic fluctuations in a matter bounce}
\label{adiabatic_fluctuations_mb} 

Instead of relying on an entropic mechanism, one may also generate a
scale-invariant spectrum directly in the adiabatic mode if the
background evolution conforms to a matter dominated phase
\cite{Finelli:2001sr}.  Cosmological models in which the contraction
phase is dominated by a dust-like fluid leading to such a spectrum
were studied in
\cite{Gasperini:1993hu,Gasperini:1994xg,Wands:1998yp,Brandenberger:2001bs,Martin:2001ue,Lyth:2001nq,Lyth:2001nv,Khoury:2001zk,Durrer:2002jn,Hwang:2001zt,Cartier:2003jz}.

To understand the origin of the scale-invariant spectrum, consider
Eq.~(\ref{eomuI}) for a single degree of freedom.  If the scale factor
evolves according to
\begin{equation}
a\propto (-t)^p,
\end{equation}
one arrives at
\begin{equation}
\nu=\frac{1-3p}{2(1-p)}.
\end{equation}
Since the scalar spectral index in (\ref{scalarspectral}) reads
\begin{equation}
n_\mathrm{s}-1=3-2|\nu|,
\end{equation}
we need $\nu=3/2$ in order to get a scale-invariant spectrum. This
corresponds to $p=2/3$, that is, a matter dominated contracting
phase. Such a phase need not be dominated by actual dust, but could be
mimicked by an oscillating scalar field in a quadratic potential. The
latter can also accommodate a slightly red spectrum by altering the
potential \cite{Wands:1998yp}. This matter bounce was first proposed
in \cite{Wands:1998yp,Finelli:2001sr,Brandenberger:2009jq}. Thus,
there are two ways of getting a scale-invariant spectrum in the
contracting phase: the matter bounce and the entropic mechanism
discussed in Sec.~\ref{absenceNG}.

Models that use the matter bounce to obtain a scale-invariant spectrum
include the Ho\v{r}ava-Lifshitz bounce, the Quintom bounce, the
ghost-condensate bounce, and the stringy S-Brane bounce of
Sec.~{\ref{stringy}}. See Sec.~\ref{overviewofbounces} for an overview
and \cite{Brandenberger:2012zb} for a review.

\subsubsection{Matching conditions}
\label{sbounceptbs}

In this section we sketch the Deruelle-Mukhanov matching conditions
for cosmological perturbations in an FLRW universe at a sharp
transition \cite{Israel:1966rt,Hwang:1991an,Deruelle:1995kd}. At the
hypersurface of the transition, the induced metric is continuous and
the extrinsic curvature jumps according to the surface tension.  In an
ever-expanding or contracting FLRW setting and a simple jump in the
equation of state parameter, they entail the continuity of the Hubble
parameter and the scale factor at the background level. Furthermore,
these conditions provide matching conditions for perturbations of the
metric, as well as the perturbed energy momentum tensor. For details
we refer the interested reader to the textbooks
\cite{Mukhanov:2005sc,PeterUzan2009}.

These matching conditions are often crucial in nonsingular as well as
singular bouncing models. For example, matching perturbations across
the singular bounce in the original ekpyrotic scenario
\cite{Khoury:2001wf} was performed in \cite{Finelli:2001sr} in the
effective 4D description.  However, since the singularity is not
resolved, ambiguities remain, which led to an extensive discussion of
the proper matching conditions for a singular set-up. For example,
\cite{Finelli:2001sr} argued, based on
\cite{Israel:1966rt,Deruelle:1995kd}, that only the decaying mode in
the contracting phase couples to the dominant one in the expanding
phase. In general, $k-$mode mixing is present in bouncing scenarios so
that both modes, the subdominant and dominant ones, of the contracting
universe determine the dominant mode (the relevant one as far as
present-day observations are concerned) in the expanding phase
\cite{Khoury:2001zk}\footnote{Following \cite{Martin:2003sf}, we call
  $k-$mode mixing the mixture of both dominant and sub-dominant
  (growing and decaying in the inflation literature) modes at fixed
  scale $k$ in order to avoid confusion with higher order mixing terms
  involving different scales.}.

In nonsingular set-ups, matching conditions are generally unnecessary
because the evolution of perturbations can be followed, at least
numerically, throughout the nonsingular bounce. However, to gain an
analytic understanding of the resulting power spectrum and its
sensitivity to model parameters, matching conditions are often invoked
to stitch together distinct analytic approximations covering the
contracting phase, the bounce and the expanding phase. The application
of the matching conditions in \cite{Deruelle:1995kd} requires the
specification of a scalar quantity that determines the position of the
transitional hypersurface.  Once this quantity is specified, no
ambiguities remain.  Applications include Refs.
\cite{Brandenberger:2001bs,Bozza:2003pr,Battefeld:2004mn,Battefeld:2005cj,Geshnizjani:2005hc}.

Recently, in \cite{Brandenberger:2013zea}, the perturbations across
the S-brane bounce have been matched via these junction conditions
\cite{Israel:1966rt,Deruelle:1995kd}. The scalar quantities
determining the location of the S-brane is the temperature, \ie the
value of the thermal scalar potential $\phi$ as established in
Sec.~\ref{stringy}, Eq.~(\ref{stringaction}). The junction conditions
entail
\begin{enumerate}
\item continuity of the scale factor of the 3-dimensional spatial
  metric in the string frame,
\item continuity of its time derivative,
\item continuity of the dilaton,
\item discontinuous jump of the dilaton's time derivative, set by the
  brane tension.
\end{enumerate}
Applications of these conditions lead, after a straightforward but
tedious computation, to the power spectrum of the curvature
perturbation after the bounce in terms of pre-bounce quantities. It
should be noted that the matching is performed during the contracting
and expanding phase respectively, not across the actual
bounce. Provided the conditions under which it applies are valid, a
point that needs be checked for each individual model independently,
this matching is therefore no different than matching perturbations
over the transition from a radiation dominated expanding universe to a
matter dominated one and does not entail the ambiguities which
hampered the original ekpyrotic scenario, see Sec.~\ref{modemixing}.

We discussed in Sec.~\ref{viabilityofptb} that certain gauges can
become ill-defined in a nonsingular bounce. As a consequence, the
computation in \cite{Brandenberger:2013zea} was performed in several
gauges and no gauge dependence was found.

\subsection{The primordial tensor spectrum in an ekpyrotic universe
  and a matter bounce}
\label{rcyclic}

\begin{figure}[tb]
\begin{center}
\includegraphics[scale=0.65]{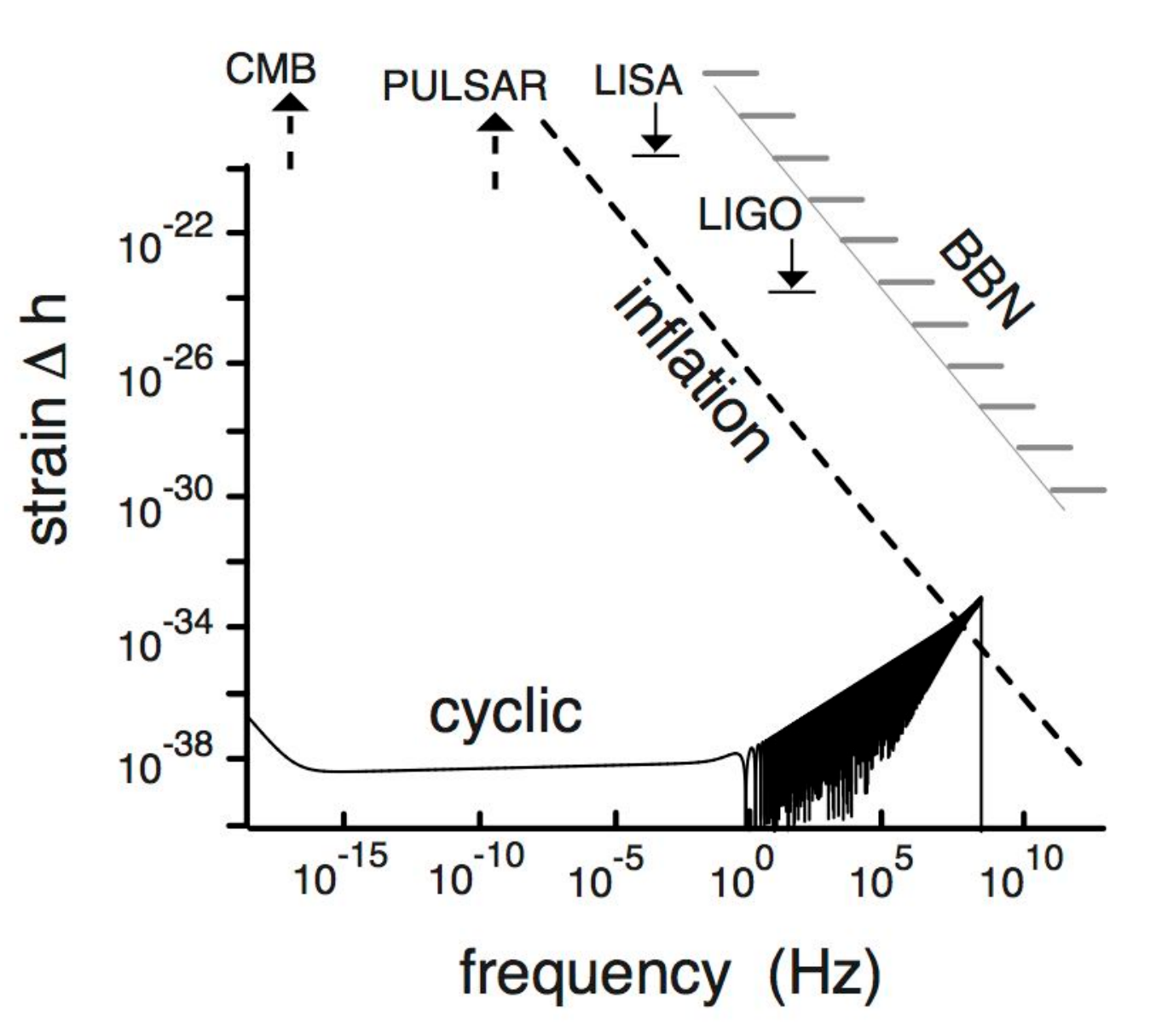}
\caption{The ekpyrotic/cyclic model's prediction of the present day
  dimensionless strain, $\Delta h(k,\tau_0)$, generated during the
  ekpyrotic phase, taken from \cite{Boyle:2003km}.  The obtained
  gravity wave density needs to be four orders of magnitude below the
  BBN bound, which leads to unobservably small tensor modes on CMBR
  scales. Hence, an observation of a primordial tensor to scalar ratio
  of $r\approx 0.2$ is incompatible with this generation
  mechanism. LISA and LIGO bounds were estimates at the time
  \cite{Boyle:2003km} was published.}
\label{real3}
\end{center}
\end{figure}

Tensor perturbations $h_{ij}$ are the gravitational degrees of
freedom, commonly called gravitational waves. They evolve
independently of linear scalar and vector perturbations.  The
corresponding perturbed FLRW space time is described by the line
element, see \cite{Bassett:2005xm} for a review,
\begin{equation}
  \dd s^2=-\dd t^2+a^2(\delta_{ij}+h_{ij})\dd x^i \dd x^j
  \label{tensor_polarization}\,.
\end{equation}
Tensor perturbations, $h_{ij}$ are tranverse $\partial^ih_{ij}=0$,
trace-free $\delta^{ij}h_{ij}=0$, and gauge invariant.  Arbitrary
tensor perturbations can be decomposed into eigenmodes of the spatial
Laplacian $\nabla^2e_{ij}=-k^2e_{ij}$, with comoving wavenumber $k$
and scalar amplitude $h(t)$,
\begin{equation}
h_{ij}=h(t)e_{ij}^{(+,\times)}(x),
\end{equation}
where $+$ and $\times$ denote the two possible polarization
states. The equation of motion for the amplitude is
\begin{equation}
\ddot h+3H\dot h+\frac{k^2}{a^2}h=0\label{tensorwe},
\end{equation}
and the tensor power spectrum is defined as
\begin{equation}
  \mathcal P_{_\mathrm{T}}\equiv |\Delta h(k,\tau)|^2\equiv2\frac{4\pi
    k^3}{(2\pi)^3}|h|^2,
\end{equation}
where the extra factor of $2$ arises from the addition of the two
independent polarizations of the graviton
(\ref{tensor_polarization}). The quantity $\Delta h(k,\tau)$ is
sometimes called the dimensionless strain. Eq.~(\ref{tensorwe}) can be
written in terms of
\begin{equation}
u=\frac{ah}{2}\label{h},
\end{equation}
leading to a mode evolution similar to that of the Mukhanov-Sasaki
variable (\ref{eomvk}), \ie
\begin{equation}
u''+\left(k^2-\frac{a''}{a}\right)u=0\label{gralsoln},
\end{equation}
where the prime denotes a derivative with respect to conformal time
$\eta$.

Following the computation in \cite{Boyle:2003km}\footnote{We fixed a
  typo in the background solution of \cite{Boyle:2003km}. } during the
ekpyrotic phase, and using the solution (\ref{scaling_solution}),
leading to (\ref{scale_factor_epsilon}) and (\ref{bar_epsilon}), we
can write the general solution of (\ref{gralsoln}) as
\begin{equation}
u(\tau)=\frac{\sqrt{y}}{2}\left[A_1(k)
H_{\nu_{_\mathrm{T}}}^{(1)}\left(y\right)+
A_2(k)H_{\nu_{_\mathrm{T}}}^{(2)}\left(y\right)\right],
\end{equation}
where $A_{1,2}(k)$ are constants, $H_{\nu_{_\mathrm{T}}}^{(1,2)}$ are
the Hankel functions of order $\nu_{_\mathrm{T}}\approx 1/2-p$, with
$p=2/c^2$, $y\equiv -k(\eta-\eta_\mathrm{ek})$ and we ignore terms of
order $1/p$; here, $\eta_\mathrm{ek}$ is the conformal time at which
the potential would have reached $-\infty$ if it would not have bent
up again, see Fig.~\ref{ekpyrotic}. The constants are determined by
matching this solution to the Minkowski vacuum, leading to
\begin{equation}
A_1(k)=\frac{1}{2}\sqrt{\frac{\pi}{k}} \ \ \ \ \hbox{and} \ \ \ \ \ A_2(k)=0.
\end{equation}
The resulting tensor power spectrum is deeply blue, opposite to that
obtained during a phase of inflation: $\mathcal P_{_\mathrm{T}}\propto
|k_{\eta}|^{3-2\nu_{_\mathrm{T}}}$ so that $n_{_\mathrm{T}}$, defined
as $n_{_\mathrm{T}}\equiv d\ln\mathcal P_{_\mathrm{T}}/d\ln k$,
becomes
\begin{equation}
n_{_\mathrm{T}}=3-2\nu_{_\mathrm{T}}=2+2p,
\end{equation}
which should be compared to the inflationary slow roll result
$n_{_\mathrm{T}}^{\mathrm{SR}}=-2\epsilon^\mathrm{SR}$.

After the ekpyrotic phase, a kinetic-driven contracting phase, the
bounce and a kinetic-driven expanding phase follow, before the
universe becomes radiation dominated. Solving the equations of motion
(\ref{gralsoln}) in these regimes, using the Deruelle-Mukhanov
matching conditions (Sec.~\ref{sbounceptbs}), as well as using the
radiation transfer functions, enables the computation of the ekpyrotic
gravitational wave spectrum today, leading to the plot in
Fig.~\ref{real3} taken from \cite{Boyle:2003km}. Contrary to the
inflationary strain, which falls off, the ekpyrotic strain becomes
more important for increasing $k$ (smaller wavelenghts) due to the
deeply blue spectrum.
 
Fig.~\ref{real3} shows that the big-bang nucleosynthesis bound is the
strongest one: requiring that the energy density in gravitational
waves leaves the successful predictions of BBN unaffected requires
that the gravitational wave contribution on CMBR scales is
unobservably small, many orders of magnitude below the inflationary
slow roll prediction.  Other bounds from direct detection experiments,
such as LIGO, are much weaker, albeit still sufficient to suppress the
gravitational wave spectrum on CMBR scales.

 Above we considered the generation of gravitational waves induced
by quantum fluctuations, which turned out to be unobservably small. In
this case, the dominant source for gravitational waves are second
order effects, whereby $h_{ij}$ is sourced by scalar fluctuations
\cite{Baumann:2007zm}. 

The B{\footnotesize ICEP2} preliminary detection of gravitational
waves \cite{Ade:2014xna,Ade:2014gua} of\footnote{This value still
  contains foregrounds, which have been argued to be subdominant
  \cite{Flauger:2014qra,Mortonson:2014bja}, but might still lower the
  primordial value by a factor of order one or more.}
\begin{equation}
r=\frac{\mathcal P_{_\mathrm{T}}}{\mathcal P_{\zeta}}=
0.2^{+0.07}_{-0.05},\label{BICEP2}
\end{equation}
along with the COBE normalization for the power spectrum
\cite{Bassett:2005xm}, would, if confirmed, lead to
\begin{equation}
\Delta h\approx 2\times 10^{-5},
\end{equation}
which is far above the ekpyrotic/cyclic contribution, see
Fig.~\ref{real3}; thus, the cyclic model and any other one using an
ekpyrotic phase would be ruled out if no other source for primordial
gravitational waves at CMBR scales were present. If future
experiments refute the B{\footnotesize ICEP2} claims and reveal no
nearly scale-invariant, first order tensor spectrum, but a measurable
scalar induced, second order tensor spectrum instead, inflation would
be in trouble and alternative models, such as the ekpyrotic/cyclic
ones would be favored \cite{Baumann:2007zm}. For a comparison of the
spectrum of scalar induced tensors with gravitational waves generated
during inflation and current and future experiments see Fig. 4 of
\cite{Baumann:2007zm}, which is not only based on analytic estimates,
but also a full numerical analysis. 

Since gravitational waves depend only on the background scale factor,
as shown in Eq.~(\ref{gralsoln}), it is hard to imagine a mechanism in
a contracting universe with an ekpyrotic phase that could lead to
such a large level of gravitational waves on CMBR scales, yet
unobservable ones on BBN scales. It has been speculated
\cite{Ijjas:2014fja} that additional gravitational waves may be
generated during phase transitions, from topological defects or during
the bounce, within the confines of cyclic/ekpyrotic cosmology, but no
concrete mechanisms have been investigated.

If fluctuations crossed the potential $a''/a$ during a matter
dominated phase, as in the matter bounce scenario of
Sec.~\ref{adiabatic_fluctuations_mb}, the spectrum of gravitational
waves is scale invariant with a considerably higher tensor-to-scalar
ratio of about $r\sim\mathcal{O} (30)$ \cite{Allen:2004vz,Cai:2008qw}.
Thus, instead of being unobservable, such matter bounce scenarios
produce gravitational waves in excess of current bounds.  Mechanisms
to lower this ratio may be possible \cite{Cai:2008qw}, but they have
not been investigated thoroughly yet.

On the other hand, a potential observation of gravitational waves at
B{\footnotesize ICEP2} levels is consistent with the simplest large
field models of single field slow roll inflation,
$r=16\epsilon^\mathrm{SR}$. See \cite{Starobinskii:1979} for the first
computation of tensor modes during inflation. 

As we will see in Sec.~\ref{sbounceptbs}, the ekpyrotic phase appears
to be necessary to prevent BKL instabilities, see Sec.~\ref{BKL}, and
enable a smooth bounce. Thus, an observation of gravitational waves is
a severe challenge for any bounce.  
\subsection{B{\footnotesize ICEP2}}\label{bicep2}
The B{\footnotesize ICEP2} collaboration
\cite{Ade:2014xna,Ade:2014gua} has created the deepest polarization
maps ever made and have detected a large-angle primordial B-mode
polarization in the CMBR. After the subtraction of the foreground, the
data shows a value of\footnote{ This is in tension with
  P{\scriptsize LANCK}'s upper bound on $r < 0.11$ at $95\%$
  confidence \cite{Ade:2013kta,Ade:2013zuv}. Though, arguments that
  B{\scriptsize ICEP2} and P{\scriptsize LANCK} are not in
  conflict have been put forward in \cite{Audren:2014cea}. }

\begin{equation}
r=\frac{\mathcal P_{_\mathrm{T}}}{\mathcal P_{\zeta}}=
0.2^{+0.07}_{-0.05},\label{BICEP2}
\end{equation}
at a $68\%$ CL. The value of $r$ in (\ref{BICEP2}) is the tensor
fraction evaluated on scales $x_\mathrm{lss}/100\lesssim k^{-1}\lesssim
x_\mathrm{lss}$ and $x_\mathrm{lss}=14$ Gpc is the distance to
the last-scattering
surface \cite{Ade:2014xna,Ade:2014gua}. The uncertainty is only
statistical and it is possible that the emission from unaccounted
foregrounds such as dust accounts for all or some of the observed
polarization.  The B{\footnotesize ICEP2} collaboration claims to have
used the least noise contaminated region in the sky; the analysis
incorporated polarized foreground mapping with an assumed maximum of
$20\%$ foreground contamination.  The implications of this result, if
confirmed, is set to have profound consequences to our understanding
of the universe. If these fluctuations are proven to have been sourced
by cosmological tensor modes, a large number of cosmological models
would be ruled out. Due to the importance of such a result, there has
been a heated debate of whether the signal is due to foreground dust
or not.  Using preliminary maps from P{\footnotesize LANCK}, Mortonson
\etal \cite{Mortonson:2014bja} and Flauger \etal
\cite{Flauger:2014qra} give an estimate of the dust polarization
contamination, which, extrapolated to the B{\footnotesize ICEP2}
patch, could account for the B-mode signal detected by B{\footnotesize
  ICEP2}. Due to the uncertainties in the amplitude of the dust
polarization at the frequencies utilized by B{\footnotesize ICEP2} it
is not possible to conclude whether the signal is due to polarized
dust or gravitational waves. In support of this claim, the
P{\footnotesize LANCK} collaboration \cite{Adam:2014bub} team
published results stating that the polarized thermal emission from
Galactic dust is the main foreground present in measurements of the
polarization of the CMBR at frequencies above $100\,$GHz. Colley and Gott
\cite{Colley:2014nna} analyzed the genus topology of the BICEP2
B-modes in the B{\footnotesize ICEP2} region from the publicly
available Q and U at $353\,$GHz preliminary P{\footnotesize LANCK} polarization maps and
concluded that they have a primordial origin. This debate is ongoing,
but should be resolved once the additional data from the future Keck
Array observations at $100\,$GHz and P{\footnotesize LANCK} observations
at higher frequencies become available.

\subsection{Non-Gaussianities}
\label{absenceNG}

\begin{figure}[tb]
\begin{center}
\includegraphics[scale=0.6]{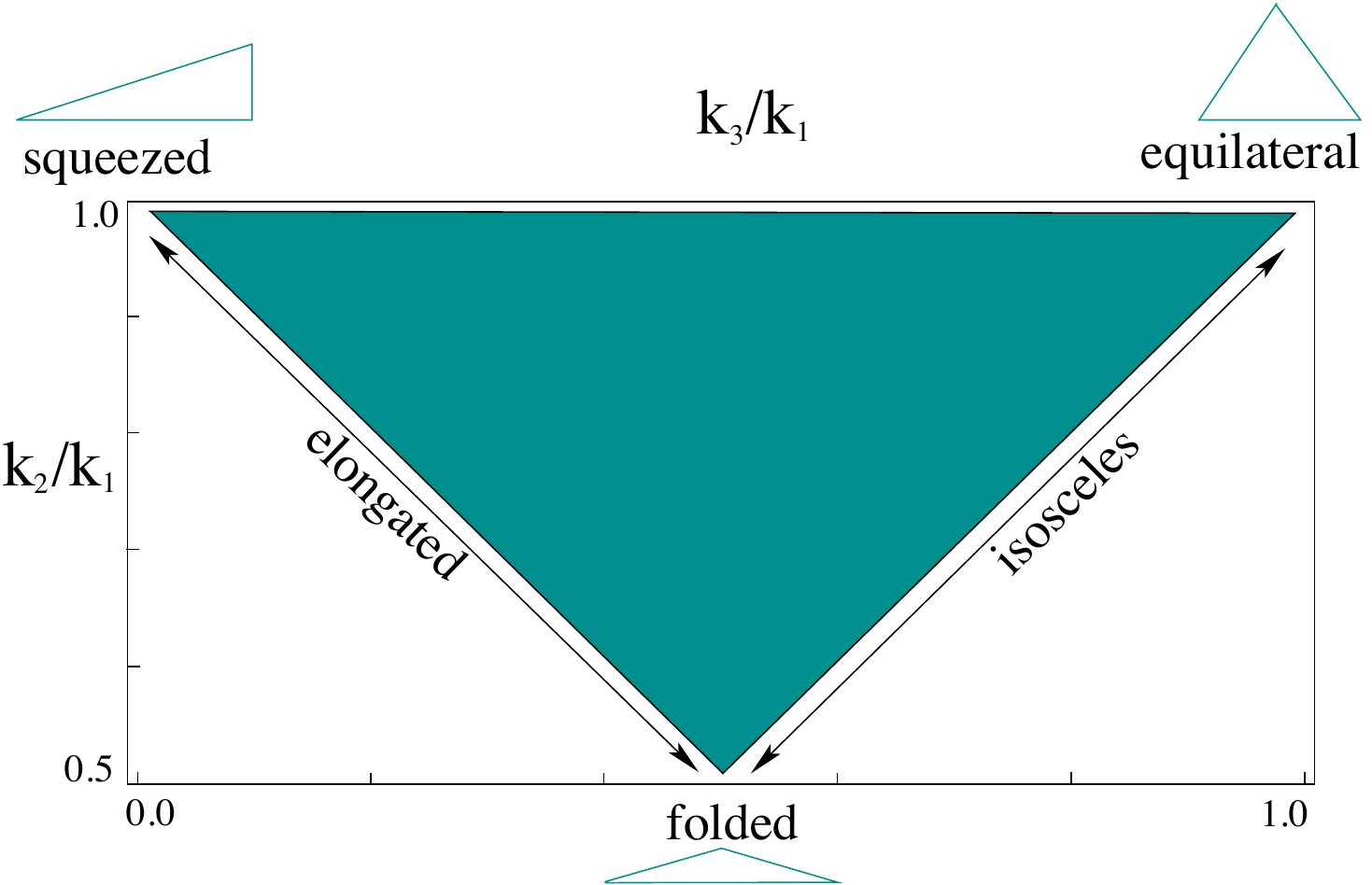} 
\caption{ Common shape configurations of the bispectrum in (\ref{bispec})
\cite{Baumann:2009ds}.  
\label{NGshape}}
\end{center}
\end{figure}

\begin{figure*}[tb]
\begin{center}
\includegraphics[scale=0.4]{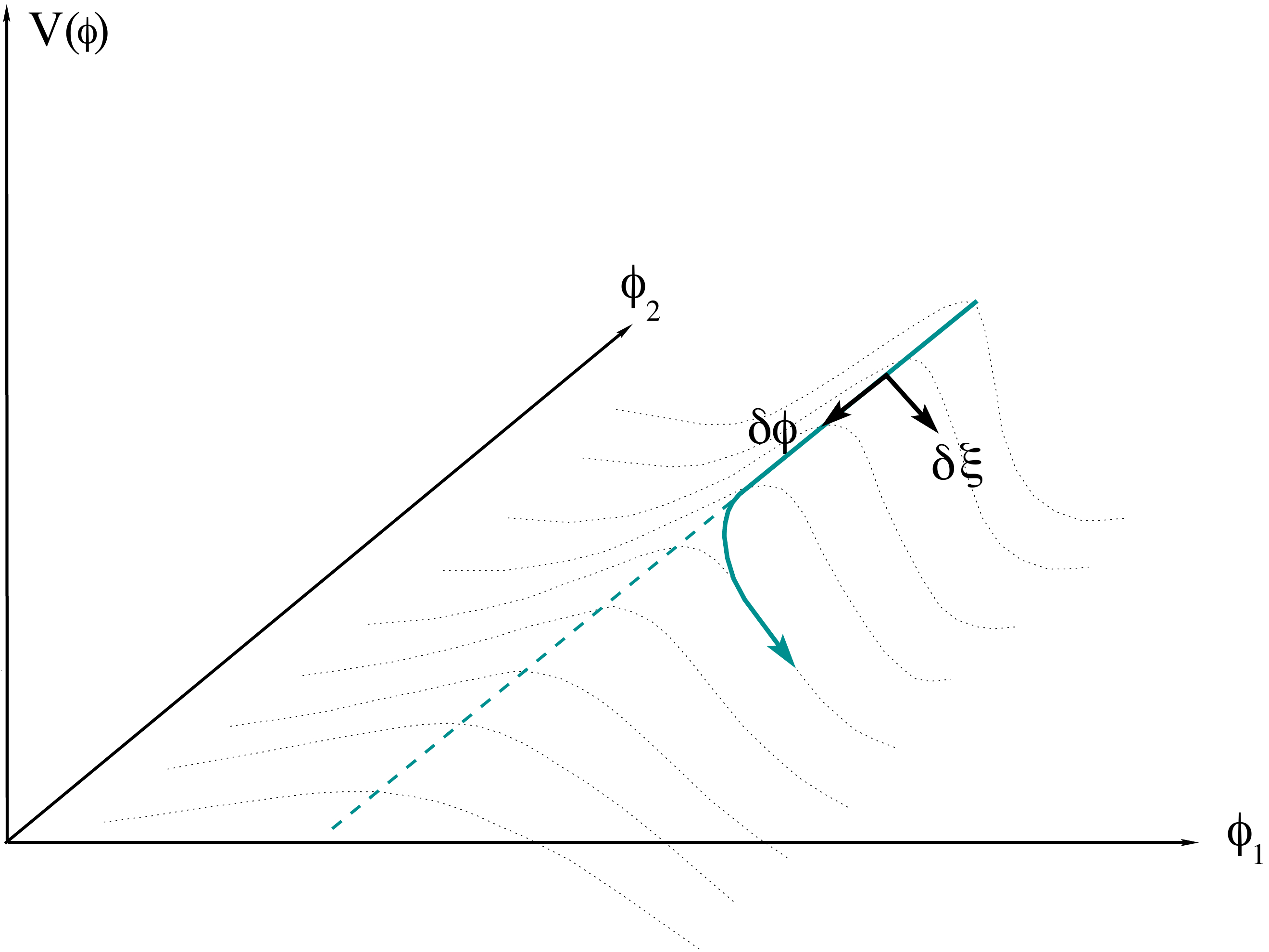}
\label{fig3} 
\caption{A turn in field space may arise due to the ekpyrotic
  potential: as the fields need to follow the ridge of the potential
  in order to yield a scale-invariant spectrum, see
  Sec.~\ref{reflection}, a bend in the trajectory can occur naturally
  by falling off such a ridge towards the end of the ekpyrotic
  phase. However, the generated non-Gaussianities in (\ref{koyama})
  are in excess of current limits
  \cite{Ade:2013uln,Ade:2013nlj}. Further, a cyclic model is
  impossible if this conversion mechanism is employed.}
\label{traj2}
\end{center}
\end{figure*}
  
Higher order correlation functions, particularly the 3-point function,
provide a measure of non-Gaussianities, see
\cite{vernizzi:2006ve,Lyth:2005fi,Battefeld:2006sz,Battefeld:2009ym}
for reviews. Since the 3-point function vanishes identically for a
Gaussian spectrum, it is ideally suited as a measure for
non-Gaussianities. The corresponding bispectrum ${\mathcal
  B_R}(k_1,k_2,k_3)$ is defined via
\begin{equation}
\langle{\mathcal R}(\bm{k}_1){\mathcal R}(\bm{k}_2){\mathcal
  R}(\bm{k}_3)\rangle=(2\pi)^3\delta\left(\bm{k}_1+\bm{k}_2+
\bm{k}_3\right){\mathcal
B_R}.\label{bispectrum}
\end{equation} 
Since the full bispectrum is not currently accessible by measurements,
bounds are commonly imposed onto certain triangle configurations of
the wave-numbers.  The amplitude of particular configurations is given
by non-linearity parameters $f_{_\mathrm{NL}}$, which can be defined
as
\begin{equation}
\langle{\mathcal R}(\bm{k}_1){\mathcal R}(\bm{k}_2){\mathcal
R}(\bm{k}_3)\rangle=
\frac{3(2\pi)^7}{10}\delta^3\left(\bm{k}\right)
f_{_\mathrm{NL}}\mathcal P_{\mathcal R}^2(k)\frac{\sum_i k_i^3}{\Pi_jk_j^3}
\label{bispec}.
\end{equation}
Common configurations include the local shape ($k_3\ll k_1,k_2$) and
the equilateral shape ($k_1=k_2=k_3$) among others
\cite{Babich:2004gb} such as the folded or orthogonal shapes
\cite{Baumann:2009ds,Battefeld:2011ut}, see Fig.~\ref{NGshape}.
Current bounds set by P{\footnotesize LANCK} are \cite{Ade:2013ydc},
\begin{eqnarray}
f_{_\mathrm{NL}}^\mathrm{local}&=&2.7\pm5.8\label{currentbounds},\\
f_{_\mathrm{NL}}^\mathrm{equil}&=&-42\pm75\nonumber\nonumber,\\
f_{_\mathrm{NL}}^\mathrm{ortho}&=&-25\pm39\nonumber\,,
\end{eqnarray}
($68\%$ CL) which are consistent with a Gaussian (all non-linearity
parameters vanish) spectrum. Such suppressed non-Gaussianities are
consistent with the prediction in canonical, single field, slow-roll
models of inflation \cite{Maldacena:2002vr,Creminelli:2003iq}. For
instance, the squeezed limit of the three-point function,
\begin{equation}
\langle{\mathcal R}(\bm{k}_1){\mathcal R}(\bm{k}_2){\mathcal R}(\bm{k}_3)\rangle
=(2\pi)^3\delta(\bm{k})
(1-n_\mathrm{s})P(k_1)P(k_3),
\end{equation}
is proportional to $n_\mathrm{s}-1$ and is therefore heavily
suppressed \cite{Creminelli:2004yq}.  Larger non-Gaussianities are
expected in bouncing models, since models are of the fast-roll type,
entropy perturbations are present, and the non-trivial bounce physics
may affect the computation.

Within the $\delta N$ formalism, the local non-linearity parameter can
be computed as
\begin{equation}
  \frac{6}{5}f_{_\mathrm{NL}}^\mathrm{local}=\frac{N_IN_JN^{IJ}}{(N_KN^K)^2},
  \label{fnllocal}
\end{equation}
which often offers a simple means of estimating non-Gaussianities. For
higher order correlation functions see \cite{Baumann:2009ds}.

We have seen in Sec.~\ref{isocurvature} how the ekpyrotic phase can
generate a nearly scale-invariant spectrum of perturbations in an
isocurvature field which may be transferred to the curvature
perturbation by curvaton-like mechanisms
\cite{Enqvist:2001zp,Lyth:2001nq,Enqvist:2005pg,Sasaki:2006kq,Enqvist:2008gk,Kawasaki:2011pd}.
Here we would like to explore whether non-Gaussianities are generated
and how they compare to current bounds, such as those of
Eq.~(\ref{currentbounds}).

Typically, the conversion mechanism in ekpyrotic/cyclic models
generate non-Gaussianity of the local type \cite{Babich:2004gb} that
is considerably larger than in inflationary models, but
non-Gaussianities below current bounds are possible
\cite{Fertig:2013kwa,Ijjas:2014fja}, see also Sec.~\ref{NGcurrent}.
The latter models, dubbed non-minimal entropic, are based on replacing
the tachyonic isocurvature field during the ekpyrotic phase, which we
focus on in the following, by a non-minimally coupled entropic field
as first proposed in \cite{Qiu:2013eoa,Li:2013hga}. These models still
encompass non-Gaussianity from the conversion mechanism.

Entropy perturbations are generated during the ekpyrotic phase when
the potentials are of the form shown in Fig.~\ref{ekpyrotic}; in a
two-field model, the single entropy perturbation $\delta \xi$
exemplifies the perturbation orthogonal to the trajectory in field
space, see Fig.~\ref{traj2} and Fig.~\ref{traj1}. In order to transfer
a scale-invariant spectrum of $\delta \xi$, see (\ref{nxi}), to the
curvature perturbation, several concrete mechanisms have been
proposed:
\begin{enumerate}
\item a bending of the trajectory caused by falling off the ridge in
  the potential in (\ref{potential}) \cite{Koyama:2007if},
\item a bending caused by the reflection on a sharp boundary in field
  space \cite{Lehners:2008my},
\item a conversion after the bounce caused by modulated instant
  preheating \cite{Battefeld:2007st}.
\end{enumerate}

In the following paragraphs we summarize the contribution due to these
mechanisms. It should be noted that predictions for non-linearity
parameters can change by factors of order one during (p)reheating if
an adiabatic regime has not been reached previously
\cite{Leung:2012ve,Leung:2013rza}; that non-gaussianities are subject
to change and often transient as long as the adiabatic regime is not
reached was pointed out in \cite{Byrnes:2009qy,Battefeld:2009ym} see
also \cite{Elliston:2011dr} for subsequent work.
 
Further, the actual bounce may also contribute to non-Gaussianities,
as pointed out in \cite{Gao:2014hea} by one of the authors of this
review and collaborators; this case study is based on a
curvature-dominated bounce. However, it can be argued that results are
more general \cite{Gao:2014heb} since the ingredients leading to a
bounce summarized in this review require fields or fluids that violate
the NEC. Generically, this intrinsic contribution appears to be far in
excess of current bounds, hence possibly providing stringent
constraints on bouncing scenarios.

\subsubsection{A bending of the trajectory caused by falling off the
  ridge in the potential}
\label{bending}

In the ekpyrotic scenario, for fast roll to occur, the exponent $c$ in
(\ref{potential}) has to be large during the ekpyrotic phase. A
conversion during the ekpyrotic phase, by means of the background
trajectory falling off the ridge of the potential, as illustrated in
Fig.~\ref{traj2}, leads to a local non-linearity parameter of order
\cite{Koyama:2007if},
\begin{equation}
f_{_\mathrm{NL}}^\mathrm{local}=-\frac{5}{15}c_i^2<0 \ \ \ \Longrightarrow
\ \ \ |f_{_\mathrm{NL}}^\mathrm{local}|\gg 1,\label{koyama}
\end{equation}
which can be computed by evaluating (\ref{fnllocal}), as well as the
particular background solution providing the volume expansion rate
$N(\xi)$; the index $i$ denotes the field that freezes in the
late-time single field solution. Such a large and negative
contribution imposes severe constraints on the relevant scenario.
 
\subsubsection{A bending caused by the reflection on a sharp boundary
  in field space}
\label{reflection}
 
A bending of the trajectory in field space in the four dimensional
effective theory that induces a conversion of entropy to curvature
perturbations occurs naturally in the cyclic model
\cite{Lehners:2007ac}. The change in the trajectory is due to a
negative-tension brane bouncing off a spacetime singularity, before
collision with a positive-tension brane \cite{Lehners:2006pu}. After
an ekpyrotic phase, an estimate of the non-Gaussianity gives
\cite{Lehners:2008my}
\begin{equation}
|f_{_\mathrm{NL}}|\sim \mathcal O(c_i)\gg 1,
\end{equation}
where $i=1,2$, and $c_i$ are the exponents in (\ref{potential}); see
Fig.~\ref{traj1} for an illustration. An updated discussion in
\cite{Lehners:2013cka} showed that values for $f_{_\mathrm{NL}}$ can be smaller
and of order $f_{_\mathrm{NL}}\sim V'''/V''$, where a $'$ denotes a derivative
in the entropy direction. Often, models such as this one are in
tension with the P{\footnotesize LANCK} data. 

A computation of the ekpyrotic trispectrum \cite{Lehners:2009ja}
generated during a conversion in the ekpyrotic phase, as in the
previous section, or a subsequent conversion via the above mechanism,
leads to a distinct large contribution, which could be used to tell
ekpyrotic models apart from inflationary ones, if it were observed.

\subsubsection{A conversion after the bounce caused by modulated
  instant preheating}
\label{modulatedp}

A conversion after a bounce via modulated preheating is also possible;
for example, perturbations can be imprinted during modulated instant
preheating \cite{Felder:1999pv,Battefeld:2007st}. The local
non-linearity parameter is set by the dependence of the coupling
constant between the bosonic preheat matter field, $\xi_\mathrm{pr}$,
and a fermionic degree of freedom on the isocurvature field,
$\mathcal{L}_\mathrm{int}=-h(\xi)\xi_\mathrm{pr}\psi\bar{\psi}$
resulting in
\begin{equation}
f_{_\mathrm{NL}}^{}=\frac{5}{9}\left(1-\frac{\gamma_2}{\gamma^2_1}\right),
\end{equation}
where
\begin{equation}
  \gamma_n\equiv\left.\frac{1}{h}\frac{\partial^nh}{\partial\xi^n}
  \right|_{t_\mathrm{kHc1}}.
\end{equation}
As a result, the non-linearity parameter is of $\mathcal{O}(1)$
without fine-tuning and certain $h(\xi)$ are already constrained by
P{\footnotesize LANCK} \cite{Ade:2013uln,Ade:2013nlj}.
\begin{figure*}[tb]
\begin{center}
\includegraphics[scale=0.35]{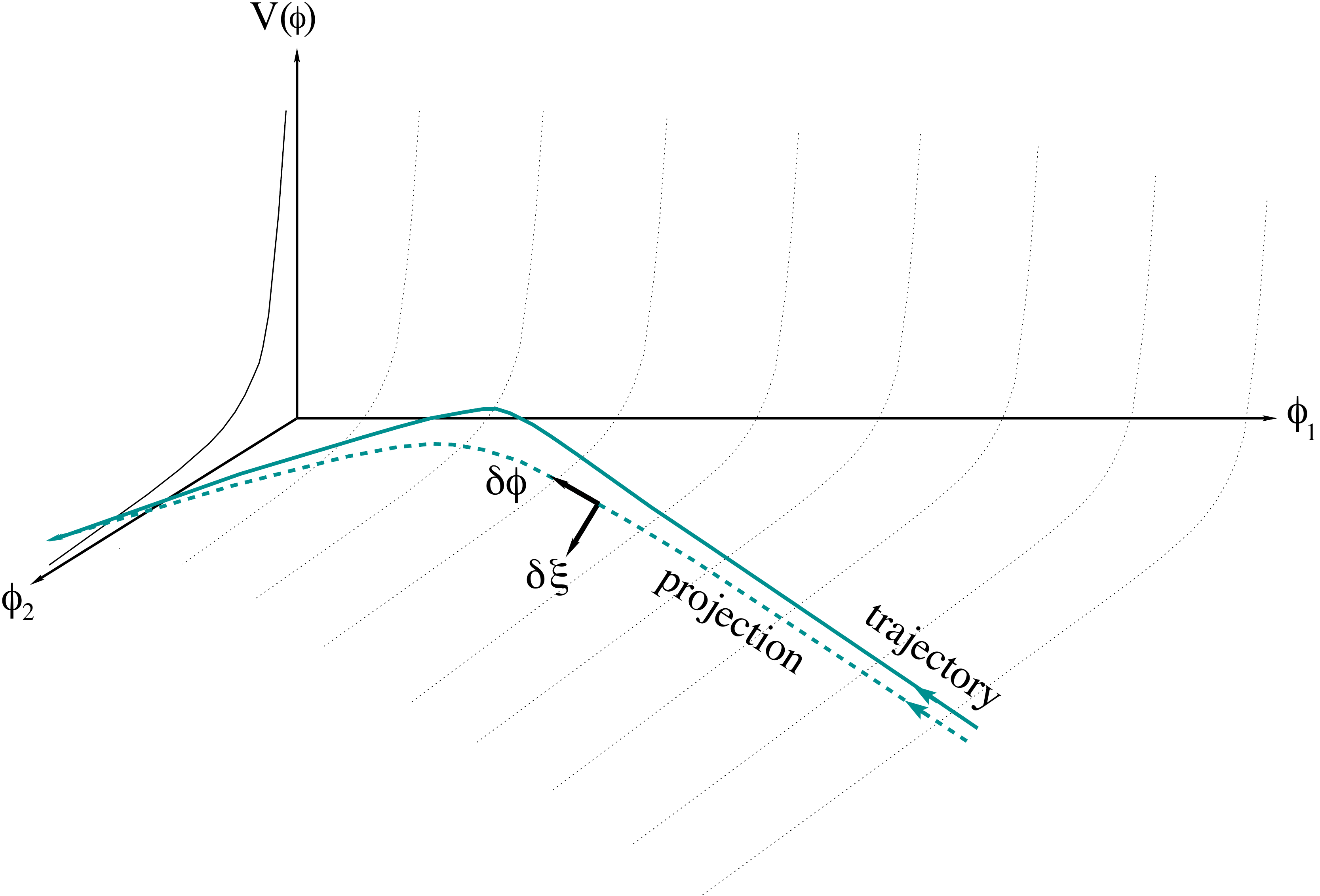}
\label{fig2} 
\caption{ Schematic of the trajectory in field space during the
  kinetic phase.  The entropy perturbation orthogonal to the
  trajectory's projection, dashed line, is denoted as $\delta\xi$; The
  sharp turn caused by a steep wall in field space
  \cite{Lehners:2008my} converts the entropy mode into perturbations
  tangential to the trajectory, namely into the adiabatic mode
  $\delta\phi$.  Non-Gaussianities are naturally of $\mathcal{O}(10)$
  \cite{Lehners:2008my} and therefore in tension with P{\footnotesize
    LANCK}'s results \cite{Ade:2013uln,Ade:2013nlj}. A cyclic model,
  as in the Ph\oe{}nix universe, is possible.}
\label{traj1}
\end{center}
\end{figure*} 

\subsubsection{Non-Gaussianities in other proposals}
\label{NGcurrent}

A model which does not generate large intrinsic non-Gaussianities is
that based on a {\it non-minimal entropic mechanism}
\cite{Fertig:2013kwa,Ijjas:2014fja}, following the model introduced in
\cite{Qiu:2013eoa,Li:2013hga}. It employs a second scalar, which does
not contribute to the potential, but is non-minimally coupled to the
ekpyrotic scalar $\phi$ in its kinetic function. In contrast to
standard entropic mechanisms, there are no non-Gaussianities generated
during the ekpyrotic phase at  leading  order. The authors
evaluate the contribution to curvature perturbations sourced by
entropy perturbations at second order within different phases: first,
the contribution to the curvature perturbation from entropy
perturbations during the ekpyrotic phase -- these are non-existent;
second, the introduction of a time-varying equation of state, as
required to end the ekpyrotic phase, gives perfectly Gaussian entropy
perturbations and vanishing curvature perturbations; third, a
conversion from entropy to curvature perturbations, including a bend
in the trajectory at the end of the ekpyrotic phase, or a conversion
at the bounce via modulated (p)reheating.  The magnitude of
non-Gaussianities is dependent on the non-linearity of the conversion
mechanism, as explained in the previous sections, and thus model
dependent, but it can be below the bounds imposed by P{\footnotesize
  LANCK}.

Non-Gaussianities in a nonsingular {\it matter bounce} were studied in
\cite{Cai:2009fn}. The {\it matter bounce} referred to here deals with
fluctuations generated as quantum vacuum perturbations which exit the
Hubble radius during a matter-dominated contracting phase
\cite{Finelli:2001sr,Wands:1998yp,Allen:2004vz,Peter:2008qz}.  The
amplitude and shape of the three-point function is computed. The local
non-linearity parameter arising from the adiabatic mode is
$f_{_\mathrm{NL}}^\mathrm{local}=-35/8\simeq -4.3$. Its large value,
compared to inflationary predictions, is caused by the growth of
adiabatic fluctuations after Hubble crossing during the contracting
phase. As with the perturbation spectrum itself, in order to compare
the non-Gaussianities, here calculated in the contracting phase, with
actual observations, one needs to evaluate how they pass through the
bounce itself. This is the subject of the following section.

\subsection{Getting perturbations through a bounce}
\label{modemixing}

In order to agree with observations, it is often assumed that a nearly
scale-invariant spectrum of curvature perturbations $\zeta$ is
generated during the contracting phase; in Sec.~\ref{sis} we discussed
several possible ways to generate such a scale-invariant spectrum.
However, the question remains whether it emerges unscathed after the
bounce.

After Hubble crossing, $\zeta$ is frozen, at least as long as the
evolution takes place during an adiabatic regime. However, as the
bounce is approached, modes become sub-Hubble again between
$\eta_\mathrm{hc-entry}$ and $\eta_\mathrm{hc}^{(2)}$, see
Fig.~\ref{spacetimeNSMB} and Fig.~\ref{fig:4}; during this interval,
modes can evolve in a way that differs from their ``frozen''
super-Hubble evolution, requiring a careful analysis of the
perturbations during the actual bounce. If the bounce is fast enough,
the naive intuition that perturbations are left unchanged can be
correct, but counter examples exist (and in particular, the idea
according to which scales larger than the bounce duration cannot be
affected for ``causality'' reasons is wrong, as the horizon is made
much larger than any relevant scale \cite{Martin:2003bp}).

To illustrate possible stumbling blocks, consider the original
ekpyrotic scenario \cite{Khoury:2001zk} as a case study, where it was
shown that the Bardeen potential $\Phi$ inherits a scale-invariant
spectrum, and it was argued that it remains unaltered during the
bounce \cite{Khoury:2001zk}. This argument was based on the fact that
the dominant mode\footnote{In an expanding or contracting universe,
  the second order differential equation of a perturbation variable
  has two solutions, a dominant and a sub-dominant one, the latter, by
  definition, becoming less and less important as time
  progresses. This is not always the case, as discussed in
  Sec.~\ref{Fatal_effects}.} of the Bardeen potential carries the
desired spectrum.  In this scenario the pre- and post-bounce solutions
are glued together at some hypersurface. The resulting matching
conditions in \cite{Khoury:2001zk} were criticized in
\cite{Martin:2002ar}. If one employs the Deruelle-Mukhanov matching
conditions \cite{Deruelle:1995kd,Hwang:1991an} on a constant field
hypersurface, one finds that the dominant mode in the contracting
phase only couples to the sub-dominant one after the bounce
\cite{Brandenberger:2001bs}. See also
\cite{Hwang:2001ga,Tsujikawa:2001ad}. As a consequence, as we have
seen in Sec.~\ref{isocurvature}, the curvature fluctuation $\zeta$
inherits a blue spectrum \cite{Lyth:2001nq}.

Evidently, the resulting spectrum depends crucially on the type of
matching surface \cite{Durrer:2002jn}. In addition, some variables may
become singular, as pointed out in \cite{Lyth:2001nv}, see
Sec.~\ref{viabilityofptb}.

In a nonsingular bounce, certain variables may still grow sufficiently
large as to eventually behave in a non-perturbative way, but the
resulting spectrum of cosmological
fluctuations is at least in principle unambiguous: one simply needs to
follow a well-behaved fluctuation variable throughout the bounce (or
perform a non-perturbative analysis). In general, the curvature
fluctuation $\zeta$ after the bounce inherits components from both
modes, dominant and sub-dominant, of the contracting phase in varying
degrees.  The detailed form of this $k-$mode mixing matrix is model
dependent, see \cite{Martin:2003sf,Martin:2004pm,Bozza:2005xs}.

This problem does not arise for tensor modes since those always remain
below their potential, \ie $a''/a$, as shown in Eq.~(\ref{gralsoln})
and Fig.~\ref{spacetimeNSMB}. Thus, the spectrum of gravitational
waves is largely insensitive to the bounce.

In the next section we discuss potentially dangerous instabilities, in
particular the possible regrowth of the sub-dominant mode during the
contracting phase in specific models.

\section{Potentially fatal effects undermining nonsingular models}
\label{Fatal_effects}

\begin{figure*}[tb]
\includegraphics[scale=0.65]{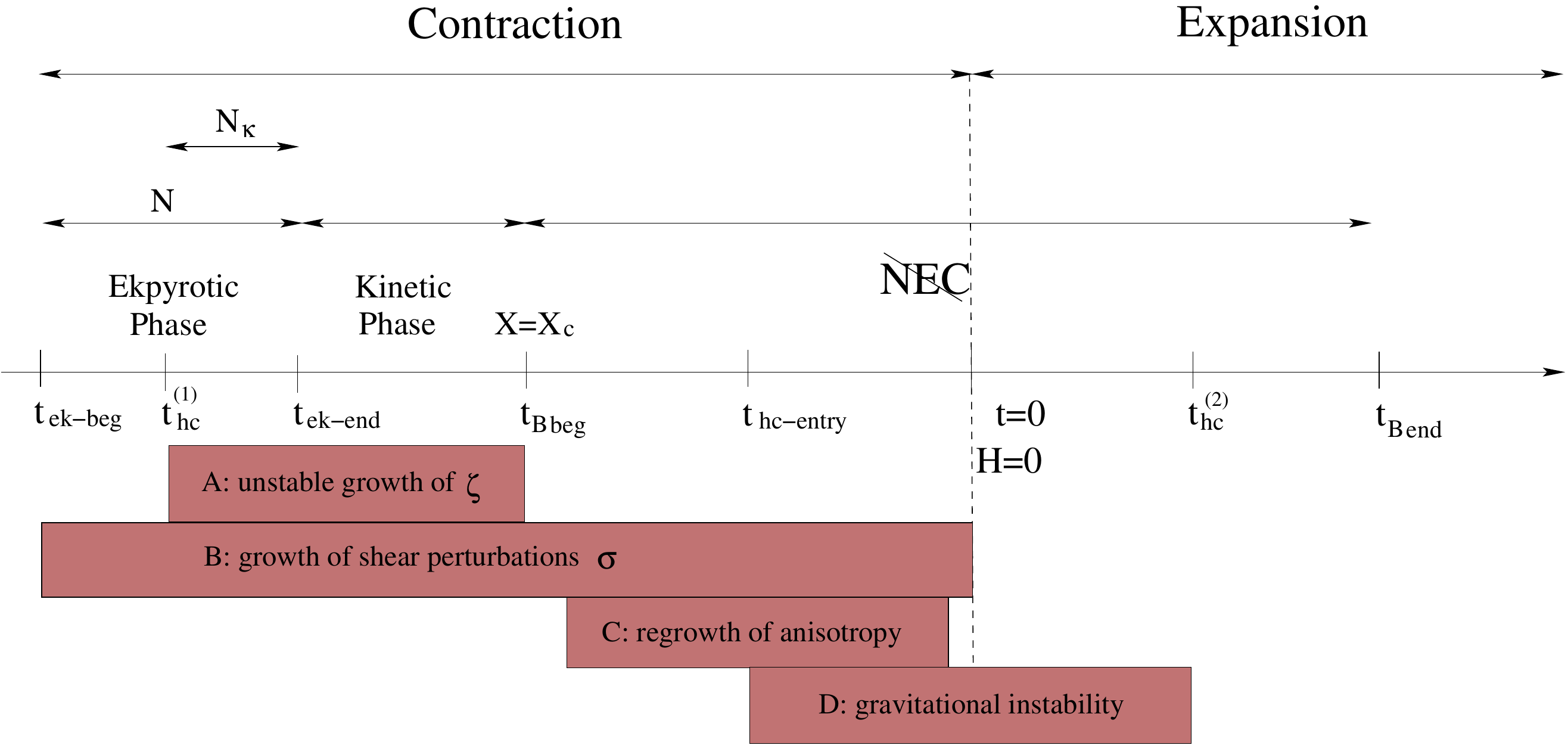}
\caption{Dangerous instabilities undermining a nonsingular bounce
  according to \cite{Xue:2011nw}. Here, $N\equiv
  \ln(a_{\mathrm{ek-beg}}/ a_{\mathrm{ek-end}})$ is the total number
  of e-folds during the ekpyrotic phase and $N_k\equiv \ln (
  a_{\mathrm{hc}}^{(1)}/ a_{\mathrm{ek-end}})$ is the number of
  e-folds of ekpyrosis left after the mode with wavenumber $k$ has
  crossed the Hubble scale. The beginning and end of the ekpyrotic
  phase are represented by $t_{\mathrm{ek-beg}}$ and
  $t_{\mathrm{ek-end}}$ respectively, while $t_{\mathrm{Bbeg}}$ and
  $t_{\mathrm{Bend}}$ denote the beginning and end of the bouncing
  phase. The modes first leave the Hubble length at
  $t_{\mathrm{hc}}^{(1)}$, re-enter at $t_{\mathrm{hc-entry}}$ and
  leave the Hubble radius for a second time at
  $t_{\mathrm{hc}}^{(2)}$.}
\label{fig:4}
\end{figure*}

We follow \cite{Xue:2010ux,Xue:2011nw,BKthesis} to investigate four
effects that can undermine the success of nonsingular models:
\renewcommand{\theenumi}{\Alph{enumi}}
\begin{enumerate}
\item unstable growth of curvature fluctuations in the adiabatic mode,
\item growth of quantum induced anisotropy for vector perturbations
  and scalar shear,
\item regrowth of initial anisotropy as sub-dominant modes become
  dominant,
\item gravitational instability during the bounce,
\end{enumerate}
\renewcommand{\theenumi}{\arabic{enumi}}

 Throughout this section, we work in the framework of the new
ekpyrotic scenario as a concrete case study and use the following
notation: $P(X)$ is the kinetic function, $X\equiv \frac12\dot\phi^2$,
$T$ is the kinetic energy, and a subscript ``c'' as in
$X_{\mathrm{c}}$ denotes the ghost condensate point. We assume that
the ghost condensation occurs for small $X$, so that the kinetic
energy is vastly smaller during the bounce than at the end of the
ekpyrotic phase.  A subscript $1$ denotes any time during the bounce,
$\zeta^\mathrm{adiab}$ and $\zeta_\mathrm{s}$ are the adiabatic
curvature perturbation and the curvature perturbation generated via
the entropic mechanism respectively. We call $N_{k}$ the number of
e-folds of ekpyrosis left after the mode with wavenumber $k$ has
crossed the Hubble radius and $N$ is the total number of e-folds
during the ekpyrotic phase. The shear is $\sigma$: $\sigma^\mathrm{v}$
is the vector perturbation arising from the shear perturbation
$\sigma^{\mathrm{s}}_{ij}$ and $\sigma^\mathrm{s}$ is the scalar part
of the shear in the synchronous gauge.  The beginning and end of a
bounce phase are represented respectively as $B_\mathrm{beg}$ and
$B_\mathrm{end}$. Similarly, ``$\mathrm{ek-beg}$'' and
``$\mathrm{ek-end}$'' denote the beginning and end of the ekpyrotic
phase, while $t_\mathrm{Bbeg}$ and $t_\mathrm{Bend}$ denote the
beginning and end of the bouncing phase. In \cite{Xue:2011nw}, the
instabilities are considered only in the contracting phase. Our
notation differs from \cite{Xue:2011nw} as we consider contraction and
expansion. The reader should take note that some of these problems can
be avoided if ghost condensation occurs for large $X$, as in the
matter bounce model described in Sec.~\ref{workingmodel}. 

\subsection{Unstable growth of curvature fluctuations}
\label{ugcp}

The potentially unstable growth of curvature perturbations in
nonsingular bouncing models may endanger their validity
\cite{Xue:2010ux,Xue:2011nw}.  This threat comes in the form of a
subdominant adiabatic mode that crosses its potential before the
ekpyrotic phase is over, grows exponentially and ultimately gives rise
to a blue spectrum, precluding the scale-invariant contribution to the
temperature fluctuations in the CMBR.  To see how this happens, we
want to compare the resulting adiabatic perturbations to the
isocurvature ones, by calculating the evolution of $\mathcal R$ in the
transition phase between the ekpyrotic and the bouncing one.  The
comoving curvature perturbations $\mathcal R_k$, labeled by the
comoving wave number $k$, obey the equations of motion,
\begin{equation}
  \mathcal{R}_k''+2\frac{z'}{z}\mathcal{R}'_k+c_{_\mathrm{S}}^2k^2
  \mathcal{R}_k=0\,,
  \label{eom}
\end{equation} 
where $z\equiv a\sqrt{2\dot H/c_{_\mathrm{S}}^2H^2}$. The sound speed
is
\begin{equation}
c_{_\mathrm{S}}^2=\frac{P_{,X}}{2XP_{,XX}+P_{,X}}\approx
1\,.\label{soundspeed}
\end{equation}
For small $k$, the solution to (\ref{eom}) is
\begin{equation}
\mathcal{R}_k=C_1(k)+C_2(k)\int\frac{\dd\eta}{z^2}\,,\label{soln}
\end{equation}
where $C_1\sim1/\sqrt{k}$ and $C_2\sim\sqrt{k}$; the first term of
(\ref{soln}) is a constant solution with a blue spectrum; the second
one is a decaying solution with a bluer spectrum and always
ignored. Following \cite{Xue:2010ux,Xue:2011nw} we look at the second,
integral term to show how this initially sub-dominant term can be
amplified to eventually dominate over isocurvature perturbations;
plugging in $z$, we have
\begin{equation}
\mathcal{R}_k^\mathrm{adiab}=C_2(k)\int{\frac{\dd\eta}{z^2}}=
C_2(k)\int{\frac{c_{_\mathrm{S}}^2H^2}{a^2(-2\dot H)}}\dd \eta\,.
\end{equation}
Using (\ref{soundspeed}) along with $\dot H=-XP_{,X}$ and
$T_{,X}=P_{,X}+2XP_{,XX}$ as detailed in \cite{Xue:2011nw}, the
adiabatic contribution to the curvature perturbation
\begin{equation}
\mathcal{R}^\mathrm{adiab}_k\approx
\frac{C_2(k)}{3a_{\mathrm{ek-end}}^3}
\frac{V_{\mathrm{c}}}{(-V_{,\phi_{\mathrm{c}}})}\frac{1}{\sqrt{2X_{\mathrm{c}}}}\,,
\end{equation}
is greatly amplified. The ratio of this mode to the curvature
perturbation $\mathcal{R}_s$ produced via the entropic mechanism is
\begin{equation}
  \frac{\mathcal{R}^\mathrm{adiab}}{\mathcal{R}_\mathrm{s}}\sim \ex^{N-2N_k},
\end{equation}
where $N_k$ is the number of e-folds of ekpyrosis left after the mode
with wavenumber $k$ has crossed the Hubble scale and $N$ is the total
number of e-folds during the ekpyrotic phase.

Since $N_k\sim 10$ for modes observable in the CMBR, this regrowth of
perturbations can be problematic for $N\gtrsim 60$.  One might hope to
alleviate this problem by making
$-V_{,\phi_{\mathrm{c}}}/V_{\mathrm{c}}$ exponentially large; however,
this is impossible for modes observed in the CMBR due to the COBE
normalization, see (\ref{COBEp}) where the exponential ekpyrotic
potential in (\ref{Pekpyrotic}) is used. Other solutions to this
problem must therefore be found.

\subsection{Growth of quantum induced anisotropy}
\label{quatuminducedanisotropy}

Starting with a homogeneous universe, do vector perturbations and
scalar shear perturbations generated by quantum fluctuations dominate
the energy density and prevent a nonsingular bounce?  Under
approximations made specifically for a bounce generated via a ghost
condensate, namely, that both $T$, the kinetic energy and
$X=\dot\phi^2/2$ are monotonically decreasing, and $V_{,\phi}\approx
V_{,\phi_{\mathrm{c}}}$ as well as $T\approx2XP_{,X}$, the Friedmann
equations during the bouncing phase
\begin{eqnarray}
3H^2&=&\rho_{\phi}+\sigma^2\approx T+V_{\mathrm{c}}+\sigma^2,\\
\dot H&=&-XP_{,X}-\sigma^2\approx\frac{-T}{2}-\sigma^2,
\end{eqnarray}
show that indeed it would be hard to get rid of shear perturbations
$\sigma^2$.  The potential energy at which the NEC is violated is
denoted by $V_{\mathrm{c}}$ and $P(X)$ is the kinetic term. Assuming
that $X$ and thus $T$ is monotonically decreasing during the bounce,
one can derive a necessary condition for a ghost condensate to occur
by requiring that $H^2$ vanishes at some point. Taking into account
anisotropy $\sigma$ one needs to have,
\begin{equation}
\sigma_1^2\lesssim
\frac{|T_1|}{2}\ex^{\frac{-2\rho_1}{|T_1|}-1},\label{condbounce}
\end{equation}
for some time $t_1$ during the bounce.  If this bound is satisfied,
the BKL-unstable behavior is avoided and a smooth bounce results,
whereas a violation most likely indicates a failed bounce. It is
expected that a similar condition has to be satisfied for any type of
nonsingular bounce \cite{Xue:2011nw}.

To investigate sources of anisotropy, consider the general perturbed
metric, containing vector and tensor modes as well as scalars,
\begin{eqnarray}
\dd s^2&=&a^2(\eta)\left\{-(1+2A)\dd\eta^2+2\left(B_{,i}+S_{i}\right)\dd\eta \dd
x^i \right.\nonumber\\
&&+\left. \left[\left(1-2\psi\right)\delta_{ij}+2E_{,ij}+2F_{ij}+h_{ij}\right]
\dd x^i\dd x^j
\right\}\,. 
\end{eqnarray}
There are two common ways in which anisotropies arise
\cite{Xue:2011nw}. The first stems from non-zero, gauge-invariant
vector perturbations arising from the shear perturbation at constant
time hypersurfaces,
\begin{equation}
\sigma_i^\mathrm{v}\equiv F_i'-S_i\,,
\end{equation}
which grow as the scale factor decreases during the ekpyrotic phase as
\cite{Battefeld:2004cd}
\begin{equation}
  \sigma_i^\mathrm{v}\propto\frac{1}{a^2}\,.\label{vector}
\end{equation}
However, if we start with a universe initially devoid of vector
perturbations and only scalars are present, this hurdle can be
overcome as scalars do not source vector perturbations
\cite{Battefeld:2004cd}. This is possible as vector perturbations, not
being dynamical in this context, do not need to have non-vanishing
initial conditions coming from, \eg quantum vacuum fluctuations.  The
second source of anisotropies is due to the equation of state
parameter $w$ passing through $-1$.  The scalar part of the shear
perturbation is
\begin{eqnarray}
\sigma_{ij}^\mathrm{s}&=&a\left[ \left( E'_{,ij}-B_{,ij}\right) - 
\frac{1}{3}\delta_{ij}\nabla^2\left(E'-B\right) \right]\nonumber\\
&\equiv& a\left( {\sigma^\mathrm{s}_{,ij} -
\frac{1}{3}\delta_{ij}\nabla^2\sigma^\mathrm{s}}\right).
\end{eqnarray}
Unlike $\sigma^\mathrm{v}$, $\sigma^\mathrm{s}$ is not gauge invariant
and its evolution is coupled to the comoving curvature pertubation
$\zeta$. This would imply that the comoving shear and curvature
perturbations feed off each other and grow from quantum fluctuations.
However, in the synchronous gauge, $\zeta$ and the shear perturbation
stay small throughout the contracting phase till near the bounce. The
comoving and synchronous shear perturbations, $\sigma^\mathrm{c}$ and
$\sigma^\mathrm{s}$, are related, so that one can translate one to the
other \cite{Xue:2011nw}.

In the synchronous gauge\footnote{The subscript $\cdot_{(\mathrm{s})}$ denotes
  the synchronous gauge, whereas, the superscript $\cdot^\mathrm{s}$ denotes the
  scalar part of the shear.} one obtains
\begin{equation}
\sigma_{(\mathrm{s})ij}=\frac{1}{a}\left[\sigma_{(\mathrm{s}),ij}-\frac{1}{3}\delta_{ij}
\nabla^2 \sigma_{(\mathrm{s})}\right]\propto\frac{1}{a^3},   
\end{equation}  
which shows similar growth as the shear from vector perturbations in
(\ref{vector}). Since scalar shear perturbations are continuously
sourced during the contraction, this shear cannot be fined tuned away.
A lower estimate of the shear anisotropy at the end of the bounce can
be computed via the self-correlation function of $\sigma^\mathrm{s}$
integrated from modes that are deeply blue to modes that are above
their potential till near the bounce \cite{Xue:2011nw},
\begin{equation}
\langle(\sigma^\mathrm{s})^2\rangle_\mathrm{Bend}
\sim\frac{a^6_\mathrm{Bbeg}}{a^6_{t=0}}
\ex^{2N}H^4_\mathrm{ek-end}. \label{shear}
\end{equation}  
Here, $t=0$ denotes the turnaround between contraction and
expansion. Consider a case study where the bounce is produced via a
ghost condensate. In order to overcome the anisotropy, the condition
for a bounce within this model is (\ref{condbounce}), which requires
that
\begin{equation}
\sigma_{t=0}^2\lesssim\frac{|T_{t=0}|^2}{2V_\mathrm{c}}\ex^{-1}\sim
\frac{V_\mathrm{c}}{2\ex}, \label{sigmaend}
\end{equation} 
which can be satisfied only if (recall we are using units in which the
reduced Planck mass is unity)
\begin{equation}
V_\mathrm{c}\lesssim \ex^{-4N}.
\end{equation}
This value is in contradiction with the potential energy required by COBE at the
ekpyrotic phase \cite{Lehners:2007ac}, 
\begin{equation}
V_\mathrm{c}\sim 3pV_\mathrm{ek-end}\sim p^2\times 10^{-6},\label{COBEp}
\end{equation}
where $p\sim 10^{-2}$ sets the exponent of the ekpyrotic potential in
(\ref{Pekpyrotic}) \cite{Khoury:2003vb}.  As such, the anisotropy
arising from the scalar shear can dominate the energy density before
the bounce, leading to BKL-like contractions, and prevent a bounce.

Working in the synchronous gauge, the computation stays perturbative
(in contrast to the comoving gauge perturbation calculation), but the final result is
still gauge dependent. In \cite{Xue:2011nw} it was checked that the
bounce is spoiled in two other gauges, the uniform Hubble gauge and
the longitudinal gauge. In the latter, shear is absent, but the bounce
is still spoiled by the appearance of large velocity
perturbations. Hence, although it appears that this problem is
physical, a full gauge-invariant computation has not been performed
yet, so no definite conclusion can be drawn.

\subsection{Regrowth of initial anisotropy}
\label{initialanisotropy}

A detailed analysis of a ghost condensate-mediated bouncing phase in
\cite{Xue:2010ux,Xue:2011nw,BKthesis} shows that the initial
anisotropy originally quelled during the ekpyrotic phase overtakes the
scalar field energy when $w<-1$ during the bouncing phase. Following
the ekpyrotic phase, the curvature and anisotropy are suppressed by
\begin{equation}
\frac{H^2_\mathrm{ek-end}}{H^2_\mathrm{ek-beg}}\equiv
\ex^{2N}.\label{ekpyroticsuppression}
\end{equation}
To quantify the duration of the bounce, Xue \etal \cite{Xue:2011nw}
studied three different stages: at the beginning of the bounce the
kinetic energy is negligible and
$|H|\approx|H_\mathrm{c}|=\sqrt{V_\mathrm{c}/3}$. Subsequently, the
friction term becomes dominant due to an increase in the negative
kinetic energy. Close to the bounce $T\approx -V_\mathrm{c}$, but
since the Hubble rate is small, friction is again negligible. Of these
three phases, the first and third ones are brief. Thus, the duration
of the bounce phase can be approximated by
\begin{equation}
\Delta T_\mathrm{b}\approx -t_\mathrm{Bbeg}\approx
\frac{N}{3H_\mathrm{c}}\,,\label{bounceduration}
\end{equation}
where $H_\mathrm{c}^2=V_\mathrm{c}/3$.  This result shows that it is
not possible to complete the bounce in just a few Hubble times,
leading to a growth of anisotropies.  Hence, the scale factor $a(t)$
scales as
\begin{equation}
a\propto|T|^{-1/3},
\end{equation}
and it contracts as
\begin{equation}
\frac{a_{t=0}}{a_\mathrm{Bbeg}}=\left|
\frac{T_\mathrm{Bend}}{T_{t=0}}\right|^{-\frac{1}{3}}\lesssim
\ex^{-\frac{1}{3}N}\,. \label{contract}
\end{equation}
Therefore anisotropies increase by a factor of
\begin{equation}
\frac{\sigma^2_{t=0}}{\sigma^2_\mathrm{Bbeg}}\gtrsim \ex^{2N}\,,\label{anis}
\end{equation}
which cancel the original anisotropy suppression experienced during
ekpyrotic phase (\ref{ekpyroticsuppression}). As such, anisotropy
persists, undermining the bouncing phase of the model at hand, unless
the initial anisotropy is fine-tuned to,
\begin{equation}
\sigma^2_\mathrm{ek-beg}\lesssim X_\mathrm{c}\,.
\end{equation}
This regrowth of anisotropy appears generic in models containing a
prolonged friction-dominated phase with $T<0$ as evident from
equations (\ref{contract}) and (\ref{anis}). See section
\ref{cyclic_ic} for an example of how the cyclic universe eradicates
the initial anisotropy via the presence of a Dark Energy phase before
the ekpyrotic one.

 \subsection{Gravitational instability}
\label{gravitationalinstability}
 
During the bounce, modes re-enter the Hubble sphere briefly, where
they may grow unstable  if $c_{_\mathrm{S}}^2<0$. The latter is
not a necessary condition for a bounce to occur, but it is often the
case. The presence of an instability becomes  evident by
considering the equation of motion of the Mukhanov-Sasaki variable
 \begin{equation}
 v_k''+\left(c_{_\mathrm{S}}^2k^2-\frac{z''}{z}\right)v_k=0\,,
 \end{equation}
 where $v_k=z\mathcal R_k$.
 If we let
 \begin{equation}
|c_{_\mathrm{S}}^2k^2|>\left|\frac{z''}{z}\right|\,,
\end{equation} 
then
\begin{equation}
v_k''+c_{_\mathrm{S}}^2k^2v_k=0\,,
\end{equation}
can be solved to
\begin{equation}
v_k\propto \ex^{|c_{_\mathrm{S}}|k\Delta\eta}v_{k_0}\,.
\end{equation}
The resulting instability could be alleviated if $|c_{_\mathrm{S}}^2|$
were sufficiently small and/or the bounce duration in conformal time
$\Delta\eta$ were not too large.

For a bounce mediated via a ghost condensate, the time interval inside
the Hubble radius,
 \begin{equation}
 |c_{_\mathrm{S}}|k\Delta\eta\sim|c_{_\mathrm{S}}|\ex^{-\frac{N}{3}-N_k}\,,
 \end{equation}
 represents a problem for modes with $N_k<N/3$.  Hence to avoid the
 instability, the speed of sound has to be
 \begin{equation}
 |c_{_\mathrm{S}}^2|\lesssim \ex^{-\frac{2}{3}N}\,.
 \end{equation}
 
\section{Topics for future research interest}

The attempts to investigate bouncing cosmologies as an alternative to
the inflationary paradigm have been riddled with difficulties,
roadblocks and no-go theorems. As a consequence, models have become
rather complicated to avoid the many pitfalls. Nevertheless, use of
newly proposed ingredients such as galileons or string gases have led
to a few models that, although not free of problems, appear promising
and thus offer some hope.

We hope to have given a critical, yet unbiased
assessment of failures and successes.  Focusing on the few surviving
candidates, again all based on a classical description of gravity, we
would like to outline a possible road-map for future research that
might be of interest to  scientists  working on bouncing
cosmologies.

\subsection{Preheating and reheating}\label{preheat}

The nature of reheating the early Universe is an active area of
research in inflationary cosmology with a long history
\cite{Kofman:1997yn,Bassett:2005xm}.  See for example
\cite{Meyers:2013gua} for recent work on perturbative reheating after
multi-field inflation and \cite{Zhang:2013asa,Underwood:2013pwa} for
preheating in DBI inflation. Reheating is a process whereby the cold
post-inflationary universe attains the high temperature needed for
nucleosynthesis. The inflaton decay can occur either perturbatively or
via instabilities and/or resonances as in preheating.  Perturbative
reheating is almost always hampered by the incomplete decay of the
inflaton which may spoil nucleosynthesis \cite{Braden:2010wd}. Most
studies of preheating focus on canonical scalar fields
\cite{Battefeld:2008bu,Battefeld:2007st}, but little attention has
been paid to the decay of non-standard fields such as scalars with
more general kinetic terms or galileons, among others. Notable
exceptions are a study of preheating in DBI inflation based on lattice
simulations \cite{Child:2013ria} as well as the rapid decay of a
galileon in galileon genesis \cite{LevasseurPerreault:2011mw}. Another
focus has been reheating in Starobinsky's model of $R^2$-inflation
\cite{Starobinsky:1980te,Rudenok:2014daa} as well as in models of
Higgs inflation \cite{Mazumdar:2010sa} and the MSSM
\cite{Allahverdi:2010xz}.

In the framework of cosmic bounces, this phase has hardly been studied
at all, since the bounce physics itself was poorly understood. Early
ideas entail the transfer of kinetic energy during the brane collision
in singular ekpyrotic models \cite{Steinhardt:2001st,Lehners:2011kr},
the transfer of the ghost condensate potential energy due to a steep
drop of the potential and the assumed coupling to other degrees of
freedom in new ekpyrosis \cite{Buchbinder:2007ad}, or the application
of inflationary reheating/preheating lore after a bounce
\cite{Battefeld:2007st}. In the context of a nonsingular bounce with a
matter dominated contracting phase caused by an oscillating scalar
field, stochastic resonance, \ie preheating, was studied in
\cite{Cai:2011ci}, where it was shown that preheating can be more
efficient: resonances can commence in the contracting, matter
dominated phase and continue throughout the bounce, in effect doubling
the period of preheating. However, only canonical scalar fields were
considered for preheating dynamics, leaving the NEC violating field
out of this process.  In the framework of a singular bounce mediated by a
brane collision, it was speculated that the presence of additional
light degrees of freedom before a singular bounce would render the
bounce nonsingular and reheat the universe subsequently via the
intermediately produced light degrees of freedom. The S-brane bounce
may be viewed as a realization of this idea.

While several mechanisms to generate a nearly scale-invariant spectrum
of perturbations are known, and models differ in this regard, we have
encountered only two distinct avenues of inducing a nonsingular
classical bounce without introducing fatal instabilities (but not
entirely devoid of problems):
\begin{enumerate}
\item via a galileon/ghost condensate, as in the matter bounce
  scenario \cite{Cai:2012va}, the super-bounce \cite{Koehn:2013upa} or
  the non-minimal entropic bounce \cite{Qiu:2013eoa,Li:2013hga},
\item via a thermal string gas in the S-brane bounce
  \cite{Brandenberger:2013zea} or the Hagedorn phase in string gas
  cosmology \cite{Nayeri:2005ck}.
\end{enumerate}
In either case, (p)reheating has not been studied, but it is possible
to identify the challenges ahead.

In the first case (employing galileons/ghost condensates), it is
necessary to ensure that any coupling to matter fields does not lead
to a premature decay of the bounce-inducing ingredient, while ensuring
that it decays sufficiently before nucleosynthesis. Since it is
unknown if preheating dynamics are operational for galileons, studies
investigating preheating should be conducted, including lattice
simulations if preheating is feasible. Further, in the presence of
additional isocuvature fields that carry the scale-invariant spectrum,
their decay products have to dominate over the galileon's. Once a
better understanding of the thermal history is available, the presence
of thermal relics and the effect onto non-Gaussianities will have to
be investigated.

In the second case, a thermal component, a string gas modeled by a
scalar in the S-brane bounce, is present, with a temperature close to
the Hagedorn one.  Naively, one should expect the production of a
large number of different particle species in this regime, not only
standard model ones. Thus, it is crucial to identify mechanisms that
predominately lead to standard model particles after the bounce. The
production of thermal relics, such as gravitinos, at these high
temperatures is an example of this challenge, see
Sec.~\ref{relics}. In addition, while it can be argued that the
thermal string gas may be modeled by a scalar field
at the background and
even the perturbed level, it is probably an oversimplification to use
such a setup to discuss (p)reheating dynamics, much like effective
single field models of inflation are insufficient to study preheating
\cite{Battefeld:2008bu}.

Evidently, reheating emerges as one of the most challenging and
ill-understood regimes of bouncing cosmologies, even more so than in
inflationary setups, offering many new challenges for future research.

\subsection{An implementation within string theory}\label{implementationST}

Great advances have been made to implement inflationary cosmology in
string theory, see. \eg
\cite{Cline:2006hu,McAllister:2007bg,Burgess:2011fa,Baumann:2014nda}
for reviews. Crucial components of any implementation are the
identification of the dynamical degrees of freedom with their
microphysical counterparts, such as a brane separation turning into an
inflaton, as well as a thorough understanding of moduli
stabilization\footnote{For a recent assessment on moduli stabilization
  in Type IIB string theory, see \cite{Dasgupta:2014pma}, where
  necessary conditions at the full $10$D level were derived.}, such as
the shape and size of extra dimensions. The inflaton potential's
sensitivity to quantum corrections emerged as a roadblock for many
setups, which led to a preference of small field models or highly
symmetric scenarios such as monodromy inflation
\cite{Silverstein:2008sg}.

For bouncing cosmologies we are just at the beginning stages of
finding implementations within string theory. Since galileon models
emerged over the last few years as promising phenomenological
candidates to provide a well-behaved cosmological bounce, the logical
next step should be to find concrete realizations of galileons in
string theory. In that regard the construction of a supersymmetric
version of a galileon setup as provided in \cite{Koehn:2013upa} is
encouraging.  Furthermore, once methods of realizing galileons in
string theory are found, suitable potentials have to be devised. Only
after such an implementation is found can questions pertaining to
moduli stabilization or quantum corrections be addressed. However,
these models do not generate measurable primordial gravitational waves
on CMBR scales and would therefore be ruled out if the B{\footnotesize
  ICEP2} result stands the test of time and no additional mechanism to
generate gravitational waves is present.

A model incorporating stringy degrees of freedom to generate a bounce
has been proposed in \cite{Brandenberger:2013zea} (the S-brane bounce)
making use of a string gas. Here the main simplification consists of a
gas approximation, which is further approximated by modeling the
thermal string gas by a scalar field.  Furthermore, the study stayed
at the level of the 4D effective theory. All of these approximations
should be relaxed. In the framework of string gas cosmology, the
dynamics of toroidal internal dimensions can be discussed in the
framework of dilaton gravity \cite{Battefeld:2004xw}. However, since
the bounce takes place in the Hagedorn regime, any discussion of
moduli stabilization is daunting\footnote{\label{stabilization}It has
  been proposed to use the string gas to stabilize internal dimensions
  as well as the dilaton
  \cite{Patil:2004zp,Patil:2005fi,Patil:2005nm,Danos:2008pv,Mishra:2011fc},
  see also
  \cite{Watson:2003gf,Watson:2004aq,Brandenberger:2005bd,Cremonini:2006sx,Liu:2011nw}. This
  proposal works within dilaton gravity and is not necessarily valid
  in the Hagedorn regime. Furthermore, since string gases redshift
  like matter, this stabilization mechanism is problematic at late
  times \cite{Battefeld:2006cn}, particularly in the presence of a
  cosmological constant \cite{Ferrer:2005hr}.}.

If bouncing cosmology is to provide an alternative to the inflationary
paradigm, and if string theory is the correct way to describe physics
at the highest energy level, then considerable improvements have to be
made to implement them in string theory.  To this end, a resolution of
the singularity should take center-stage. Only if such a resolution is
achieved will we be able to assess whether a galileon bounce, a string
gas bounce, or a singular antigravity bounce can provide a good
description of the early Universe.

\subsection{Spatial curvature and non-Gaussianities}\label{K1}

Spatial curvature is often neglected, and indeed, we mostly assumed
$\Ka=0$ throughout this review. The usual argument for this choice,
apart from the observational fact that curvature {\sl is} negligible
today, is that it should not dominate during a bouncing phase.  The
equivalent energy density being $\Ka / a^2$, curvature is always
subdominant in the Friedmann Eqs.~(\ref{FriedCosm}) when compared to
dust ($\rho_\mathrm{m} \propto a^{-3}$), radiation ($\rho_\mathrm{r}
\propto a^{-4}$) and in particular the shear ($\rho_\theta \propto
a^{-6}$), as discussed in Sec.~\ref{ekpyroticphase},  when $a\to
0$.  Now let the energy density near the bounce consist in two pieces,
$\rho_+>0$ and $\rho_-<0$ say, where $\rho_-$ denotes the component
whose negative contribution during the bouncing phase allows for the
bounce to actually take place in the framework of GR. The Friedmann
equation for the background (\ref{FriedCosm}), if valid until the
bounce, shows two possibilities: either $|\rho_-| \simeq \rho_+ \gg
\Ka/a^2$, in which case the curvature contribution is indeed
negligible at all times, including the bounce itself, or $\Ka/a^2
\simeq \rho_+ \simeq |\rho_-| \gg $, \ie all terms are of the same
order of magnitude. In the latter case, the curvature contribution
cannot be neglected, and the Friedmann equation then serves to
determine the actual value of the scale factor at the bounce. Although
perhaps very contrived and fine-tuned, this is the case we now
consider for the sake of completeness. 

Even though it can be written as a perfect fluid with equation of
state parameter $w=-1/3$ in the background equations, curvature is
actually not a fluid and cannot be perturbed; as a consequence, it
does not entail any dynamical degrees of freedom at the perturbed
level, but if $\Ka\not=0$, conservation laws, such as that of the
curvature perturbation $\zeta$, are not generically valid. As a
consequence, new variables should be considered instead, such as the
BST curvature variable $\zeta_{_\mathrm{BST}}$ \cite{Wands:2000dp}:
although the ``pure'' curvature perturbation $\zeta$ is no longer
conserved in the presence of spatial curvature, the new quantity,
which reduces to the former in the limit $\Ka\to 0$, can be conserved
on super Hubble scales under specific conditions which do not
necessarily hold in a bouncing scenario
\cite{Abramo:2007mp,Martin:2003bp}. One can note at this point that in
the so-called $K-$bounce scenario in \cite{Abramo:2007mp}, for
instance, the bounce
is explicitly performed in a spatially curved universe, the $\Ka\to 0$
limit being assumed regular; it might not necessarily be a valid
assumption depending on the specific form of the action
\cite{DeSantiago:2012nk}.

How spatial curvature can drastically modify a model's prediction is
illustrated in \cite{Martin:2003sf,Falciano:2008gt}: here, a simple
bouncing model is considered, based on a scalar field and
curvature. The former has a potential whose maximum is reached at the
bounce, which is canceled by the curvature contribution, such that
$H\to 0$ without violating the NEC. As a result, many instabilities
are avoided by construction. The effective potential entering the
equations of motion for perturbations is strongly
model-dependent. Furthermore, the amplitude and the spectral index of
curvature fluctuations strongly depend on dynamics during the
bounce. Thus, the mere presence of a curvature term can imply that the
observed spectrum is not necessarily the one produced during the
contracting phase.

The situation can be even more problematic if non-linear terms in the
perturbative expansion are considered: for the same model, an almost
de Sitter bounce can be a strong producer of non-Gaussianities, as
shown by one of the authors of this review and collaborators in
\cite{Gao:2014hea}. As a result, the prediction of non-Gaussianities
based on pre-bounce physics, as in \cite{Cai:2009fn}, may not be
sufficient. The latter would indicate that non-Gaussianities from the
matter bounce are usually small, while the former would predict
exactly the opposite.

As we have seen, curvaton-like models already produced reasonably
large non-Gaussianities from the conversion mechanism. If their
amplitude remains unaffected by the bounce physics, they can still be
affected by the subsequent phase of (p)reheating, which can change not
only the amplitude but also the sign of non-linearity parameters
\cite{Leung:2012ve,Leung:2013rza,Elliston:2014zea}. Thus, as
highlighted above, a thorough understanding of preheating is essential
for any comparison of predictions with current and forthcoming
high-precision data.

\subsection{Primordial gravitational waves}

Primordial gravitational waves have been propagating undisturbed to us
since their formation and particularly after decoupling, some
$380,000$ years after the big bang. On their way, they have imprinted
a particular signal on the CMBR, which induces vorticity in the
polarization field. This polarization pattern consists of a B-mode
component at angular degree that cannot be generated by primordial
density perturbations. The amplitude of this signal is given by the
tensor to scalar ratio, $r$, a function of the energy scale of
inflation. The B{\footnotesize ICEP2} experiment \cite{Ade:2014xna}
claims to have observed such gravitational waves, and if the
interpretation of this result is confirmed independently by other
experiments at different frequencies and on different patches on the
sky, this would support inflation as the standard paradigm of the
early Universe, see Sec. \ref{bicep2} for a summary of the ongoing
debate. Indeed, inflation predicts gravitational waves with amplitudes
determined by the energy scale at which inflation happened
\cite{Bassett:2005xm,Starobinskii:1979},
\begin{equation}
\rho^{1/4}=2.2\times 10^{16}\mathrm{GeV}\left(\frac{r}{0.2}\right )^{1/4},
\end{equation}
where $\rho$ is the energy density at the time of inflation. Hence,
according to this observation, inflation took place around the GUT
scale. In addition, however, via the Lyth bound \cite{Lyth:1996im},
the arc length of the inflationary trajectory would have to be
super-Planckian
\begin{equation}
\Delta\phi > M_{_\mathrm{Pl}},
\end{equation}
thus potentially raising theoretical questions regarding the use
classical GR at this stage.

As we have seen in Sec.~\ref{rcyclic}, the primordial gravitational
wave spectrum generated during an ekpyrotic phase, that is a slow
contraction, is blue with a strongly suppressed amplitude. Since all
of the promising proposals include such a contracting phase, with the
exception of the S-brane bounce and the Hagedorn bounce by T-duality
in string gas cosmology, see Sec.~\ref{stringy}, they would be ruled
out in the absence of any additional mechanism to yield such a
primordial gravitational wave spectrum. While an ekpyrotic phase was
not incorporated in the S-brane bounce, the latter suffers from
instability problems, see Sec.~\ref{BKL}, exactly due to the absence
of an ekpyrotic phase. If we ignore these problems and focus on a
matter dominated contracting phase, see
Sec.~\ref{adiabatic_fluctuations_mb}, the spectrum of gravitational
waves is scale invariant but can be in excess of current bounds
$r\sim\mathcal {O} (30)$ \cite{Allen:2004vz,Cai:2008qw}. Thus, only a
few models are potentially in agreement with the
{B\footnotesize{ICEP2}} experiment: the string gas model (the Hagedorn
bounce, not the S-brane bounce) sketched in Sec.~\ref{stringy} (see
however \cite{Kaloper:2006xw}), the matter bounce curvaton
\cite{Cai:2011zx} and the new matter bounce \cite{Cai:2014xxa}, since
they can accommodate the indicated amplitude of gravitational waves.
String gas cosmology also predicts that the tensor spectral index is
blue, opposite to the red one in inflationary cosmology, so that one
may be able to discriminate between these frameworks. The Hagedorn
bounce however has its own problems for the reasons previously
discussed (flatness, relic problem, \etc).

\vskip1cm 

All in all, it appears that most classical gravity based bouncing
models might be in trouble for one reason or another (\eg the observed
red tilt in the spectrum, the lack of non-Gaussianities, primordial
tensor modes, some questions of stability and background
compatibility), so that the search for a complete model is
ongoing. Quantum gravity effects, viewed from this perspective, could
provide more effective solutions in the future.

To alleviate the difficulties highlighted in this review often
requires complicating the model.  The fact that a working model is
complicated does not necessarily mean that it is incorrect vis-\`a-vis
the way the Universe actually evolved; although Ockham's razor demands
that we look for a simple theoretical framework, it does not imply
that a model's veracity relies on its simplicity. For example, as
Copernicus proposed the heliocentric model in lieu of Ptolemic's
geocentric one, he included more epicycles due to his insistence on
using circular orbits. Thus, his model could have been the victim to
Ockham's razor. Only subsequent improvements to his model,
particularly the use of ellipses, led to the simple Keplerian model we
know today, which is still an approximation to the full solution in
General Relativity. Bouncing cosmologies may be at a similar stage,
where simplicity, if present, is not yet apparent. Thus, we should
strive to extract distinct predictions of bouncing cosmologies and
confront them with experiments, while simultaneously aiming to improve
the conceptual underpinning. We hope the present review of pros and
cons can be helpful in achieving these goals.

\section*{Acknowledgments}
  It is a pleasure to thank Thorsten Battefeld, 
  Martin Bojowald, Robert Brandenberger,
  Yi-Fu Cai, Jean-Luc Lehners, Andrei Linde, Lubo\v{s} Motl, Jérôme
  Martin, Hermann Nicolai, Suboth Patil, Nelson Pinto-Neto, Paul Steinhardt, Sandro
  Vitenti and BinKan Xue for comments, suggestions and discussions. We
  would also like to thank M\'onica Forte, Miguel Sousa Costa, Jorge Luis
  Cervantes Cota, Anupam Mazumdar, Marco Peloso, Yun-Song Piao,
  Parampreet Singh, Martin S. Sloth and Gregory Vereshchagin for comments
  on the draft.


\end{document}